\documentclass[onecollarge]{svjour2}       
\usepackage{graphicx}
\usepackage{amssymb}
\usepackage{amsmath}
\usepackage{color}
\def\defeq{\;\buildrel\hbox{\small def}\over{\,=\;}}

\begin{document}
\title{Tidal synchronization of an anelastic multi-layered body: Titan's synchronous rotation.}
\subtitle{\today\ }
\author{Hugo Folonier \textperiodcentered \ Sylvio Ferraz-Mello}

\institute{H. A. Folonier and S. Ferraz-Mello \at \emph{Instituto de Astronomia Geof\'isica e Ci\^encias
Atmosf\'ericas, Universidade de S\~ao Paulo, S\~ao Paulo, Brasil} \\
\email{folonier@usp.br} and \email{sylvio@iag.usp.br}\\}

\titlerunning{}

\maketitle
\begin{abstract}

Tidal torque drives the rotational and orbital evolution of planet-satellite and star-exoplanet systems. This paper presents one analytical tidal theory for 
a viscoelastic multi-layered body with an arbitrary number of homogeneous layers. Starting with the static equilibrium figure, modified to include tide and 
differential rotation, and using the Newtonian creep approach, we find the dynamical equilibrium figure of the deformed body, which allows us to calculate 
the tidal potential and the forces acting on the tide generating body, as well as the rotation and orbital elements variations. In the particular case of 
the two-layer model, we study the tidal synchronization when the gravitational coupling and the friction in the interface between the layers is added. For 
high relaxation factors (low viscosity), the stationary solution of each layer is synchronous with the orbital mean motion ($n$) when the orbit is circular, 
but the rotational frequencies increase if the orbital eccentricity increases. This behavior is characteristic in the classical Darwinian theories and in 
the homogeneous case of the creep tide theory. For low relaxation factors (high viscosity), as in planetary satellites, if friction remains low, each layer 
can be trapped in different spin-orbit resonances with frequencies $n/2,n,3n/2,2n,\ldots$. When the friction increases, attractors with differential 
rotations are destroyed, surviving only commensurabilities in which core and shell have the same velocity of rotation. We apply the theory to Titan. The 
main results are: i) the rotational constraint does not allow us confirm or reject the existence of a subsurface ocean in Titan; and ii) the 
crust-atmosphere exchange of angular momentum can be neglected. Using the rotation estimate based on Cassini's observation (Meriggiola et al. in Icarus 
275:183-192, 2016), we limit the possible value of the shell relaxation factor, when a deep subsurface ocean is assumed, to $\gamma_s\lesssim10^{-9}$ {\rm 
s$^{-1}$}, which correspond to a shell's viscosity $\eta_s\gtrsim10^{18}$ {\rm Pa s}, depending on the ocean's thickness and viscosity values. In the case 
in which a subsurface ocean does not exist, the maximum shell relaxation factor is one order of magnitude smaller and the corresponding minimum shell's 
viscosity is one order higher.

\end{abstract}

\section{Introduction}{\label{sec1}}

Tidal torque is a key physical agent controlling the rotational and orbital evolution of systems with close-in bodies and may give important clues on the 
physical conditions in which these systems originated and evolved. The viscoelastic nature of a real body causes a non-instantaneous deformation, and the 
body continuously tries to recover the equilibrium figure corresponding to the varying gravitational potential due to the orbital companion. In standard 
Darwin's theory (e.g. Darwin, 1880; Kaula, 1964; Mignard, 1979; Efroimsky and Lainey, 2007; Ferraz-Mello et al., 2008), the gravitational potential of the 
deformed body is expanded in Fourier series, and the viscosity is introduced by means of \emph{ad hoc} phase lags in the periodic terms or, alternatively, 
an \textit{ad hoc} constant time lag.

All these theories predict the existence of a stationary rotation. If the lags are assumed to be proportional to the tidal frequencies, the stationary 
rotation has the frequency $\Omega_{stat}\simeq n(1+6e^2)$, where $n$ is the mean motion and $e$ is the orbital eccentricity. The synchronous rotation is 
only possible when the orbit is circular, but the stationary rotation becomes super-synchronous in the non-zero eccentricity case. In these theories, the 
excess of rotation $6ne^2$ does not depend on the rheology of the body. However, this prediction is not confirmed for Titan, where the excess provided by 
the theory is $\sim 38^\circ$ per year, and the Cassini mission, using radar measurement, has not shown discrepancy from synchronous motion larger than 
$\sim0.02^\circ$ per year (Meriggiola, 2012; Meriggiola et al., 2016). Standard theories circumvent this difficulty by assuming that the satellite has an 
\textit{ad hoc} triaxiality, which is permanent and not affected by the tidal forces acting on the body.

Recently, a new tidal theory for viscous homogeneous bodies has been developed by Ferraz-Mello (2013, 2015a) (hereafter FM13 and FM15, respectively). A 
Newtonian creep model, which results from a spherical approximate solution of the Navier-Stokes equation for fluids with very low Reynolds number, is used 
to calculate the surface deformation due to an anelastic tide. This deformation is assumed to be proportional to the stress, and the proportionality 
constant $\gamma$, called the relaxation factor, is inversely proportional to the viscosity of the body. In the creep tide theory, the excess of synchronous 
rotation is roughly proportional to $6n\gamma^2e^2/(n^2+\gamma^2)$. This result reproduces the result obtained with Darwin's theory in the limit 
$\gamma >> n$ (gaseous bodies), but tends to zero when $\gamma \rightarrow 0$ reproducing the almost synchronous rotation of stiff satellites, without the 
need of assuming an \textit {ad hoc} permanent triaxiality. The asymmetry created by the tidal deformation of the satellite is enough to create the torques 
responsible for its almost synchronous rotation.

Tidal theories founded on  hydrodynamical equations were also developed by Zahn, (1966) and  Remus et al. (2012).

A planar theory using a Maxwell viscoelastic rheology and leading to similar results was developed by Correia et al. (2014) and generalized later to the 
spatial case by Bou\'e et al. (2016). Despite the different methods used to introduce the elasticity of the body, this approach is virtually equivalent to 
the creep tide theory (Ferraz-Mello, 2015b).  Other general rheologies were studied by Henning et al. (2008) and Frouard et al. (2016).

However, real celestial bodies are quite far from being homogeneous, and how the tide influences its dynamic evolution is not entirely clear yet. 
Differentiation is common in our Solar System, and several satellites present evidence of a subsurface liquid ocean. We may cite, for instance, Europa (Wahr 
et al., 2006; Khurana et al., 1998) and Enceladus (Porco et al., 2006; Nimmo et al., 2007). One paradigmatic case is Titan, where, in addition, the exchange 
of a certain amount of angular momentum between the surface and the atmosphere may be important (Tokano and Neubauer, 2005; Richard et al., 2014); in addition, 
the presence of an internal ocean (Tobie et al., 2005; Lorenz et al., 2008, Sohl et al., 2014) may decouple rotationally the crust from the interior 
(Karatekin et al., 2008). The rotation of the crust has been studied by Van Hoolst et al. (2008) using the static tide and internal effects as gravitational 
coupling and pressure torques. They found that the crust rotation is influenced mainly by the atmosphere and the Saturn torque and claim that the viscous 
crust deformation and the non-hydrostatic effects could play an important role in the amplitude of the crust oscillation.

Here, we extend the planar creep tide theory to the case of a viscoelastic body formed by $N$ homogeneous layers and study the stationary rotation of the 
particular case $N=2$. Adapting the multi-layered Roche static figure, given by Folonier et al. (2015), to include differential rotation, we solve the creep 
tide equation for each layers interface. Moreover, we add the gravitational coupling and the friction in the interface between the layers. 

The layout of the paper is as follows: in Sect. 2 we present the creep tide model for a multi-layered body, using the static equilibrium figure of Folonier 
et al. (2015), adapted to include the differential rotation. In Sect. 3, we compute the disturbing potential of the deformed body. The forces and toques are 
calculated in Sects. 4 and 5. In Sect. 6 we calculate the work done by the tidal forces acting on the bodies. The variations in semi-major axis and 
eccentricity are shown in Sect. 7. In Sect. 8, we develop the two-layer model, adding the interaction torques between the core and the shell. In Sect. 9, we 
compare the two-layer model with the homogeneous theory. In Sect. 10, we apply to Titan, and, finally, the conclusions are presented in Sect. 11. The paper 
is completed by several appendices where are given technical details of some of the topics presented in the forthcoming sections. In addition, an Online 
Supplement is provided with further details, not worthy of inclusion in the paper but useful for the reproduction of several developments. 

\section{Non-homogeneous Newtonian creep tide theory}{\label{sec2}}

Let us consider one differentiated body $\tens{m}$ of mass $m_T$, disturbed by one mass point $\tens{M}$ of mass $M$ orbiting at a distance $r$ from the 
center of $\tens{m}$. We assume that the body is composed of $N$ homogeneous layers of densities $d_i$ ($i=1,\ldots,N$) and angular velocities 
$\vec{\Omega}_i$, perpendicular to the orbital plane.

The outer surface of the $i$th layer is $\zeta_i(\widehat{\varphi}_i,\widehat{\theta}_i,t)$, where $\zeta_i$ is the distance of the surface points to the 
center of gravity of $\tens{m}$ and the angles $\widehat{\varphi}_i, \widehat{\theta}_i$ are their longitudes and co-latitudes in a fixed reference system. 
At each instant, we assume that the static equilibrium figure of each layer under the action of the tidal potential and the rotation may be approximated by 
a triaxial ellipsoidal equilibrium surface $\rho_i(\widehat{\varphi}_i,\widehat{\theta}_i,t)$, whose semi-major axis is oriented towards \tens{M}.

The adopted rheophysical approach is founded on the simple law
\begin{equation}
  \dot{\zeta}_i = \gamma_i(\rho_i-\zeta_i),
 \label{eq:creep}
\end{equation}
where $\gamma_i$ is the relaxation factor at the outer surface of the $i$th layer. This is a radial deformation rate gradient related to the viscosity 
through (see Appendix 1)
\begin{equation}
\gamma_i = \frac{(d_i-d_{i+1})g_iR_i}{2\eta_i},
\label{eq:gamma_i}
\end{equation}
where $R_i$ and $g_i$ are the equatorial mean radius and the gravity acceleration at the outer surface of the $i$th layer. $\eta_i$ is the viscosity of the 
inner layer (assumed to be larger than that of the outermost layers).

Although the creep equation is valid in a reference system co-rotating with the body, we can use the coordinates in a fixed reference system. This is 
because only relative positions appear in the right-hand side of the creep equation. If $\widehat{\varphi}_F$ is the longitude of a point in one frame fixed 
in the body, then we have
\begin{equation}
 \widehat{\varphi}_i = \widehat{\varphi}_F + \Omega_i t.
\end{equation}

\subsection{The static equilibrium figure}{\label{sec2.1}}

The static equilibrium figure of one body composed by $N$ homogeneous layers, under the action of the tidal potential and the non-synchronous rotation, when 
all layers rotate with the same angular velocity, was calculated by Folonier et al. (2015).\footnote{Although considering first-order deformations (linear 
theory for the flattenings) the results of Folonier et al. (2015) are in excellent agreement with the results founded on a high-order perturbative method of 
Wahl et al. (2017).}In this work, we need, beforehand, to extend these results to the case in which each layer has one different angular velocity. 

We assume that each layer has an ellipsoidal shape with outer semiaxes $a_i$, $b_i$ and $c_i$, where the axis $a_i$ is pointing towards $\tens{M}$ and $c_i$ 
is the axis of rotation. Then, the equatorial prolateness $\epsilon_{\rho}^{(i)}$ and polar oblateness $\epsilon_z^{(i)}$ of the outer surface of the $i$th 
layer can be written as
\begin{equation}
 \epsilon_{\rho}^{(i)} = \frac{a_i-b_i}{R_i} = \mathcal{H}_i\epsilon_\rho; \ \ \ \ \ \ \ \ \epsilon_{z}^{(i)} = \frac{b_i-c_i}{R_i} = \mathcal{G}_i\overline{\epsilon}_z,
\label{eq:ep-i}
\end{equation}
where $R_i=\sqrt{a_ib_i}$ is the outer equatorial mean radius of the $i$th layer, $\epsilon_\rho$ is the flattening of the equivalent Jeans homogeneous 
spheroid and $\overline{\epsilon}_z$ is the flattening of the equivalent MacLaurin homogeneous spheroid in synchronous rotation:
\begin{equation}
\epsilon_\rho = \frac{15MR_N^3}{4m_Tr^3}; \ \ \ \ \ \ \ \ \ \ \overline{\epsilon}_z = \frac{5n^2R_N^3}{4Gm_T}.
\label{eq:ep-homo1}
\end{equation}
Here, $G$ is the gravitation constant, $R_N$ is the equatorial mean radius of \tens{m} and $n$ is the mean motion of \tens{M}. The Clairaut's coefficients 
$\mathcal{H}_i$ and $\mathcal{G}_i$ depend on the internal structure and are (see Appendix 2)
\begin{equation}
\mathcal{H}_i = \sum_{j=1}^N (\tens{E}^{-1})_{ij}x_j^3;\ \ \ \ \ \ \ \ \ \ \ \mathcal{G}_i = \sum_{j=1}^N (\tens{E}^{-1})_{ij}x_j^3\left(\frac{\Omega_j}{n}\right)^2,
\label{eq:Clairaut-coe}
\end{equation}
where $\displaystyle(\tens{E}^{-1})_{ij}$ are the elements of the inverse of the matrix $\tens{E}$, whose elements are
\begin{equation}
(\tens{E})_{ij} = \left\{\begin{array}{lll}\displaystyle - \frac{3}{2f_N}(\widehat{d}_j-\widehat{d}_{j+1})x^3_i,                                              & \ \ \ \ \ i<j \\
                                           \displaystyle - \frac{3}{2f_N}(\widehat{d}_i-\widehat{d}_{i+1})x^3_i + \frac{5}{2}- \frac{5}{2f_N}\sum_{k=i+1}^{N} (\widehat{d}_k-\widehat{d}_{k+1}) (x_k^3-x_i^3), & \ \ \ \ \ i=j \\
                                           \displaystyle - \frac{3}{2f_N}(\widehat{d}_j-\widehat{d}_{j+1})\frac{x^5_j}{x^2_i},                                              & \ \ \ \ \ i>j \end{array}\right.
\end{equation}
where $x_i=R_i/R_N$ and $\widehat{d}_i=d_i/d_1$ are the normalized mean equatorial radius and density, respectively, and $f_N=3\int_0^1\widehat{d}(z)z^2\ dz$.

Finally, the static ellipsoidal surface equation of the outer boundary of the $i$th layer, to first order in the flattenings, can be written as
\begin{equation}
\rho_i = R_i\left(1+\frac{1}{2}\epsilon_{\rho}^{(i)} \sin^2{\widehat{\theta}}\cos{(2\widehat{\varphi}_i-2\varphi_M)}-\left(\frac{1}{2}\epsilon_{\rho}^{(i)}+\epsilon_{z}^{(i)}\right) \cos^2{\widehat{\theta}}\right),
\label{eq:static-surface}
\end{equation}
(see Section A in the Online Supplement), where $\varphi_M$ is the longitude of $\tens{M}$ in the same fixed reference system used to define 
$\widehat{\varphi}_i$.

\subsection{The creep equation}{\label{sec2.2}}

Using the static equilibrium surface (\ref{eq:static-surface}), the creep equation (\ref{eq:creep}) becomes
\begin{equation}
 \dot{\zeta}_i + \gamma_i\zeta_i = \gamma_i R_i\left(1+\frac{1}{2}\epsilon_{\rho}^{(i)} \sin^2{\widehat{\theta}}\cos{(2\widehat{\varphi}_i-2\varpi-2v)}-\left(\frac{1}{2}\epsilon_{\rho}^{(i)}+\epsilon_{z}^{(i)}\right) \cos^2{\widehat{\theta}}\right),
\end{equation}
where $\varpi$ and $v$ are the longitude of the pericenter and the true anomaly, respectively.

For resolving the creep differential equation, we proceed in a similar way as FM13 and FM15. We consider the two-body Keplerian motion. The equations of the 
Keplerian motion of \tens{M} are
\begin{equation}
 r=\frac{a(1-e^2)}{1+e\cos{v}},
\end{equation}
and
\begin{equation}
 v = \ell + \left(2e-\frac{e^3}{4}\right)\sin{\ell}+\frac{5e^2}{4}\sin{2\ell}+\frac{13e^2}{12}\sin{3\ell}+\mathcal{O}(e^4),
\end{equation}
where $a$, $e$, $\ell$ are the semi-major axis, the eccentricity, and the mean anomaly, respectively. 

The resulting equation is then an ordinary differential equation of first order with forced terms that may be written as
\begin{equation}
 \dot{\zeta}_i + \gamma_i\zeta_i = \gamma_i R_i\left(1+\sum_{k\in \mathbb{Z}} \left(\mathcal{C}_{ik}\sin^2{\widehat{\theta}}\cos{\Theta_{ik}}+\mathcal{C}''_{ik}\cos^2{\widehat{\theta}}\cos{\Theta''_{ik}} \right)\right),
\label{eq:zeta2}
\end{equation}
where the arguments of the cosines $\Theta_{ik}$, $\Theta''_{ik}$ are linear functions of the time
\begin{eqnarray}
 \Theta_{ik} = 2\widehat{\varphi}_i-2\varpi+(k-2)\ell;\ \ \ \ \ \ \ \ \ \ \ \Theta''_{ik} = k\ell.
\end{eqnarray}

The constants $\mathcal{C}_{ik},\mathcal{C}''_{ik}$ are
\begin{eqnarray}
\mathcal{C}_{ik} = \frac{1}{2}\mathcal{H}_i\overline{\epsilon}_\rho E_{2,k};\ \ \ \ \ \ \ \ \ \ \ \mathcal{C}''_{ik} = - \frac{1}{2}\mathcal{H}_i\overline{\epsilon}_\rho E_{0,k} - \delta_{0,k}\mathcal{G}_i\overline{\epsilon}_z,
\label{eq:C_ik}
\end{eqnarray}
where $\delta_{0,k}$ is the Kronecker delta ($\delta_{0,k}=1$ when $k=0$ and $\delta_{0,k}=0$ when $k\neq0$), the constant $\overline{\epsilon}_\rho$ is
\begin{equation}
 \overline{\epsilon}_\rho = \frac{15MR_N^3}{4m_Ta^3},
\label{eq:ep-homo2}
\end{equation}
and $E_{q,p}$ are the Cayley functions (Cayley, 1861), defined by
\begin{equation}
 E_{q,p}(e) = \frac{1}{2\pi}\int_0^{2\pi}\left(\frac{a}{r}\right)^3\cos{(qv+(p-q)\ell)\ d\ell}.
 \label{eq:cayley}
\end{equation}

After integration, we obtain the forced terms
\begin{equation}
\delta \zeta_i = R_i\sum_{k\in \mathbb{Z}} \left(\mathcal{C}_{ik}\sin^2{\widehat{\theta}}\cos{\sigma_{ik}}\cos{(\Theta_{ik}-\sigma_{ik})}+\mathcal{C}''_{ik}\cos^2{\widehat{\theta}}\cos{\sigma''_{ik}}\cos{(\Theta''_{ik}-\sigma''_{ik})} \right).
\label{eq:delta-zeta}
\end{equation}
The phases $\sigma_{ik}$ and $\sigma''_{ik}$ are
\begin{align}
\tan{\sigma_{ik}}   & = \frac{\nu_i+kn}{\gamma_i}; & \cos{\sigma_{ik}}   &= \frac{\gamma_i}{\sqrt{\gamma_i^2+(\nu_i+kn)^2}};   & \sin{\sigma_{ik}}   &= \frac{\nu_i+kn}{\sqrt{\gamma_i^2+(\nu_i+kn)^2}} \nonumber\\
\tan{\sigma''_{ik}} & = \frac{kn}{\gamma_i};       & \cos{\sigma''_{ik}} &= \frac{\gamma_i}{\sqrt{\gamma_i^2+(kn)^2}};         & \sin{\sigma''_{ik}} &= \frac{kn}{\sqrt{\gamma_i^2+(kn)^2}},
 \label{eq:sigma_i}
\end{align}
where 
\begin{equation}
\nu_i=2\Omega_i-2n, 
\end{equation}
is the semi-diurnal frequency. These phases are introduced during the exact integration of the creep equation (\ref{eq:zeta2}).

If we define the angles
\begin{eqnarray}
 \delta_{ik} = 2\varpi-(k-2)\ell+\sigma_{ik};\ \ \ \ \ \ \ \ \ \ \ \delta''_{ik} = k\ell-\sigma''_{ik},
\label{eq:delta_ik}
\end{eqnarray}
and the equatorial and polar flattenings
\begin{equation}
 \epsilon_\rho^{(ik)} = 2\mathcal{C}_{ik}\cos{\sigma_{ik}}; \ \ \ \ \ \ \ \epsilon_z^{(ik)} = -\mathcal{C}''_{ik}\cos{\sigma''_{ik}}\cos{\delta''_{ik}}-\frac{\epsilon_\rho^{(ik)}}{2},
\label{eq:epsilon_ik}
\end{equation}
the solution (\ref{eq:delta-zeta}) can be written as
\begin{equation}
 \delta \zeta_i = R_i\sum_{k\in \mathbb{Z}} \left(\frac{1}{2}\epsilon_\rho^{(ik)}\sin^2{\widehat{\theta}}\cos{(2\widehat{\varphi}_i-\delta_{ik})}-\left(\frac{1}{2}\epsilon_\rho^{(ik)}+\epsilon_z^{(ik)}\right)\cos^2{\widehat{\theta}} \right),
\label{eq:delta-zeta2}
\end{equation}
which has a simple geometric interpretation: using Eq. (\ref{eq:static-surface}), we can identify each term of the Fourier expansion of the height 
$\delta\zeta_i$, with the boundary height of one ellipsoid, with equatorial and polar flattenings $\epsilon_\rho^{(ik)}$ and $\epsilon_z^{(ik)}$, 
respectively, rotated at an angle $\delta_{ik}/2$, with respect to the axis $x$.

\section{The disturbing potential}{\label{sec3}}

The potential of the $i$th layer of \tens{m} at a generic point $\tens{M}^*(r^*,\theta^*,\varphi^*)$ external to this layer, can be written as the 
potential of one spherical shell of outer and inner radii $R_i$ and $R_{i-1}$, respectively, plus the disturbing potential due to the mass excess or deficit 
corresponding to the outer and the inner boundary heights $\delta\zeta_i$ and $\delta\zeta_{i-1}$. It is important to note that since these excesses or 
deficits are very small, we may calculate the contribution of each term of the Fourier expansion separately and then sum them to obtain the total 
contribution.

In this way, we assume that the $i$th layer has outer and inner boundary heights given by the $k$th term of the Fourier expansion. The equatorial 
and polar flattenings of the outer boundary, $\epsilon_\rho^{(ik)}$ and $\epsilon_z^{(ik)}$, are given by Eq. (\ref{eq:epsilon_ik}), and the bulge is 
rotated at an angle $\delta_{ik}/2$ with respect to the axis $x$. Similarly, the inner boundary height $\delta\zeta_{i-1}^{(1)}$, can be identified with the 
boundary height of one ellipsoidal surface, with equatorial and polar flattenings
\begin{equation}
 \epsilon_\rho^{(i-1,k)} = 2\mathcal{C}_{i-1,k}\cos{\sigma_{i-1,k}}; \ \ \ \ \ \ \ \epsilon_z^{(i-1,k)} = -\mathcal{C}''_{i-1,k}\cos{\sigma''_{i-1,k}}\cos{\delta''_{i-1,k}}-\frac{\epsilon_\rho^{(i-1,k)}}{2},
\end{equation}
rotated at an angle $\delta_{i-1,k}/2$, with respect to the axis $x$.

The disturbing potential at an external point $\tens{M}^*(r^*,\theta^*,\varphi^*)$, due to the mass excess or deficit, corresponding to the $k$th term of 
the Fourier expansion of the outer and the inner boundary heights $\delta\zeta_i$ and $\delta\zeta_{i-1}$, is
\begin{eqnarray}
 \delta U_{ik}(\vec{r}^*) &=& -\displaystyle\frac{3GC_i}{2r^{*3}}\sin^2{\theta^*}\frac{\Delta\big(R_i^5\mathcal{C}_{ik}\cos{\sigma_{ik}}\cos{(2\varphi^*-\delta_{ik})}\big)}{R_i^5-R_{i-1}^5}\nonumber\\
                          & & -\frac{G C_i}{2r^{*3}}(3\cos^2{\theta^*}-1)\frac{\Delta\big(R_i^5\mathcal{C}''_{ik}\cos{\sigma''_{ik}}\cos{\delta''_{ik}}\big)}{R_i^5-R_{i-1}^5},
\end{eqnarray}
where $C_i$ is the axial moment of inertia of the $i$th layer (see Section A in the Online Supplement) and $\Delta(f_i) = f_i-f_{i-1}$, denotes the 
increment of one function $f_i$, between the inner and the outer boundaries of this layer.

Taking into account that the total disturbing potential of the $i$th layer, can be approximated by the sum of the contribution of each term of the Fourier 
expansion, we obtain
\begin{equation}
 \delta U_i(\vec{r}^*)=\sum_{k\in \mathbb{Z}}\delta U_{ik}(\vec{r}^*).
\end{equation}

\section{Forces and torques}{\label{sec4}}

To calculate the force and torque due to the $i$th layer of $\tens{m}$, acting on one mass $M^{*}$ located in $\tens{M}^{*}(r^{*},\theta^{*},\varphi^{*})$, 
we take the negative gradient of the potential of the $i$th layer at the point $\tens{M}^{*}$ and multiply it by the mass placed in the point, that is, 
$\vec{F}_i=-M^{*}\nabla_{\vec{r}^*} \delta U_i$
\begin{eqnarray}
 F_{1i} &=& -\frac{9GM^*C_i}{2r^{*4}}\sin^2{\theta^*}\sum_{k\in \mathbb{Z}}\frac{\Delta\big(R_i^5\mathcal{C}_{ik}\cos{\sigma_{ik}}\cos{(2\varphi^*-\delta_{ik})}\big)}{R_i^5-R_{i-1}^5}\nonumber\\
        & & -\frac{3GM^*C_i}{2r^{*4}}(3\cos^2{\theta^*}-1)\sum_{k\in \mathbb{Z}}\frac{\Delta\big(R_i^5\mathcal{C}''_{ik}\cos{\sigma''_{ik}}\cos{\delta''_{ik}}\big)}{R_i^5-R_{i-1}^5}\nonumber\\
 F_{2i} &=& \frac{3GM^*C_i}{2r^{*4}}\sin{2\theta^*}\sum_{k\in \mathbb{Z}}\frac{\Delta\big(R_i^5\mathcal{C}_{ik}\cos{\sigma_{ik}}\cos{(2\varphi^*-\delta_{ik})}\big)}{R_i^5-R_{i-1}^5}\nonumber\\
        & &-\frac{3GM^*C_i}{2r^{*4}}\sin{2\theta^*}\sum_{k\in \mathbb{Z}}\frac{\Delta\big(R_i^5\mathcal{C}''_{ik}\cos{\sigma''_{ik}}\cos{\delta''_{ik}}\big)}{R_i^5-R_{i-1}^5} \nonumber\\
 F_{3i} &=& -\frac{3GM^*C_i}{r^{*4}}\sin{\theta^*}\sum_{k\in \mathbb{Z}}\frac{\Delta\big(R_i^5\mathcal{C}_{ik}\cos{\sigma_{ik}}\sin{(2\varphi^*-\delta_{ik})}\big)}{R_i^5-R_{i-1}^5}.
\end{eqnarray}

The corresponding torque is $\vec{M}_i=\vec{r}^*\times\vec{F}_i$, or since $\vec{r}^*=(r^*,0,0)$,
\begin{equation}
 M_{1i}=0;  \ \ \ \  M_{2i}=-r^{*}F_{3i};  \ \ \ \  M_{3i}=r^{*}F_{2i},
 \label{eq:torque_esfericas}
\end{equation}
that is
\begin{eqnarray}
 M_{2i} &=& \frac{3GM^*C_i}{r^{*3}}\sin{\theta^*}\sum_{k\in \mathbb{Z}}\frac{\Delta\big(R_i^5\mathcal{C}_{ik}\cos{\sigma_{ik}}\sin{(2\varphi^*-\delta_{ik})}\big)}{R_i^5-R_{i-1}^5}\nonumber\\
 M_{3i} &=& \frac{3GM^*C_i}{2r^{*3}}\sin{2\theta^*}\sum_{k\in \mathbb{Z}}\frac{\Delta\big(R_i^5\mathcal{C}_{ik}\cos{\sigma_{ik}}\cos{(2\varphi^*-\delta_{ik})}\big)}{R_i^5-R_{i-1}^5}\nonumber\\
        & &-\frac{3GM^*C_i}{2r^{*3}}\sin{2\theta^*}\sum_{k\in \mathbb{Z}}\frac{\Delta\big(R_i^5\mathcal{C}''_{ik}\cos{\sigma''_{ik}}\cos{\delta''_{ik}}\big)}{R_i^5-R_{i-1}^5}.
\end{eqnarray}

\section{Forces and torques acting on $\tens{M}$}{\label{sec5}}

Since we are interested in the force acting on $\tens{M}$ due to the tidal deformation of the $i$th layer of $\tens{m}$, we must substitute 
$(M^{*},r^{*},\theta^{*},\varphi^{*})$ by $(M,r,\frac{\pi}{2},\varpi+v)$. Replacing the angles $\delta_{ik}$ and $\delta''_{ik}$ given their definitions 
(Eq. \ref{eq:delta_ik}), the forces, then are
\begin{eqnarray}
 F_{1i} &=& -\frac{9GMC_i}{2r^4}\sum_{k\in \mathbb{Z}}\frac{\Delta\big(R_i^5\mathcal{C}_{ik}\cos{\sigma_{ik}}\cos{(2v+(k-2)\ell-\sigma_{ik})}\big)}{R_i^5-R_{i-1}^5}\nonumber\\
        & & +\frac{3GMC_i}{2r^4}\sum_{k\in \mathbb{Z}}\frac{\Delta\big(R_i^5\mathcal{C}''_{ik}\cos{\sigma''_{ik}}\cos{(k\ell-\sigma''_{ik})}\big)}{R_i^5-R_{i-1}^5}\nonumber\\
 F_{2i} &=& 0 \nonumber\\
 F_{3i} &=& -\frac{3GMC_i}{r^4}\sum_{k\in \mathbb{Z}}\frac{\Delta\big(R_i^5\mathcal{C}_{ik}\cos{\sigma_{ik}}\sin{(2v+(k-2)\ell-\sigma_{ik})}\big)}{R_i^5-R_{i-1}^5}.
\label{eq:force_i}
\end{eqnarray}
and the corresponding torques are
\begin{eqnarray}
 M_{2i} &=& \frac{3GMC_i}{r^3}\sum_{k\in \mathbb{Z}}\frac{\Delta\big(R_i^5\mathcal{C}_{ik}\cos{\sigma_{ik}}\sin{(2v+(k-2)\ell-\sigma_{ik})}\big)}{R_i^5-R_{i-1}^5}\nonumber\\
 M_{3i} &=& 0,
\end{eqnarray}
After Fourier expansion, the torque along to the axis $z$ ($M_{zi}=-M_{2i}$), can be written as
\begin{eqnarray}
 M_{zi} = \frac{3GMC_i}{a^3}\sum_{k,j\in \mathbb{Z}}E_{2,k+j}\frac{\Delta\big(R_i^5\mathcal{C}_{ik}\cos{\sigma_{ik}}\sin{(j\ell+\sigma_{ik})}\big)}{R_i^5-R_{i-1}^5}.
\label{eq:torque-creep}
\end{eqnarray}

Finally, replacing the coefficient $\mathcal{C}_{ik}$ given by Eq. (\ref{eq:C_ik}), the time average of the tidal torque over one period 
$\langle M_{zi}\rangle = \frac{1}{2\pi}\int_0^{2\pi}M_{zi}d\ell$ is
\begin{equation}
\langle M_{zi}\rangle = \frac{45GM^2R_N^3C_i}{16m_Ta^6}\sum_{k\in \mathbb{Z}}E_{2,k}^2\frac{\Delta\big(\mathcal{H}_iR_i^5\sin{2\sigma_{ik}}\big)}{R_i^5-R_{i-1}^5}.
 \label{eq:torque_z-creep-prom}
\end{equation}

The above expression for the time average, which is equivalent to take into account only the terms with $j=0$, is only valid if $\nu_i$ is constant. This 
condition is satisfied, for example, by homogeneous bodies with $\gamma\gg n$, as stars and giant gaseous planets, where the stationary rotation is 
$\sim6n\gamma e^2/(n^2+\gamma^2)$. However, the final rotation of the homogeneous rocky bodies, with $\gamma\ll n$, as satellites and Earth-like planets, is 
dominated by a forced libration $\sim B_1\cos{(\ell+\phi_1)}$ with the same period as  the orbital motion of the system (see Chap. 3 of FM15). In this case, 
any time average that involves the rotation, should also take into account this oscillation. It is worth emphasizing that in this paper we calculate the 
time average of some quantities, as the work done by the tidal forces and the variations in semi-major axis and eccentricity, assuming that $\nu_i$ is 
constant, which is valid only for bodies with low viscosity. The applications to Titan in this paper were done using the complete equations, where the 
distinction between these extreme cases is not necessary.

\section{Work done by the tidal forces acting on \tens{M}}{\label{sec6}}

The time rate of the work done by the tidal forces due to the $i$th layer is $\dot{W}_i=\vec{F}_i\cdot\vec{v}$, where $\vec{v}$ is the relative 
velocity vector of the external body\footnote{The definition of power (the time derivative of work) used in this section is the most general definition of 
the power done by the force couple formed by the disturbing force $\mathbf{F}_i$ acting on the external body and its reaction $-\mathbf{F}_i$ acting on the 
$i$th layer of the deformed body. It may be written as $\mathbf{F}_i \cdot \mathbf{V}_\tens{M} + (-\mathbf{F}_i) \cdot \mathbf{V}_i$ where 
$\mathbf{V}_\tens{M}$ and $\mathbf{V}_i$ are, respectively, the velocities of the body $\tens{M}$ and of the $i$th layer of the body $\tens{m}$, w.r.t. a 
fixed reference frame. It is equivalent to $\mathbf{F}_i \dot (\mathbf{V}_\tens{M} - \mathbf{V}_i)$, that is, $\mathbf{F}_i \cdot \mathbf{v}$ (see Scheeres, 
2002; Ferraz-Mello et al., 2003).} whose components in spherical coordinates are
\begin{equation}
 v_1 = \frac{nae\sin{v}}{\sqrt{1-e^2}}; \ \ \ \ \ v_2 = 0;  \ \ \ \ \  v_3 = \frac{na^2\sqrt{1-e^2}}{r}.
\label{eq:vel}
\end{equation}
Using the tidal force, given by the Eq. (\ref{eq:force_i}), the rate of the work corresponding to the $i$th layer is
\begin{eqnarray}
\frac{dW_i}{dt} &=& -\frac{3GMC_in}{2a^3}\sum_{k\in \mathbb{Z}}\frac{1}{R_i^5-R_{i-1}^5}\Delta\Bigg(R_i^5\mathcal{C}_{ik}\cos{\sigma_{ik}}\nonumber\\
                & & \times\Bigg[\cos{\sigma_{ik}}\Big(\frac{3e}{\sqrt{1-e^2}}\frac{a^4}{r^4}\sin{v}\cos{(2v+(k-2)\ell)}+\frac{a^5}{r^5}2\sqrt{1-e^2}\sin{(2v+(k-2)\ell)}\Big)\nonumber\\
                & & +\sin{\sigma_{ik}}\Big(\frac{3e}{\sqrt{1-e^2}}\frac{a^4}{r^4}\sin{v}\sin{(2v+(k-2)\ell)}-\frac{a^5}{r^5}2\sqrt{1-e^2}\cos{(2v+(k-2)\ell)}\Big)\Bigg]\Bigg)\nonumber\\
                & & +\frac{GMC_in}{2a^3}\frac{a^4}{r^4}\sum_{k\in \mathbb{Z}}\frac{1}{R_i^5-R_{i-1}^5}\Delta\Bigg(R_i^5\mathcal{C}''_{ik}\cos{\sigma''_{ik}}\frac{3e}{\sqrt{1-e^2}}\sin{v}\cos{(k\ell-\sigma''_{ik})}\Bigg),
\end{eqnarray}
or after Fourier expansion\footnote{For the details of the calculation see the Online Supplement of FM15.}
\begin{eqnarray}
\frac{dW_i}{dt} &=& -\frac{3GMC_in}{2a^3}\sum_{k,j\in \mathbb{Z}}(k+j-2)E_{2,k+j}\frac{\Delta\big(R_i^5\mathcal{C}_{ik}\cos{\sigma_{ik}}\sin{(j\ell+\sigma_{ik})}\big)}{R_i^5-R_{i-1}^5}\nonumber\\
                          & & +\frac{GMC_in}{2a^3}\sum_{k,j\in \mathbb{Z}}(k+j)E_{0,k+j}\frac{\Delta\big(R_i^5\mathcal{C}''_{ik}\cos{\sigma''_{ik}}\sin{(j\ell+\sigma''_{ik})}\big)}{R_i^5-R_{i-1}^5}.
\label{eq:rate-work}
\end{eqnarray}
The time-average over one period is
\begin{eqnarray}
\left\langle \frac{dW_i}{dt}\right\rangle &=& \frac{45GM^2R_N^3C_in}{32m_Ta^6}\sum_{k\in \mathbb{Z}}\left((2-k)E_{2,k}^2\frac{\Delta(\mathcal{H}_iR_i^5\sin{2\sigma_{ik}})}{R_i^5-R_{i-1}^5}-\frac{k}{3}E_{0,k}^2\frac{\Delta(\mathcal{H}_iR_i^5\sin{2\sigma''_{ik}})}{R_i^5-R_{i-1}^5}\right).
 \label{eq:dwi_orb(tide)}
\end{eqnarray}

The average of the term involving $\delta_{0,k}\mathcal{G}_i\overline{\epsilon}_z$ in the last term of Eq. (\ref{eq:rate-work}), for $k=0$, is
\begin{equation}
\frac{1}{2\pi}\int_0^{2\pi}n^2\mathcal{L}'_i\left(\frac{a}{r}\right)^4\sin{v}\ d\ell = \sum_{j=1}^N \frac{\Delta\big(R_i^5(\tens{E}^{-1})_{ij}x_j^3\big)}{R_i^5-R_{i-1}^5}  \frac{1}{2\pi}\int_0^{2\pi}\Omega_j^2\left(\frac{a}{r}\right)^4\sin{v}\ d\ell=0,
\label{eq:integral-zero}
\end{equation}
(see Section C in the Online Supplement).

\section{Variations in semi-major axis and eccentricity}{\label{sec7}}

In this section, we calculate the variation in semi-major axis and eccentricity. As in FM13 and FM15, we use the energy and angular momentum definitions.\footnote{We use the conservation laws of the two-body problem because they are universally known. However, it is worth emphasizing that the results 
obtained are the same obtained if instead of them we use the Lagrange variational equations.} If we differentiate the equation 
$$W=-\frac{GMm_T}{2a},$$
where $a$ is the semi-major axis of the relative orbit, we obtain the equation for the variation in semi-major axis:
\begin{equation}
\dot{a}= \frac{2a^2\dot{W}}{GMm_T}.
\end{equation}

Replacing $\dot{W}$ by the Eq. (\ref{eq:rate-work}) and summing over all layers, we obtain
\begin{eqnarray}
\dot{a} &=& -\sum_{i=1}^N\frac{3C_in}{m_Ta}\sum_{k,j\in \mathbb{Z}}(k+j-2)E_{2,k+j}\frac{\Delta\big(R_i^5\mathcal{C}_{ik}\cos{\sigma_{ik}}\sin{(j\ell+\sigma_{ik})}\big)}{R_i^5-R_{i-1}^5}\nonumber\\
        & & +\sum_{i=1}^N\frac{C_in}{m_Ta}\sum_{k,j\in \mathbb{Z}}(k+j)E_{0,k+j}\frac{\Delta\big(R_i^5\mathcal{C}''_{ik}\cos{\sigma''_{ik}}\sin{(j\ell+\sigma''_{ik})}\big)}{R_i^5-R_{i-1}^5}.
\label{eq:dot_a}
\end{eqnarray}

After the averaging over one period, we obtain
\begin{eqnarray}
\langle \dot{a}\rangle &=& \sum_{i=1}^N\frac{45MR_N^3C_in}{16m_T^2a^4}\sum_{k\in \mathbb{Z}}\left((2-k)E_{2,k}^2\frac{\Delta\big(\mathcal{H}_iR_i^5\sin{2\sigma_{ik}}\big)}{R_i^5-R_{i-1}^5}-kE_{0,k}^2\frac{\Delta\big(\mathcal{H}_iR_i^5\sin{2\sigma''_{ik}}\big)}{R_i^5-R_{i-1}^5}\right).
\label{eq:dot_a-aver}
\end{eqnarray}

In the same way, if we differentiate the total angular momentum equation
$$L=\frac{Mm_T}{M+m_T}na^2\sqrt{1-e^2}=\frac{GMm_T}{na}\sqrt{1-e^2},$$
where $e$ is the eccentricity of the relative orbit, and use $\dot{n}/n=-3\dot{a}/2a$, we obtain the equation for the variation in eccentricity
\begin{equation}
\frac{e\dot{e}}{1-e^2}= \frac{\dot{a}}{2a}-\frac{\dot{L}}{L},
\end{equation}
where $\dot{L}=\mathcal{M}_z$ is the total torque exerted by the tidal forces. The interaction torques between the layers do not affect the orbital motion, 
because they are action-reaction pairs (that is $\mathcal{M}_{ij}=-\mathcal{M}_{ji}$, $\forall\ i,j=1,\ldots,N$ and $i\neq j$), then they mutually cancel 
themselves.

Replacing $\dot{W}$ and $\mathcal{M}_z$ by the Eqs. (\ref{eq:torque-creep}) and (\ref{eq:rate-work}), and summing over all layers, we obtain
\begin{eqnarray}
\dot{e} &=& -\sum_{i=1}^N\frac{3C_in}{m_Ta^2}\frac{(1-e^2)}{2e}\sum_{k,j\in \mathbb{Z}}\Big(\frac{2}{\sqrt{1-e^2}}+(k+j-2)\Big)E_{2,k+j}\frac{\Delta\big(R_i^5\mathcal{C}_{ik}\cos{\sigma_{ik}}\sin{(j\ell+\sigma_{ik})}\big)}{R_i^5-R_{i-1}^5}\nonumber\\
        & & +\sum_{i=1}^N\frac{C_in}{m_Ta^2}\frac{(1-e^2)}{2e}\sum_{k,j\in \mathbb{Z}}(k+j)E_{0,k+j}\frac{\Delta\big(R_i^5\mathcal{C}''_{ik}\cos{\sigma''_{ik}}\sin{(j\ell+\sigma''_{ik})}\big)}{R_i^5-R_{i-1}^5}.
\label{eq:dot_e}
\end{eqnarray}

After the time-average over one period, we obtain that the variation in eccentricity are
\begin{eqnarray}
\langle \dot{e}\rangle &=& \sum_{i=1}^N\frac{45MR_N^3C_in}{16m_T^2a^4}\frac{(1-e^2)}{2ae}\nonumber\\
                       & & \times\sum_{k\in \mathbb{Z}}\left(\Big((2-k)-\frac{2}{\sqrt{1-e^2}}\Big)E_{2,k}^2\frac{\Delta\big(\mathcal{H}_iR_i^5\sin{2\sigma_{ik}}\big)}{R_i^5-R_{i-1}^5}-kE_{0,k}^2\frac{\Delta\big(\mathcal{H}_iR_i^5\sin{2\sigma''_{ik}}\big)}{R_i^5-R_{i-1}^5}\right).
\label{eq:dot_e-aver}
\end{eqnarray}

\section{The two-layer model}{\label{sec8}}

In the previous sections we have studied the tidal effect on one body composed of $N$ homogeneous layers. However, in contrast with a homogeneous body, in 
one differentiated body we must also take into account the interaction between the different layers. In this paper we consider the gravitational coupling of 
the layers and the friction that occurs at each interface of two layers in contact. An important point to keep of mind is that the number of free 
parameters increases significantly as the number of layers increases.

In this section, we study the simplest non-homogeneous problem: one body formed by two independent rotating parts. The inner layer, or \textit{core}, is 
denoted with the subscript \textit{c} and the outer layer, or \textit{shell}, is denoted with the subscript \textit{s}. Despite its simplicity, the 
two-layer model allows us to study the main features of the stationary rotations, introducing a minimum number of free parameters. 

\subsection{The tidal torques}{\label{sec8.1}}

The tidal torques due to the core and the shell, along the axis $z$, are (see Eq. \ref{eq:torque-creep})
\begin{eqnarray}
 M_{zc} &=&T_{cc}C_c\mathcal{T}_c\nonumber\\
 M_{zs} &=&T_{ss}C_s\mathcal{T}_s-T_{sc}C_s\mathcal{T}_c,
\label{creep:torque-s}
\end{eqnarray}
where the function $\mathcal{T}_i$ ($i=c,s$) is
\begin{equation}
\mathcal{T}_i = \sum_{k,j\in \mathbb{Z}} E_{2,k}E_{2,k+j}\frac{\gamma_i(\nu_j+kn)\cos{j\ell}+\gamma_i^2\sin{j\ell}}{\gamma_i^2+(\nu_i+kn)^2},
 \label{eq:T_i}
\end{equation}
the constants $T_{ij}$ are
\begin{equation}
 T_{cc} = \mathcal{T}\mathcal{H}_c; \ \ \ \ \ T_{sc} = \frac{\mathcal{T}\mathcal{H}_cR_c^5}{R_s^5-R_c^5}; \ \ \ \ \ T_{ss} = \frac{\mathcal{T}\mathcal{H}_sR_s^5}{R_s^5-R_c^5},
 \label{eq:T_ij}
\end{equation}
and the tidal parameter $\mathcal{T}$, is defined as
\begin{equation}
\mathcal{T} = \frac{45GM^2R_s^3}{8m_Ta^6} \approx \frac{3n^2\overline{\epsilon}_\rho}{2},
 \label{eq:T}
\end{equation}
$R_c$, $C_c$ are the mean outer radius and moment of inertia of the core, and $R_s$, $C_s$ are the mean outer radius and moment of inertia of the shell. The 
parameters $\mathcal{H}_c$, $\gamma_c$ are the Clairaut parameter and the relaxation factor at the core-shell interface and $\mathcal{H}_s$, $\gamma_s$ are 
the Clairaut parameter and the relaxation factor at the body's surface.

\subsection{The gravitational core-shell coupling}{\label{sec8.2}}

When the principal axes of inertia of two layers are not aligned, a restoring torque appears which tends to align these axes again. This torque was 
calculated by several authors (e.g. Buffett, 1996; Van Hoolst et al., 2008; Karatekin et al., 2008; Callegari et al., 2015) when the layers are rigid.

Here, we use one similar expression for this torque adapted to a body formed by two layers whose boundaries are prolate ellipsoids, whose flattenings are 
defined by the composition of the main elastic and anelastic tidal components. If we follow the same composition adopted in FM13, these flattenings are
\begin{equation}
 \epsilon'_c = \mathcal{H}_c\overline{\epsilon}_\rho \sqrt{\lambda_c^2 + \cos^2{\sigma_{c0}}(1 + 2\lambda_c)}; \ \ \ \ \ \ \ \ \epsilon'_s = \mathcal{H}_s\overline{\epsilon}_\rho\sqrt{\lambda_s^2 + \cos^2{\sigma_{s0}}(1 + 2\lambda_s)},
 \label{eq:real flattenings}
\end{equation}
where $0<\lambda_c<1$ and $0<\lambda_s<1$ are the relative measurements of the actual maximum heights of the elastic tides of the core and the shell, 
respectively. The geodetic lags of these two ellipsoidal surfaces, when one elastic component is added are
\begin{equation}
 \vartheta_c = \frac{1}{2}\tan^{-1}{\left(\frac{\sin{2\sigma_{c0}}}{1+2\lambda_c+\cos{2\sigma_{c0}}} \right)}; \ \ \ \ \ \ \ \ \vartheta_s = \frac{1}{2}\tan^{-1}{\left(\frac{\sin{2\sigma_{s0}}}{1+2\lambda_s+\cos{2\sigma_{s0}}} \right)}.
\end{equation}

In this case, the torques, along the axis $z$, are
\begin{eqnarray}
\Gamma_c &=& K \sin{2\xi} \nonumber\\
\Gamma_s &=&-K \sin{2\xi},
\end{eqnarray}
where $\xi=\vartheta_s-\vartheta_c$ is the offset of the geodetic lags of the two ellipsoidal boundaries and the constant of gravitational coupling $K$ is
\begin{equation}
 K = \frac{32\pi^2 G}{75}\epsilon'_c \epsilon'_s d_cd_s  R_c^5,
\end{equation}
(see Appendix 3 for more details).

We may pay attention to the sign of these torques. If $\vartheta_s > \vartheta_c$, the motion of the shell is braked, while the motion of the core is 
accelerated. This is consistent with the signs of the above equations.

\subsection{Linear drag}{\label{sec8.3}}

The model considered here also assumes that a linear friction occurs between the two contiguous layers. For the two-layer model, the torques acting 
on the core and the shell, along the axis $z$, are
\begin{eqnarray}
\Phi_c &=& \mu (\Omega_s-\Omega_c)\nonumber\\
\Phi_s &=&-\mu (\Omega_s-\Omega_c),
\label{eq:torque-fric}
\end{eqnarray}
where the friction coefficient $\mu$ is an undetermined ad-hoc constant that comes from assuming that a linear friction occurs between two contiguous layers.

When we consider that the body \tens{m} has solid layers, but not rigid, we can assume that between the core and the shell exists one thin fluid boundary 
with viscosity $\eta_o$ and thickness $h$. If this interface is a Newtonian fluid, the Eq. (\ref{eq:torque-fric}) is the law corresponding to liquid-solid 
boundary for low speeds, and $\mu$ can be written as
\begin{equation}
\mu = \frac{8\pi}{3}\ \frac{\eta_o}{h}R_c^4,
\label{eq:mu}
\end{equation}
(see Appendix 4 for more details).

\subsection{Rotational equations}{\label{sec8.4}}
Putting together all contributions to the torque, we obtain the rotational equations
\begin{eqnarray}
  C_c\dot{\Omega}_c &=& M_z^{core}  = -M_{zc}+\Gamma_c+\Phi_c\nonumber\\
  C_s\dot{\Omega}_s &=& M_z^{shell} = -M_{zs}+\Gamma_s+\Phi_s,
\label{eq:system-rotation}
\end{eqnarray}
where $M_z^{core}$ and $M_z^{shell}$ are the $z$-components of the total torque acting on the core and on the shell. These torques 
include the reaction of the tidal torque $M_{zi}$, the gravitational coupling $\Gamma_i$ and the friction $\Phi_i$.

\section{Comparison with the homogeneous case}{\label{sec9}}

In this section, we compare some of the main features of the homogeneous creep tide theory, developed in FM15, with the non-homogeneous creep tide 
theory for the two-layer model developed in this article. The main difficulty lies in the number of free parameters in these approaches. In the homogeneous 
case, with a suitable choice of dimensionless variables, the final state of rotation depends only on the ratio $n/\gamma$ and on the eccentricity $e$ (Eq. 
42 of FM15). 
\begin{figure}[h]
\begin{center}
\includegraphics[scale=0.45]{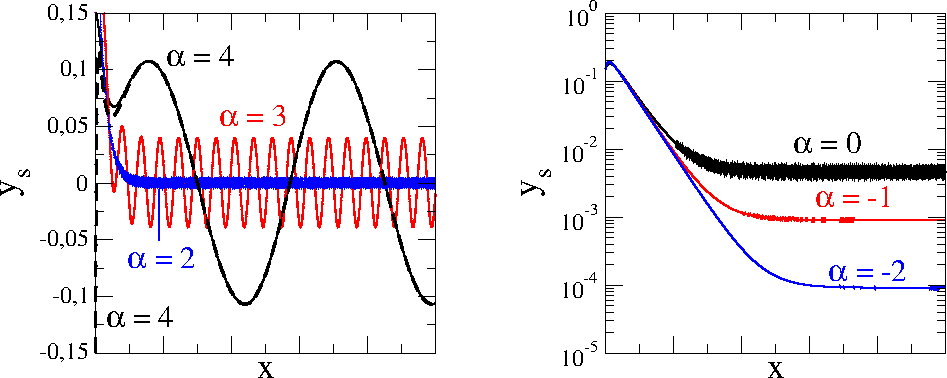}
\caption{Evolution of $y_s$ for the case $\gamma_c=\gamma_s$ with initial conditions $y_c=0.3$ and $y_s=0.15$ and several values of $\alpha=\log_{10}{(n/
\gamma_c)}=\log_{10}{(n/\gamma_s)}$. For  $\alpha=4$, we also plot the initial conditions $y_s=0$ and $y_s=-0.15$. \textit{Left}: $\alpha=4,3,2$. 
\textit{Right}: $\alpha=0,-1,-2$.}
\label{fig01}
\end{center}
\end{figure}
However, even in the most simple non-homogeneous case (the two-layer model), we need to set 12 free parameters. In order to proceed, we use 
the typical values for Titan and also Titan's eccentricity $e=0.028$ (see Tables \ref{Data-tab}-\ref{Table-parameters-2l-I} in Sect. \ref{sec10.1}), and let 
as free parameters, only $n/\gamma_i$, $e$ and $\mu$.

Following FM15, we introduce the adimensional variables $y_i=\nu_i/\overline{\gamma}$ and the scaled time $x=\ell/\overline{\gamma}$, where $\overline{\gamma}=2\gamma_c
\gamma_s(\gamma_c+\gamma_s)^{-1}$. If we consider the case in which $\gamma_c=\gamma_s$, the behavior of the evolutions of $y_c$ and $y_s$ is similar to 
that observed in the homogeneous case. Figure \ref{fig01} shows the time evolution of $y_s$, with initial conditions $y_c=0.3$, $y_s=0.15$ and differents 
values of $\alpha=\log_{10}{(n/\gamma_c)}=\log_{10}{(n/\gamma_s)}$. When $\gamma_i\ll n$ (i.e. rocky bodies), after a transient, the solution oscillates 
around zero, independently of the initial conditions (\textit{left} panel), and the amplitude of oscillation decreases when $\alpha$ decreases. In the case 
where $\alpha=4$, we also plot the solution with initial conditions $y_c=0.3$ and $y_s=-0.15$ (dashed black line). This solution increases quickly, becoming 
indistinguishable from the solution with initial value $y_s=0.15$.  When $\gamma_i\sim n$, the stationary solution becomes a super-synchronous rotation with 
the amplitude of oscillation tending to zero, and, finally, when $\gamma_i\gg n$, the stationary solution of $y_s$ becomes closer zero (\textit{right} 
panel). The evolution of $y_c$ is very similar, and the friction does not have any relevant role.
\begin{figure}[h]
\begin{center}
\includegraphics[scale=0.45]{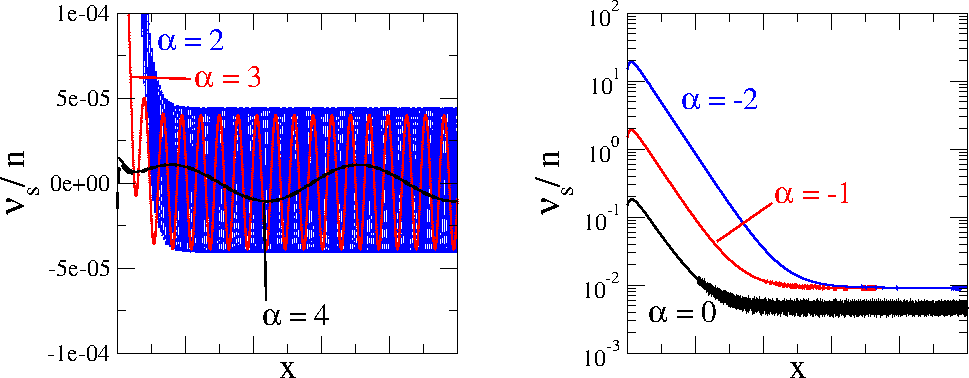}
\caption{Same as Fig. \ref{fig01}, showing $\nu_s/n$ instead of $y_s$.}
\label{fig02}
\end{center}
\end{figure}

However, when we analyze the time evolution of $\nu_s$ instead of $y_s$, we observe that when $\gamma_i\ll n$, the solution oscillates around zero 
and the amplitude of the oscillation of $\nu_s$ increases when $\alpha$ decreases (\textit{left} panel of Fig. \ref{fig02}). When $\gamma_i\gg n$. this 
amplitude decreases when $\alpha$ decreases and $\nu_s$ tends to $12ne^2$, independently of the value of $\alpha$ (\textit{right} panel of Fig. \ref{fig02}).
\begin{figure}[h]
\begin{center}
\includegraphics[scale=0.45]{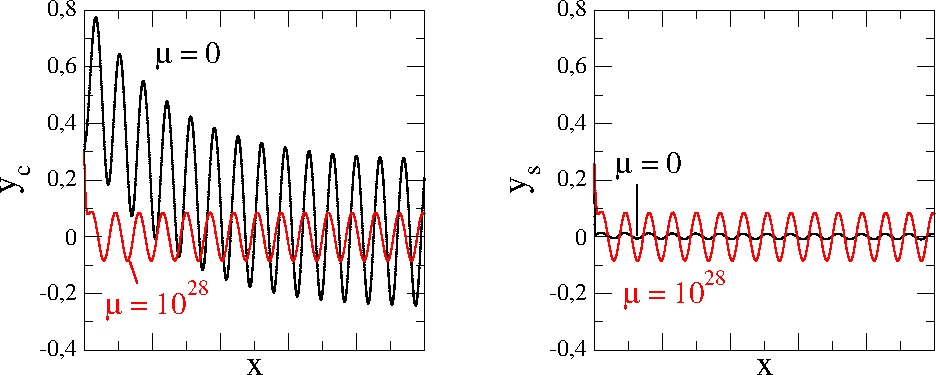}
\caption{Evolution of $y_c$ and $y_s$ for initial conditions $y_c=0.3$ and $y_s=0.15$, relaxation factors such that $\log_{10}{(n/\gamma_c)}=2$ and
$\log_{10}{(n/\gamma_s)}=4$, and two values of the friction parameter: $\mu=0$ (\textit{black}) and $\mu=10^{28}$ {\rm kg km$^2$s$^{-1}$} (\textit{red}).}
\label{fig03}
\end{center}
\end{figure}

When $\gamma_c\neq\gamma_s$, we can have different core and shell rotation behavior. In Fig. \ref{fig03}, we show the core and shell rotation (\textit{left} 
and \textit{right}, respectively) for $\log_{10}{(n/\gamma_c)}=2$ and $\log_{10}{(n/\gamma_s)}=4$. We also set two very different values for the friction: 
the frictionless case $\mu=0$ (black) and a very high value of friction $\mu=10^{28}$ {\rm kg km$^2$s$^{-1}$} (red lines), larger than the 
expected value in the case of Titan ($\mu=10^{11}-10^{13}$ {\rm kg km$^2$s$^{-1}$}), which corresponds to a typical ocean viscosity $\eta_o=\eta_{H_2O}
\approx10^{-3}$ {\rm Pa s} and a large range for the ocean thickness $h$ (see Eq. \ref{eq:mu}). In the frictionless case, we can observe the differential 
rotation between the core and the shell. After a transient, both solutions oscillate around zero with very different amplitudes, depending on the value of 
$\gamma$ of each surface. For very high friction parameter, both layers rotate with the same angular velocity. The core and the shell have the same 
amplitude of oscillation and phase, keeping the relative velocity equal to zero.
\begin{figure}[h]
\begin{center}
\includegraphics[scale=0.5]{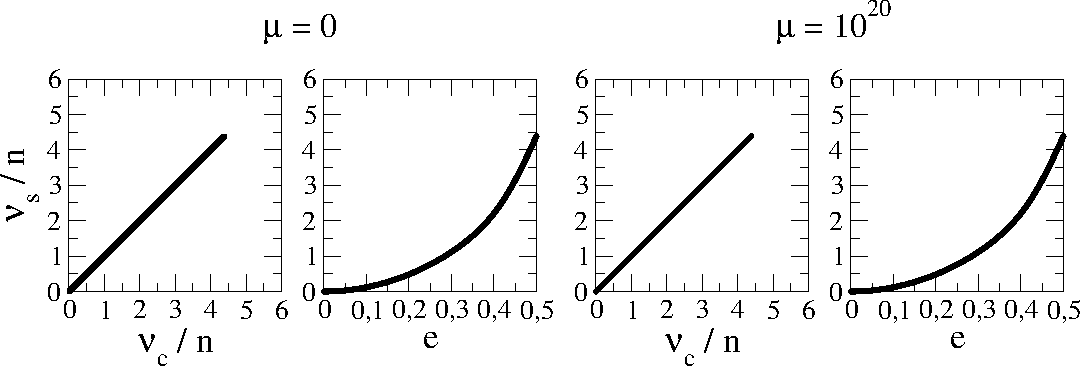}
\caption{Stationary super-synchronous family with relaxation factors equal and such that $n/\gamma_c=n/\gamma_s=0.01$ and $0\leq e\leq 0.5$. \textit{Left}: 
$\mu=0$. \textit{Right}: $\mu=10^{20}$ {\rm kg km$^2$s$^{-1}$}.}
\label{fig04}
\end{center}
\end{figure}

Finally, we study the dependence of the stationary solutions on the eccentricity. For that sake, we choose a grid of initial conditions $\nu_c/n$ and 
$\nu_s/n$, and integrate the system (\ref{eq:system-rotation}) until the stationary solution is reached. When $n/\gamma_c=n/\gamma_s\ll 1$, all initial 
conditions lead to the same equilibrium point (a super-synchronous rotation), independently of the value of the friction parameter. The value of the excess 
of rotation depends only on the eccentricity. In the \textit{left} panels of Fig. \ref{fig04}, we show the family of stationary solutions, where each point 
corresponds to a different eccentricity value in $0\leq e\leq 0.5$. If the eccentricity is zero, the rotations are synchronous to the orbital motion. When 
the eccentricity increases, the rotations become super-synchronous, and the excess of rotation $\nu_i/n$ is proportional to $e^2$ (\textit{right} panels). 

\begin{figure}[h]
\begin{center}
\includegraphics[scale=0.5]{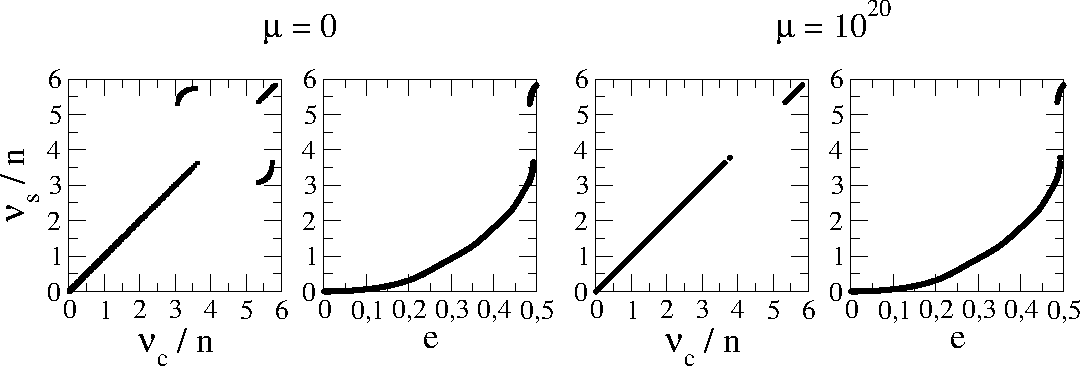}
\caption{Same as Fig. \ref{fig04} for $n/\gamma_c=n/\gamma_s=1$ and $0\leq e\leq 0.5$.}
\label{fig05}
\end{center}
\end{figure}
When $n/\gamma_c$ and $n/\gamma_s$ increase, that is, when the viscosities increase, the excess in the super-synchronous rotation decreases. If the 
eccentricity is low, the only attractor is the super-synchronous solution. When the eccentricity increases, captures in other attractors $\nu_i\simeq n,2n,3n,\dots$ 
appear gradually. This behavior is the same studied by in FM15 and also in Correia et al. (2014) in the case of homogeneous bodies.

\begin{figure}[h]
\begin{center}
\includegraphics[scale=0.45]{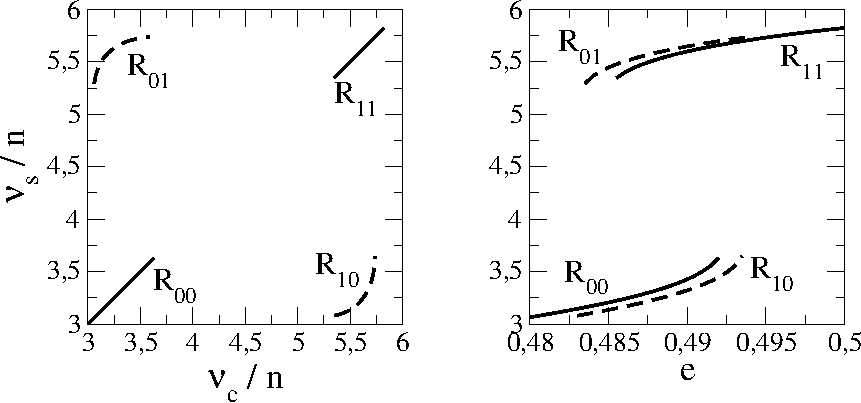}
\caption{Families of stationary rotations with equal relaxation factors such that $n/\gamma_c=n/\gamma_s=1$, $0\leq e\leq 0.5$ and $\mu=0$. Labels $R_{pq}$ 
indicates the two frequencies: $\nu_c=pn$ and $\nu_s=qn$.}
\label{fig06}
\end{center}
\end{figure}

Figure \ref{fig05} shows the families of stationary rotation for $n/\gamma_c=n/\gamma_s=1$, $0\leq e\leq 0.5$ and two values of the friction parameter: the 
frictionless case, with $\mu=0$ (\textit{top} panels), and a very high friction case, with $\mu=10^{20}$ {\rm kg km$^2$s$^{-1}$} (\textit{bottom} panels). 
In the frictionless case, when the eccentricity is smaller than $\sim0.48$, only the super-synchronous solution is possible. If the eccentricity is larger 
than $0.48$, besides the super-synchronous solution, three new stationary configurations appear: The core and the shell in the 3/2 commensurability 
($\nu_c\simeq n$ and $\nu_s\simeq n$), the core in super-synchronous rotation and the shell in the 3/2 commensurability ($\nu_c\simeq 0$ and $\nu_s\simeq n$), 
and the core in the 3/2 commensurability and the shell in super-synchronous rotation ($\nu_c\simeq n$ and $\nu_s\simeq 0$). Figure \ref{fig06} shows in more 
detail these stationary solutions. The labels $R_{pq}$ denote the stationary families indicating the resonances $\nu_c=pn$ and $\nu_s=qn$. It is important 
to note that the excesses in the rotations are large because the eccentricity is high. In the high friction case (\textit{bottom} panels of Fig. \ref{fig05}), 
only the stationary solutions with the same commensurabilities survive because in these configurations, the relative velocity of rotation between the core 
and the shell is zero.

\begin{figure}[h]
\begin{center}
\includegraphics[scale=0.5]{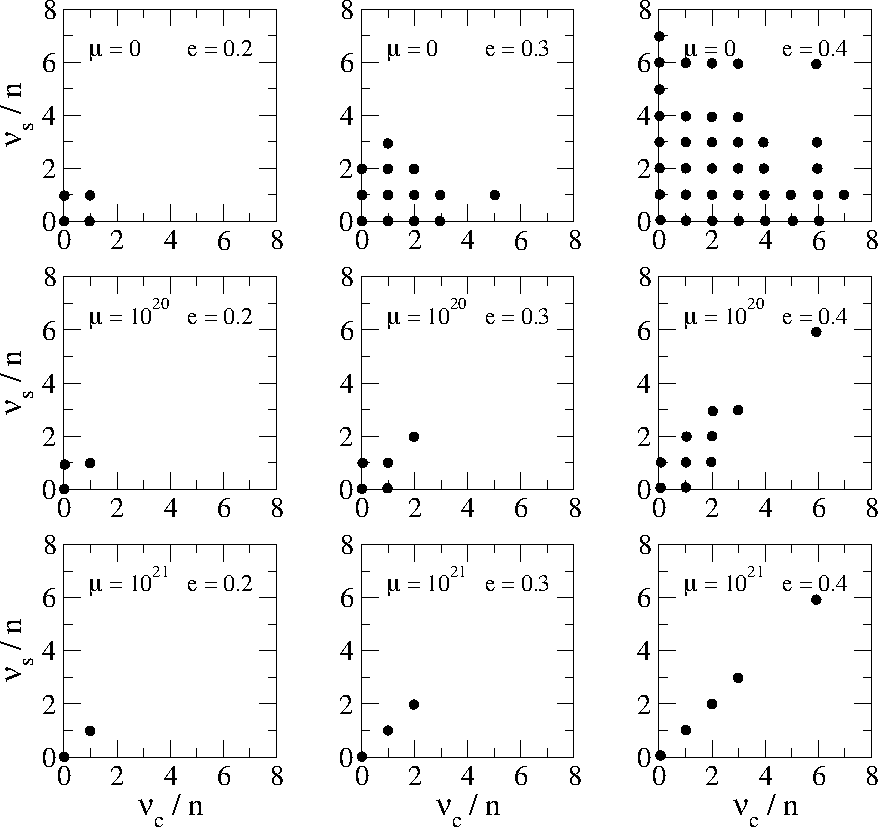}
\caption{Attractors when the relaxation factors are equal such that $n/\gamma_c=n/\gamma_s=10$. The friction $\mu$ increases from \textit{top} to 
\textit{bottom} and the eccentricity $e$ increases from \textit{left} to \textit{right}. The units of $\mu$ are {\rm kg km$^2$s$^{-1}$}.}
\label{fig07}
\end{center}
\end{figure}

If $n/\gamma_c$ and $n/\gamma_s$ continue to increase and the friction parameter is low (not necessarily zero), the core and the shell may tend to different 
resonances, depending on the eccentricity. If the friction increases, the attractors with higher differential rotation, begin to disappear, until eventually, 
as from a certain value limit of $\mu$ only survive the attractors with differential rotation zero Fig. \ref{fig07}.

\section{Application to Titan's rotation}\label{sec10}

\subsection{The model}\label{sec10.1}
 
\begin{table}[h]
\caption{Basic data}\label{Data-tab}
\begin{tabular}{lccl}
\hline
Mass (10$^{22}$kg)$^{(a)}$                            & $m_T$                                      & 13.45                              \\
Eccentricity$^{(b)}$                                  & $e$                                        & 0.028                              \\
Semi-major axis (AU)$^{(c)}$                          & $a$                                        & 0.00816825                         \\
Mean motion (deg/day)$^{(a)}$                         & $n$                                        & 22.5769768                         \\
(id.) ($10^{-6}$ {\rm s}$^{-1}$)                      &                                            & 4.560678013                        \\ 
Orbital period (day)$^{(a)}$                          & $2\pi/n$                                   & 15.9454476                         \\
Differential Rotation (deg/yr)                        & $\Omega_s-n$                               & $0.122\pm 0.040$ $^{(d)}$          \\
                                                      &                                            & $0.00_{-0.02}^{+0.02}$ $^{(e)}$    \\
Titan's ellipsoid semi-major axes (km)$^{(f)}$        & $a$                                        & 2575.152 $\pm$ 0.048               \\
                                                      & $b$                                        & 2574.715 $\pm$ 0.048               \\
                                                      & $ c$                                       & 2574.406 $\pm$ 0.044               \\
Titan's mean equatorial radius (km)$^{(f)}$           & $R_s$                                      & 2574.933 $\pm$ 0.033               \\
Titan's equatorial prolateness ($10^{-4}$)$^{(f)}$    & $\epsilon'_s$                              & 1.70 $\pm$ 0.26                    \\
Saturn's mass ($10^{26}$ kg)$^{(g)}$                  & $M$                                        & 5.68326                            \\
Saturn's mean-motion ($10^{-9}$ s$^{-1}$)$^{(g)}$     & $n_\odot$                                  & 6.713428                           \\
Titan's tidal parameter ($10^{-15}$ s$^{-2}$)$\dag$   & $\mathcal{T}$                              & 4.63                               \\
\hline
\multicolumn{3}{l}{$^{(a)}$Seidelmann et al. (2007); $^{(b)}$Iess et al. (2012); $^{(c)}$TASS 1.8 (see Vienne and Duriez, 1995) (Jan.1,2000);}\\
\multicolumn{3}{l}{$^{(d)}$Stiles et al. (2010); $^{(e)}$Meriggiola et al. (2016); $^{(f)}$Mitri et al. (2014);} \\
\multicolumn{3}{l}{ $^{(g)}$Jacobson et al. (2006); $\dag$ calculated parameter.} \\
\end{tabular}\vspace*{3mm}
\end{table}

Titan's interior was largely discussed in many papers (e.g. Tobie et al., 2005; Castillo-Rogez and Lunine, 2010; McKinnon and Bland, 2011; Fortes, 2012). 
The existing general data of the Titan-Saturn system is given in Table \ref{Data-tab}. In this section, we assume the interior model given by Sohl et al. 
(2014) (hereafter \textit{reference model}), which is given in Table \ref{Table-sohl2014}. In this model, Titan is formed by four homogeneous layers: i) an 
inner hydrated silicate core (inner core); ii) a layer of high-pressure ice (outer core); iii) a subsurface water-ammonia ocean and iv) a thin ice crust. 
For the sake of simplicity, we construct one two-layer equivalent model, where the \emph{core} is a layer formed by the inner core and the high-pressure 
ice layer, and the \emph{shell} is a layer formed by the subsurface ocean and the ice crust, but keeping some features of the four-layer model (e.g. axial 
moments of inertia and Clairaut numbers). In this way, we can use the rotational equations (\ref{eq:system-rotation}), retaining the main features of the 
realistic reference model. This simplified model is given in Table \ref{Table-2l-model}, and some calculated parameters of each layer are listed in Table 
\ref{Table-parameters-2l-I}. The existence of relative translational motions due to the non-coincidence of the barycenters of the several layers, as 
discussed by Escapa and Fukushima (2011) in the case of an icy body with an internal ocean and solid constituents, has not been taken into account.

\begin{table}[h]
\caption{Titan's four-layer reference model}\label{Table-sohl2014}
\begin{tabular}{lcccl}
\hline 
Layer              & Outer radius ({\rm km}) & Density (g/cm$^3$) & Mass (10$^{22}${\rm kg}) & Viscosity$\dag$ ({\rm Pa s})\\
\hline
Ice I shell        & 2575              & \phantom{1}0.951   &  \phantom{0}0.84         & $10^{14}-10^{16}$                 \\
Ocean              & 2464              & 1.07               &  \phantom{0}1.36         & $10^{-3}-10^{9}$ $\S$ \\
High-pressure ice  & 2286              & 1.30               &  \phantom{0}1.58         & $10^{15}-10^{20}$                 \\
Rock and iron core & 2084              & 2.55               &  \phantom{0}9.67         & $10^{20}$                         \\
\hline 
\multicolumn{5}{l}{$\dag$ Mitri et al. 2014; $\S$ adopted values.}
\end{tabular}\vspace*{3mm}
\end{table}

\begin{table}[h]
\caption{Titan's two-layer equivalent model}\label{Table-2l-model}
\begin{tabular}{lccc}
\hline 
Layer                         & Outer radius ({\rm km}) & Density (g/cm$^3$) & Mass (10$^{22}${\rm kg})  \\
\hline
Shell (crust + ocean)         & 2575              & 1.02               &  \phantom{0}2.19           \\
Core (rock + HP ice mantle)   & 2286              & 2.25               &  11.26                     \\
\hline 
\end{tabular}\vspace*{3mm}
\end{table}

In order to estimate the relative height of the elastic tide $\lambda_s$, we assume that the difference between the observed surface flattening $\epsilon'_s$ 
with the tidal flattening $\epsilon_s=\mathcal{H}_s\overline{\epsilon}_\rho E_{2,0}\cos{\sigma_{s0}}\approx\mathcal{H}_s\overline{\epsilon}_\rho\cos{\sigma_{s0}}$ 
(calculated) is due to the existence of an elastic component, with flattening $\epsilon_s^{(el)}=\lambda_s\mathcal{H}_s\overline{\epsilon}_\rho$ (see Appendix 
3 for more details). If we use Eq. (\ref{eq:real flattenings}), and assume that near the synchronous rotation $\cos^2{\sigma_{s0}} \approx 1$, we obtain 
\begin{equation}
 \lambda_s  \approx \frac{\epsilon'_s}{\mathcal{H}_s\overline{\epsilon}_\rho} - 1.
\end{equation}
For the relative heights of the elastic tide $\lambda_c$, we assume $\lambda_c \approx \lambda_s\defeq\lambda$.

\subsection{Atmospheric influence on Titan's rotation}{\label{sec10.2}}

The seasonal variation in the mean and zonal wind speed and direction in Titan's lower troposphere causes the exchange of a substantial amount of angular 
momentum between the surface and the atmosphere. The variation calculated from the observed zonal wind speeds shows that the atmosphere angular momentum 
undergoes a periodic oscillation between $3 \times 10^{18}$ and $3 \times 10^{19} {\,\rm kg} {\,\rm km}^{2}{\rm s}^{-1}$ (Tokano and Neubauer, 2005, 
hereafter TN05) with a period equal to half Saturn's orbital period and maxima at Titan's equinoxes (when the Sun is in the satellite's equatorial plane).

The angular momentum of the atmosphere may be written as $L_{\rm atm}=L_0+L_1\cos 2\alpha_\odot$ where $L_0=1.65\times10^{19}$ {\rm kg km$^2$ s$^{-1}$}, 
$L_1=1.35\times10^{19}$ {\rm kg km$^2$ s$^{-1}$} and $\alpha_\odot$ is the Saturnian right ascension of the Sun. The variation of the angular momentum is 
$\dot{L}_{\rm atm}=-2L_1 n_\odot \sin 2\alpha_\odot $. If we neglect external effects (as atmospheric tides), this variation may be compensated by an equal 
variation in the shell's angular momentum: $\delta\dot{L}_s= -\dot{L}_{\rm atm}$, which corresponds to an additional shell acceleration 
\begin{equation}
\delta\dot{\Omega}_s=\frac{2L_1 n_\odot}{C_k} \sin 2\alpha_\odot=A_\odot\sin 2\alpha_\odot.
\label{eq:atmosphere}
\end{equation}

We must emphasize that we have considered in these calculations the moment of inertia of the ice crust $C_k$, since the winds are acting on the crust and do 
not have direct action on the liquid part of the shell.

In a more recent work, Richard et al. (2014) (hereafter R14) re-calculate the amplitude of the variation of the angular momentum with the Titan IPSL GCM 
(Institut Pierre-Simon Laplace General Circulation models) (Lebonnois et al., 2012). They obtain $L_1=8.20\times10^{17}$ {\rm kg km$^2$ s$^{-1}$}, which is 
$\sim16.5$ times less than the TN05 value.
\begin{table}[h]
\caption{Titan's calculated parameters in the two-layer model}\label{Table-parameters-2l-I}
\begin{tabular}{lccc}
\hline 
                                                            &                      & Core    & Shell \\
\hline 
Clairaut number                                             & $\mathcal{H}_i$      & 0.772   & 0.806 \\
Axial moment of inertia (10$^{29}${\rm kg km$^2$})          & $C_i$                & 2.183   & 0.866 \\
Equatorial flattening (tidal) (10$^{-4}$)                   & $\epsilon_i$         & 1.15    & 1.20  \\
Relative height of the elastic tide                         & $\lambda_i$          & 0.42    & 0.42  \\
Gravitational coupling constant (10$^{-15}${\rm s$^{-2}$})  & $K/C_i$              & 2.65    & 6.69  \\
Friction parameter (10$^{-17}${\rm s$^{-1}$}) $\S$          & $\mu/C_i$            & 0.59    & 1.48  \\
Atmospheric parameter (10$^{-18}${\rm s$^{-2}$})$\dag$      & $2L_1 n_\odot/C_k$   & -       & 5.08  \\
                                                            & $2L_1 n_\odot/C_s$   & -       & 0.31  \\
\hline
\multicolumn{4}{l}{$\S$ Assuming $\eta_o=10^{-3}$ {\rm Pa s}; $\dag$ $L_1=1.35\times 10^{19}$ {\rm kg km$^2$ s$^{-1}$} (Tokano and Neubauer, 2005).}
\end{tabular}\vspace*{4mm}
\end{table}

\subsection{The results}{\label{sec10.3}}

We fix the outer radius of the inner core $R_{\rm{ic}}$ and the outer radius of the high-pressure ice layer $R_{\rm{oc}}$, the densities of the inner and outer 
cores $d_{\rm{ic}}$ and $d_{\rm{oc}}$ and the density of the crust $d_k$, to the reference model values in Table \ref{Table-sohl2014}. The density of the 
inner core is calculated so as to verify the value of Titan's mass $m_T=13.45\times10^{22}$ kg.
\begin{figure}[h]
\begin{center}
\includegraphics[scale=0.5]{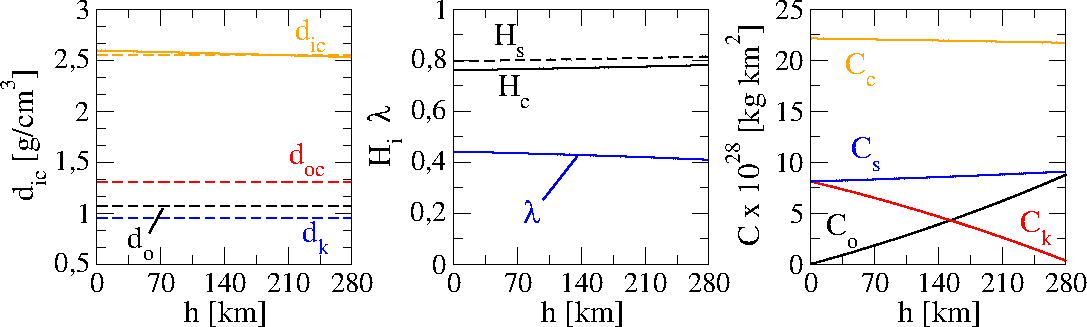}
\caption{Dependence of some parameters on the thickness of the ocean $h$. \textit{Left}: Density of the inner core $d_{\rm{ic}}$ (\textit{solid orange}) and 
the densities of the reference model (dashed lines). \textit{Middle}: Clairaut parameters $\mathcal{H}_i$ (\textit{black}) and the maximum relative height 
of the elastic tide $\lambda$ (\textit{blue}). \textit{Right}: The axial moments of inertia of the ocean $C_o$ (\textit{black}), the crust $C_k$ 
(\textit{red}), the shell $C_s=C_o+C_k$ (\textit{blue}) and the core $C_c=C_{\rm{ic}}+C_{\rm{oc}}$ (\textit{orange}).}
\label{fig08}
\end{center}
\end{figure}
Figure \ref{fig08} shows the weak dependence of the parameters on the thickness of the ocean $h$: the density of the inner core $d_{\rm{ic}}$ (solid orange 
line) and densities of the reference model (\textit{left} panel); the Clairaut numbers $\mathcal{H}_c$, $\mathcal{H}_s$ (\textit{middle} panel); and the 
axial moments of inertia $C_c$ and $C_s$ (\textit{right} panel).

\begin{figure}[h]
\begin{center}
\includegraphics[scale=0.5]{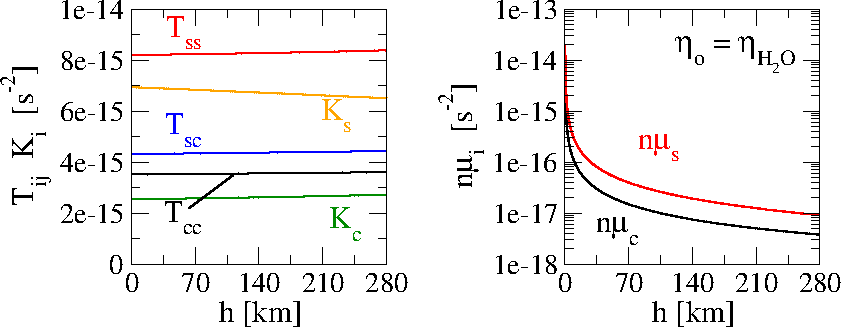}
\caption{Dependence of some parameters with the thickness of the ocean $h$. \textit{Left}: Tidal parameter $T_{ij}$ and gravitational coupling constant 
$K_i=K/C_i$. \textit{Right}: The coefficient $n\mu_i$, where $\mu_i=\mu/C_i$, for a typical ocean viscosity $\eta_o=\eta_{H_2O}\approx10^{-3}$ 
{\rm Pa s}.}
\label{fig09}
\end{center}
\end{figure}

The main consequence of the weak dependence of these parameters with the thickness of the subsurface ocean, is that both the effect of the tide and the 
gravitational coupling parameter also depend weakly on $h$. The strength of the acceleration of the rotation, due to the tide, is given by the product 
$T_{ij}\mathcal{T}_k$ (see Eqs. \ref{eq:T_i} and \ref{eq:T_ij}). While the parameter $T_{ij}$ only depends on the internal structure of Titan, the 
function $\mathcal{T}_k$ do not depend on $h$. The \textit{left} panel of Fig. \ref{fig09} shows $T_{ij}$ and the gravitational coupling amplitude 
$K_i=K/C_i$, as function of $h$. We also observe that the thickness of the ocean does not have any relevant role. Then, for the tide and the gravitational 
coupling, the rotational evolution is driven by the ratios $n/\gamma_c$, $n/\gamma_s$ and the orbital eccentricity $e$.

The \textit{right} panel of Fig. \ref{fig09} shows the quantity $n\mu_i=n\mu/C_i$ as function of the thickness $h$, when we consider the realistic ocean 
viscosity $\eta_o=\eta_{H_2O}\approx 10^{-3}$ {\rm Pa s}. The rotational acceleration of each layer, due to the friction, is $\mu_i(\Omega_s-\Omega_c)$. In 
super-synchronous rotation, the excess of rotation of each layer is of order $ne^2$, then
$$\mu_i(\Omega_s-\Omega_c)\ll n\mu_i\ll T_{ij},K_i.$$
Therefore, in Titan's case, the friction term is negligible compared with the tide and the gravitational coupling terms, independently of the $h$ value.

\begin{figure}[h]
\begin{center}
\includegraphics[scale=0.4]{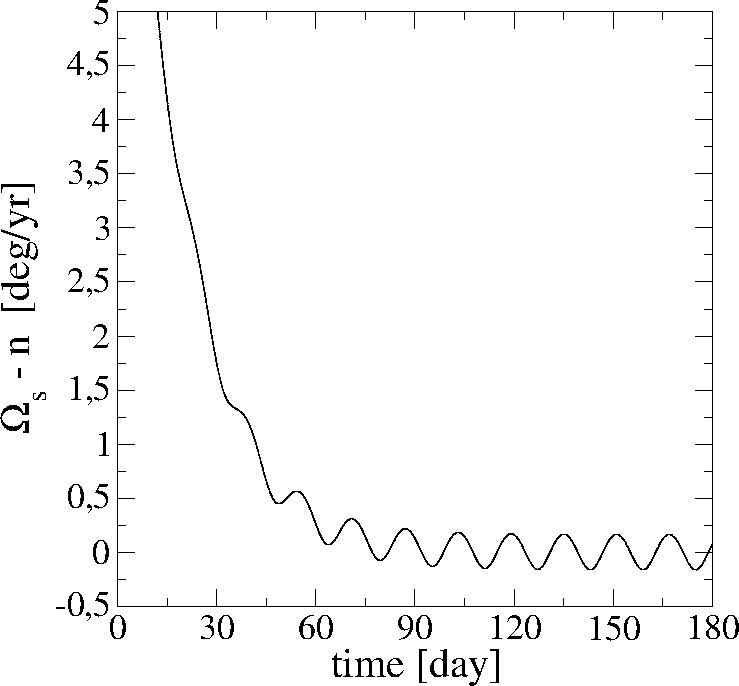}
\caption{Time evolution of $\Omega_s-n$, when $\gamma_c=\gamma_s=10^{-8}$ {\rm s}$^{-1}$, $\eta_o=10^{-3}$ {\rm Pa s} and $h=178$ {\rm km}.}
\label{fig10}
\end{center}
\end{figure}

Equations (\ref{eq:system-rotation}) and (\ref{eq:atmosphere}), allow us to calculate the velocities of rotation of the shell and of the core of Titan 
for a wide range of relaxation factors $\gamma_c$ and $\gamma_s$, when different effects are considered. For that sake, we have to adopt the values of the 
involved parameters. We use four different values for the viscosity of the subsurface ocean: a realistic value $\eta_o=\eta_{H_2O}=10^{-3}$ {\rm Pa s}, a 
moderate value $\eta_o=10^{0}$ {\rm Pa s} and two very high values $\eta_o=10^{6}$ {\rm Pa s} and $\eta_o=10^{9}$ {\rm Pa s}. For the thickness of the ocean, 
we use the values $h=15,$ 178 and 250 {\rm km}, and for the variation of the atmospheric angular momentum, we use the values given by Tokano and Neubauer 
(2005) and Richard et al. (2014). When we integrate the rotational equations, assuming the values of relaxation factor typical for rock bodies ($\gamma_i<n$), 
the results show that the excess of rotation of the shell is damped quickly and the final state is an oscillation around the synchronous motion with a period 
of $\sim15$ days (a periodic attractor), equal to the orbital period (Fig. \ref{fig10}). The amplitude of this oscillation depends on the relaxation 
factors and the ocean thickness.

\begin{figure}[h]
\begin{center}
\includegraphics[scale=0.5]{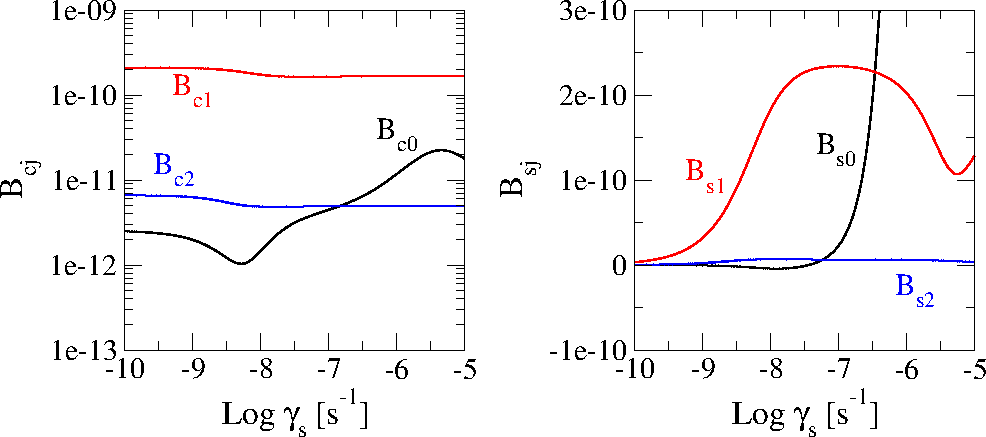}
\caption{Tidal drift and amplitudes of oscillation of the periodic terms of the Titan's core and shell in function of the shell relaxation factor $\gamma_s$. 
The core relaxation factor is $\gamma_c=10^{-8}$ {\rm s}$^{-1}$ and the ocean's viscosity and thickness are $\eta_o=10^{-3}$ {\rm Pa s} and $h=178$ {\rm km}, 
respectively. \textit{Left}: Core's parameters $B_{c0}$, $B_{c1}$ and $B_{c2}$. \textit{Right}: Shell's parameters $B_{s0}$, $B_{s1}$ and $B_{s2}$.}
\label{fig11}
\end{center}
\end{figure}

The periodic attractor of the spin rate $\nu_i$ of each layer can be approximated by the trigonometric polynomial
\begin{equation}
\nu_i\simeq B_{i0} + B_{i1}\cos{(\ell+\phi_{i1})} + B_{i2}\cos{(2\ell+\phi_{i2})}, 
\end{equation}
where the constants $B_{ij}$ and the phases $\phi_{ij}$, depend on the relaxation factors. The tidal drift $B_{i0}$ also depends on $e^2$, while the 
amplitude of oscillation $B_{ij}$, depends on $e^j$. The coefficients $B_{ij}$ and $\phi_{ij}$ gives rise to intricate analytical expressions, but are easy 
to calculate numerically (an analytical construction of these constants is presented in the Section B of the Online Supplement). Figure \ref{fig11} shows 
one example for Titan's core and shell constants $B_{cj}$ and $B_{sj}$, as a function of the shell relaxation factor, when the core relaxation factor is 
$\gamma_c=10^{-8}$ {\rm s}$^{-1}$, and the ocean's viscosity and thickness are $\eta_o=10^{-3}$ {\rm Pa s} and $h=178$ {\rm km}, respectively. We can observe 
that if $\gamma_s\gtrsim10^{-7.5}$ {\rm s}$^{-1}$, the shell oscillates around the super-synchronous rotation. When $\gamma_s\lesssim10^{-7.5}$ {\rm s}$^{-1}$, 
the tidal drift $B_{s0}$ tends to zero and the shell oscillates around the synchronous rotation, with a period of oscillation equal to the orbital period. 
Finally, if $\gamma_s\lesssim10^{-8}$ {\rm s}$^{-1}$, the amplitude of the shell rotation decreases, tending to zero when $\gamma_s$ decreases. On the other 
hand, the core oscillates around the synchronous rotation, with a period of oscillation equal to the orbital period, independently of the shell relaxation 
factor.

\begin{figure}[h]
\begin{center}
\includegraphics[scale=0.6]{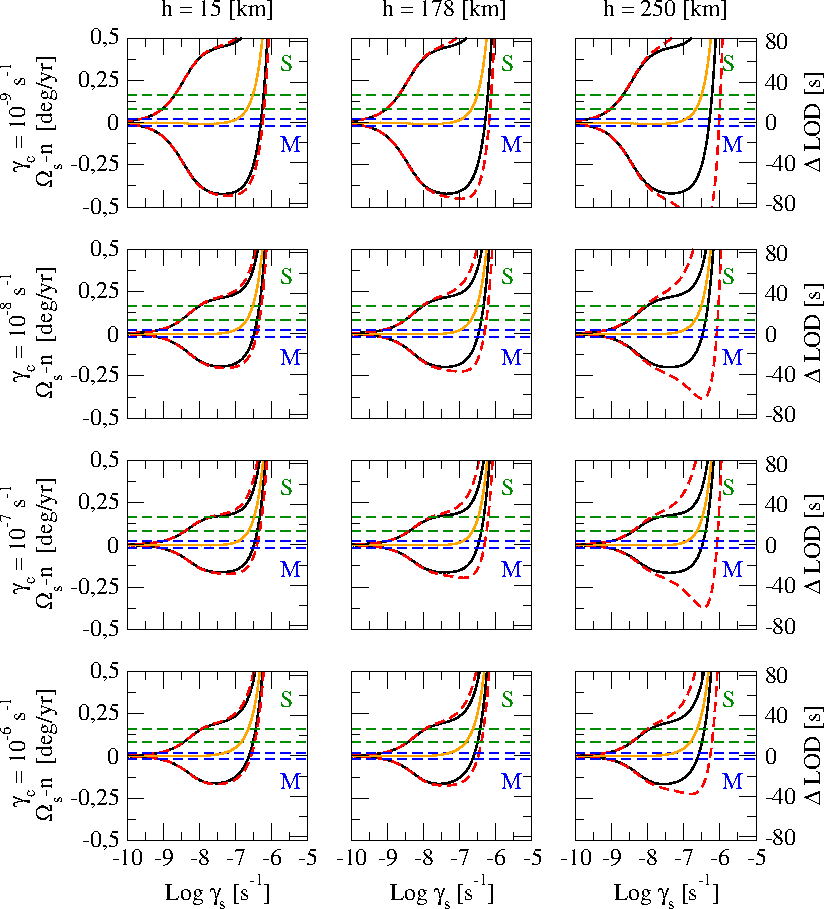}
\caption{Shell rotation and corresponding length-of-day variation of Titan in function of the relaxation factors, when $\eta_o=10^{-3}$ {\rm Pa s} and 
$L_1=1.35\times10^{19}$ {\rm kg km$^2$ s$^{-1}$}. The core relaxation factor $\gamma_c$ increases from \textit{top} to \textit{bottom} and the ocean 
thickness $h$ increases from \textit{left} to \textit{right}. We consider two dynamical models: the pair of solid black lines indicate the maximum and 
minimum of the shell rotation when the tidal forces, the gravitational coupling and the linear friction are taken into account, and the pair of dashed red 
lines indicate the maximum and minimum of the shell rotation when the angular momentum exchange with the atmosphere is added. The orange solid line 
indicates the analytical stationary rotation $B_{s0}$. The horizontal dashed lines show the confidence interval of the observed values, as determined by 
Meriggiola (2012) (\textit{blue}) and by Stiles et al. (2010) (\textit{green}).}
\label{fig12}
\end{center}
\end{figure}

In Fig. \ref{fig12}, fixing $\eta_o=10^{-3}$ {\rm Pa s} and $L_1=1.35\times10^{19}$ {\rm kg km$^2$ s$^{-1}$} (TN05), we plot the resulting maximum and 
minimum of the final oscillation of the shell rotation $\Omega_s-n$, or, equivalently, the length-of-day variation 
\begin{equation}
\Delta\ \textnormal{LOD}= \frac{2\pi}{n}-\frac{2\pi}{\Omega_s}, 
\end{equation}
in function of $\gamma_s$, for two dynamical models: i) tidal forces, gravitational coupling and linear friction (solid black lines); and ii) tidal forces, 
gravitational coupling, linear friction and the atmospheric influence (dashed red lines). The horizontal lines show the intervals corresponding to $1\sigma$ 
uncertainties of the observed values: the blue dashed lines, labelled \texttt{M}, correspond to Meriggiola (2012) and Meriggiola et al. (2016) and green 
dashed lines, labelled \texttt{S}, correspond to the Stiles et al. (2010). The core relaxation factor $\gamma_c$ increases from $\gamma_c = 10^{-9}$ {\rm 
s$^{-1}$} (\textit{top} panels) to $10^{-6}$ {\rm s$^{-1}$} (\textit{bottom} panels) and the ocean thickness $h$ increases from 15 {\rm km} (\textit{left} 
panels) to 250 {\rm km} (\textit{right} panels).

Figure \ref{fig12} shows that if $\gamma_s<10^{-7}$ {\rm s$^{-1}$}, the shell's rotation oscillate around the synchronous motion and the amplitude of 
oscillation depends on the relaxation factors and the ocean thickness. The average rotation (central orange line) is synchronous; it only becomes 
super-synchronous for relaxation values larger than $\sim10^{-6.5}$ {\rm s$^{-1}$}. We also observe that when $\gamma_s<10^{-8}$ {\rm s$^{-1}$}, 
independently of the values of $\gamma_c$ and $h$, the amplitude of oscillation of the shell tends to zero when the relaxation factor $\gamma_s$ decreases. 
Particularly, if $\gamma_s<10^{-9}$ {\rm s}$^{-1}$, the amplitude of the oscillation of the excess of rotation reproduces the dispersion of the $\Omega_s$ 
value of $\pm 0.02$ {\rm deg/yr} around the synchronous value, observed as reported by Meriggiola (2012) and Meriggiolla et al. (2016). The results are not 
consistent with the previous drift reported by Stiles et al. (2008, 2010). We note that for larger values of the relaxation, e.g. $10^{-8}$ {\rm s$^{-1}$}, 
the large short period oscillation due to the tide would be much larger than the reported values and would introduce big dispersion in the measurements, 
much larger than the reported dispersion due to the difficulties in the precise localization of Titan's features. On the other hand, the effect of the 
atmospheric torque is completely negligible in the range of possible $\gamma_s$ that reproduces the observed values of the shell rotation, even for the high 
value of $L_1$ given Tokano and Neubauer (2005). When we consider the amplitude of the variation of the angular momentum given by Richard et al. (2014), the 
contribution to the rotation variations tends to zero. 

\begin{figure}[h]
\begin{center}
\includegraphics[scale=0.6]{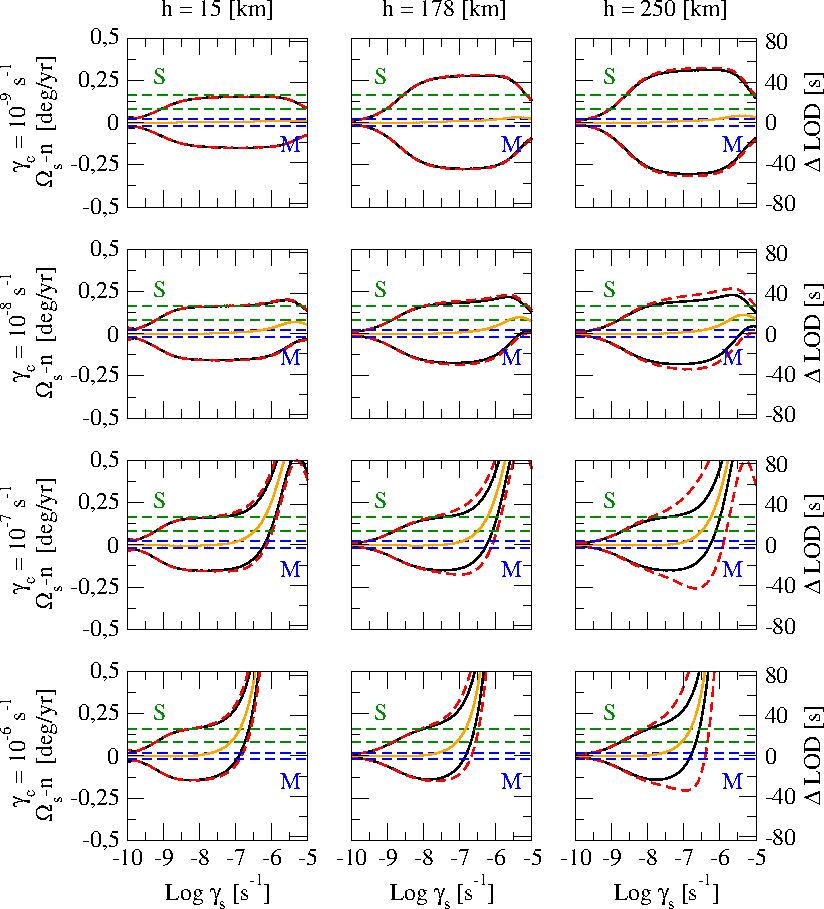}
\caption{Same as Fig. \ref{fig12} for $\eta_o=10^{9}$ {\rm Pa s}.}
\label{fig13}
\end{center}
\end{figure}

The results shown in Fig. \ref{fig12} remain virtually unchanged when the ocean viscosity is increased up to a value of $\eta_o= 10^6$ {\rm Pa s}. But if 
the ocean viscosity is increased to $\eta_o= 10^9$ {\rm Pa s}, the transfer of angular momentum between the shell and the core induces in the shell 
accelerations of the same order as the rotational acceleration due to the others forces. As a consequence, the shell rotation will follow the core rotation 
closely (which is shown in Fig. \ref{fig13}). This high value of $\eta_o$ can be interpreted as the ocean thickness tending to zero. In this case, to 
obtain the dispersion of Titan's observed rotation as determined by Meriggiola et al. (2016) we should have a value of $\gamma_s$ smaller than the values 
obtained in the previous cases, where a low viscosity ocean was assumed between the shell and the core. It is worth noting yet that, in this case, 
the observed dispersion could also be obtained taking for $\gamma_c$ an extremely low value ($10^{-9}$ {\rm s$^{-1}$}) and for $\gamma_s$ a much larger and 
unexpected value ($10^{-5}$ {\rm s$^{-1}$}). 

\begin{figure}[h]
\begin{center}
\includegraphics[scale=0.6]{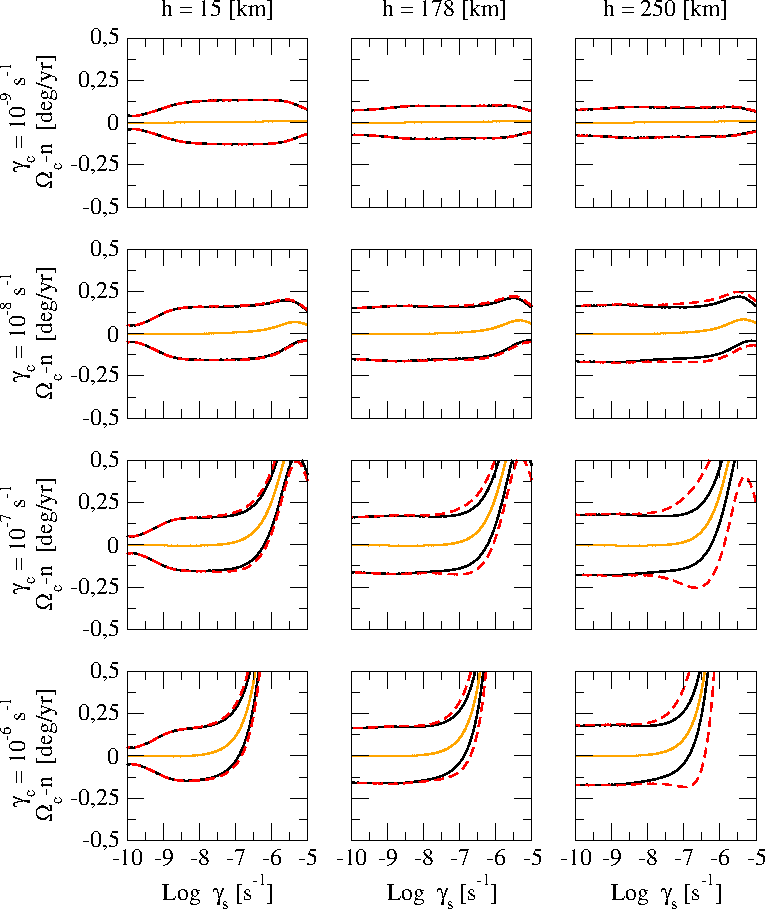}
\caption{Core rotation of Titan in function of the relaxation factors, when $\eta_o=10^{9}$ {\rm Pa s} and $L_1=1.35\times10^{19}$ {\rm kg km$^2$ s$^{-1}$}. 
The core relaxation factor $\gamma_c$ increases from \textit{top} to \textit{bottom} and the ocean thickness $h$ increases from \textit{left} to 
\textit{right}. We consider two dynamical models: the pair of solid black lines indicate the maximum and minimum of the core rotation when the tidal 
forces, the gravitational coupling and the linear friction are taken in account, and the pair of dashed red lines indicate the maximum and minimum of the 
core rotation when the angular momentum exchange with the atmosphere is added. The orange solid line, indicates the analytical stationary rotation $B_{c0}$.}
\label{fig14}
\end{center}
\end{figure}

It is important to note that, in any case, the rotational constraint does not allow us to estimate the value of the core relaxation factor $\gamma_c$. For 
realistic values of the ocean viscosity ($\eta_o=10^{-3}-10^6$ {\rm Pa s}), the shell relaxation factor may be such that $\gamma_s\lesssim10^{-9}$ 
{\rm s$^{-1}$}. The actual value will depend on the values of $h$ and $\gamma_c$ and on the interpretation of the dispersion determined by Meriggiola, which 
may include the forced short-period oscillation of $\Omega_s$. Equivalently, using Eq. (\ref{eq:gamma_i}), the shell viscosity may be such that $\eta_s
\gtrsim10^{18}$ {\rm Pa s}. These values remain without significant changes if $\eta_o<10^{9}$ {\rm Pa s}. For the case in which a subsurface ocean does not 
exist, the shell relaxation factor may be such that $\gamma_s\lesssim10^{-10}$ {\rm s$^{-1}$}, one order less than when an ocean is considered. Equivalently, 
the shell viscosity may be such that $\eta_s\gtrsim10^{19}$ {\rm Pa s}. It is worth noting that in this case, when $\gamma_s\lesssim10^{-7}$ {\rm s$^{-1}$}, 
the rotation of the core remains stuck to the rotation of the shell even when $\gamma_c$ is larger, notwithstanding the larger moment of inertia of the core 
(Fig. \ref{fig14}).

\section{Conclusion}{\label{sec11}}

In this article we extended the static equilibrium figure of a multi-layered body, presented in Folonier et al. (2015), to the viscous case, adapting it to 
allow the differential rotation of the layers. For this sake, we used the Newtonian creep tide theory, presented in Ferraz-Mello (2013) and Ferraz-Mello 
(2015a). Once solved the creep equations for the outer surface of each layer, we obtained the tidal equilibrium figure, and thereby we calculated the 
potential and the forces that act on the external mass producing the tide.

In order to apply the theory to satellites of our Solar System, we calculated the explicit expression in the particular case of one body formed by two 
layers. We may remember that the number of free parameters and independent variables increases quickly when the number of layers increases. The simplest 
version of the non-homogeneous creep tide theory (the two-layer model), allows us to obtain the main features due to the non-homogenity of the body, by 
introducing a minimal quantity of free parameters. In the used model, we have also calculated the tidal torque, which acts on each layer and also the 
possible interaction torques, such as the gravitational coupling and the friction at the interface between the contiguous layers (general development of 
these effects are given in Appendices 3 and 4). The friction was modeled assuming two homogeneous contiguous layers separated by one thin Newtonian fluid 
layer. This model of friction is particularly appropriate for differentiated satellites with one subsurface ocean, as are various satellites of our Solar 
System (e.g. Titan, Enceladus, and Europa).

The two-layer case was compared with the homogeneous case. For that sake, we fixed the free parameters of Titan and studied the main features of 
the stationary solution of this model in function of a few parameters, such as the relaxation factors $\gamma_i$, the friction parameter $\mu$ and the 
eccentricity $e$. When $\gamma_c\approx \gamma_s$, the behavior of the stationary rotations turned out to be identical to the homogeneous case. When 
$\gamma_c\approx \gamma_s\ll n$, the stationary solutions oscillate around the synchronous rotation. When $\gamma_c$ and $\gamma_s$ increase, the 
oscillation tends to zero. Finally, if $\gamma_c\approx \gamma_s\gg n$, the stationary solution is damped to super-synchronous rotation. We have also 
calculated the possible attractors when the eccentricity and the friction parameter $\mu$ are varied. We recovered the resonances trapping in 
commensurabilities $\Omega_c\approx\Omega_s\approx \frac{2+k}{2}n$ (where $k=1,2,3,\ldots \in  \mathbb{N}$) as shown in Ferraz-Mello (2015a) and Correia et 
al. (2014) for the homogeneous case, and we found that if friction remains low, the non-zero differential rotation commensurabilities 
$\Omega_c\sim\frac{2+i}{2}n$ and $\Omega_s\sim\frac{2+j}{2}n$, with $i,j=1,2,3,\ldots \in \mathbb{N}$ and $i\neq j$, are possible. When the friction 
increases, the resonances with higher differential rotation are destroyed. If $\mu$ continues increasing, only the resonances in which core and shell have 
the same rotation survive. This behavior is also observed in the non-homogeneous Darwin theory extension, when one particular \emph{ad hoc} geodetic lag 
and one dynamical Love number for each layer are chosen (Folonier, 2016).

The two-layer model was applied to Titan, but adding to it the torques due to the exchange of angular momentum between the surface and the atmosphere, as 
modeled by Tokano and Neubauer (2005) and by Richard et al. (2014), and the results were compared to the determinations of Titan's rotational velocity as 
determined from Cassini observations by Stiles et al. (2010) and Meriggiola et al. (2016). These comparisons allowed us to constraint the relaxation factor 
of the shell to $\gamma_s\lesssim10^{-9}$ {\rm s$^{-1}$}. The integrations show that for $\gamma_s\lesssim10^{-7.5}$ {\rm s$^{-1}$} the shell may oscillate 
around the synchronous rotation, with a period of oscillation equal to the orbital period, and the amplitude of this oscillation depends on the relaxation 
factors $\gamma_c$ and $\gamma_s$ and the ocean's thickness and viscosity. The tidal drift tends to zero and the rotation is dominated by the main periodic 
term.

The main result was that the rotational constraint does not allow us to confirm or reject the existence of a subsurface ocean on Titan. Only the maximum 
shell's relaxation factor $\gamma_s$ can be determined, or equivalently, the minimum shell's viscosity $\eta_s$, because the icy crust is rotationally 
decoupled from the Titan's interior. When a subsurface ocean is considered, the maximum shell's relaxation factor is such that $\gamma_s\lesssim10^{-9}$ 
{\rm s$^{-1}$}, depending on the ocean's thickness and viscosity values considered. Equivalently, this maximum value of $\gamma_s$, corresponds with a 
minimum shell's viscosity $\eta_s\gtrsim10^{18}$ {\rm Pa s}, some orders of magnitude higher than the modeled by Mitri et al. (2014). When the non-ocean 
case is considered, the maximum shell's relaxation factor is such that $\gamma_s\lesssim10^{-10}$ {\rm s$^{-1}$} and the corresponding minimum shell's 
viscosity is $\eta_s\gtrsim10^{19}$ {\rm Pa s}. For these values of $\gamma_s$, the amplitude of the oscillation of the excess of rotation reproduces the 
dispersion of the $\Omega_s$ value of $\pm 0.02$ {\rm deg/yr} around the synchronous value, observed as reported by Meriggiola (2012) and Meriggiolla et al. 
(2016). It is important to note that in all the cases studied, the influence of the atmosphere can be neglected, since it does not affect the results in the 
ranges of $\gamma_c$ and $\gamma_s$ where the excess of rotation calculated is compatible with the excess of rotation observed.

\begin{acknowledgements}
We wish to thank Michael Efroimsky and one anonymous referee for their comments and suggestions that helped to improve the manuscript. we also thank to 
Nelson Callegari for our fruitful discussions about the gravitational coupling. This investigation was supported by the National Council for Scientific and 
Technological Development, CNPq 141684/2013-5 and 302742/2015-8, FAPESP 2016/20189-9, and by INCT Inespa\c{c}o procs. FAPESP 2008/57866-1 and CNPq 
574004/2008-4.
\end{acknowledgements}

\vfill\eject

\section*{Appendix 1: Relaxation factor}{\label{ap_A}}

\begin{figure}[h]
\begin{center}
\includegraphics[scale=0.5]{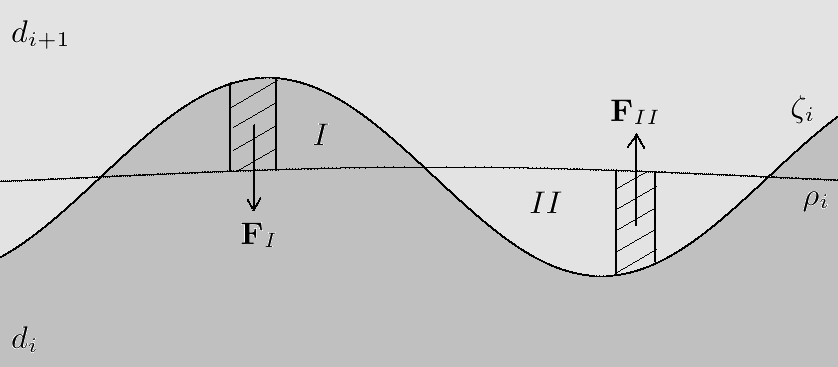}
\caption{Interface between two adjacent homogeneous layers of \tens{m} whose densities are $d_i$ (inner) and $d_{i+1}$ (outer). $\zeta_i(\phi,\theta)$ and 
$\rho_i(\phi,\theta)$ are the actual and the equilibrium surfaces, respectively, of the outer boundary of the $i$th layer. Region I (resp. II) is where the 
actual surface is above (resp. below) the equilibrium surface. $\vec{F}_I$ (resp. $\vec{F}_{II}$) is the force acting on one small element of the 
equilibrium surface in the region I (resp. region II) due to the pressure surplus (resp. pressure deficit).}
\label{fig15}
\end{center}
\end{figure}

Let us consider the equilibrium surface $\rho_i(\phi,\theta)$ between two adjacent homogeneous layers of the body \tens{m} whose densities are $d_i$ (inner) 
and $d_{i+1}$ (outer). We consider that at a given instant, the actual surface between the two layers $\zeta_i(\phi,\theta)$ does not coincide with the 
equilibrium surface (Fig. \ref{fig15}). In some parts, the separation surface is above the equilibrium surface (as in region I) and in other parts it is 
below the equilibrium surface (as in region II). Let us now consider one small element of the equilibrium surface in region I. The pressure in the base 
of this element is positive because the weight of the column above the element is larger than its weight in the equilibrium configuration. Note that the 
column is now partly occupied by the fluid with density $d_i$ and $d_i > d_{i+1}$. The pressure surplus is given by
\begin{equation}
p_I=\Delta{w} h,
\end{equation}
where $\Delta{w} = (d_i-d_{i+1})g$ is the difference of the specific weight of the two columns in the neighborhood of the separation surface, and $h$ is the 
distance of the element of the equilibrium surface to the actual separation surface. $g$ is the local acceleration of gravity.

The radial flow in the considered element is ruled by the Navier-Stokes equation:
\begin{equation}
 0=F_{ext} - \nabla p_I + \eta_i \mathbf{\Delta} \vec{u}
\end{equation}
where $F_{ext}$ is the external force per unit volume, $\vec{u}$ is the radial velocity 
and $\eta_i$ is the viscosity of the layer $i$ (assuming $\eta_i>\eta_{i+1}$). We notice that $\bf{\Delta}$ is operating on a vector, contrary to the usual 
$\Delta$. Actually, in this pseudo-vectorial notation, the formula refers to the components of $\vec{u}$ and means the  vector formed by the operation of the 
classical $\Delta$ on the three components of the vector $u$. 

We assume that the flow is orthogonal to the equilibrium surface. We remind that, by the definition of the equilibrium surface, the tangential component of 
the resultant forces\footnote{The forces considered in the determination of the equilibrium surface are the self-gravitation, the tidal forces acting on the 
body and the inertial forces due to the rotation of the body.} acting on the fluid vanishes at the equilibrium surface. Since the equilibrium surface is an 
almost spherical ellipsoid, we may consider in a first approximation that the motion of the fluid in that region is a radial flow.

If we consider that $F_{ext}=0$ (no other external forces are acting on the fluid) and restricting $\vec{u}$ to its radial component $u_r$, there follows
\begin{equation}
 0 \approx \Delta w + \eta_i\ \nabla^2u_r.
\end{equation}
Hence, 
\begin{equation}
 \nabla^2u_r=\frac{\partial^2 u_r}{\partial r^2}+\frac{2}{r}\frac{\partial u_r}{\partial r}-\frac{2u_r}{r^2} = -\frac{\Delta w}{\eta_i}.
\end{equation}

The general solution of this equation is 
\begin{equation}
 u_r(r) =  C_1r + \frac{C_2}{r^2}-\frac{\Delta w}{4\eta_i}r^2,
\end{equation}
where $C_1$ and $C_2$ are integration constants. The task of interpreting and determining its integration constants becomes easier if the solution is 
linearized in the neighborhood of $r=\rho_i$ (i.e. $h=0$):
\begin{equation}
u_r(r)=u_r(\rho_i)+u_r'(\rho_i)(r-\rho_i)+\frac{1}{2} u_r''(\rho_i)(r-\rho_i)^2 +\ldots
\end{equation}

Hence, $u_r(\rho_i)=0$, that is, there is no pressure surplus (or deficit) when the actual separation surface coincides with the equilibrium and the linear 
approximation of the solution is obtained when we assume $u_r''(\rho_i)=0$.

Therefore,
\begin{eqnarray}
C_1&=&\rho \Delta w/6 \eta_i \nonumber\\
C_2&=&\rho^4 \Delta w/ 12 \eta_i.
\end{eqnarray}

Hence, $u_r'(\rho_i)=\rho_i \Delta w/2 \eta_i$, and the linear approximation corresponding to the Newtonian creep of the fluid is 
\begin{equation}
u_r(r)=\gamma_i (r-\rho_i),
\end{equation}
where
\begin{equation}
\gamma_i = u_r'(\rho_i) =\frac{\Delta w\rho_i}{2 \eta_i}.
\end{equation}

In the region II, the calculation is similar; however, instead of a pressure surplus we have a pressure deficit because the equilibrium assumes one fluid 
with density $d_i$ below the equilibrium surface, which is now occupied by  fluid of density $d_{i+1} < d_i$. The equations are the same as above. We note 
that in the new equations, the adopted viscosity continues being $\eta_i$ since we assumed it larger than $\eta_{i+1}$. The relaxation of the surface to the 
equilibrium is governed by the larger of the viscosities of the two layers.

In the homogeneous case we have one layer body ($N=1$). If we consider $d_{N+1}=0$ (neglected the density of the atmosphere), we recover the expression of 
the relaxation factor given by Ferraz-Mello (2013; 2015a)
\begin{equation}
\gamma_N \approx\frac{w R_N}{2\eta_N},
\end{equation}
where $w= d_N g$ is the specific weight and $\rho_N\approx R_N$.

\section*{Appendix 2: Equilibrium ellipsoidal figures}{\label{ap_B}}

In this appendix we calculate the equatorial and the polar flattenings of the equilibrium ellipsoidal figures for differentiated non-homogeneous  bodies in 
non-synchronous rotation when each layer has a different angular velociy. For this, we extend the results obtained by Folonier et al. (2015), where the 
rigid rotation hypothesis has been assumed.

\begin{figure}[h]
\begin{center}
\includegraphics[scale=0.4]{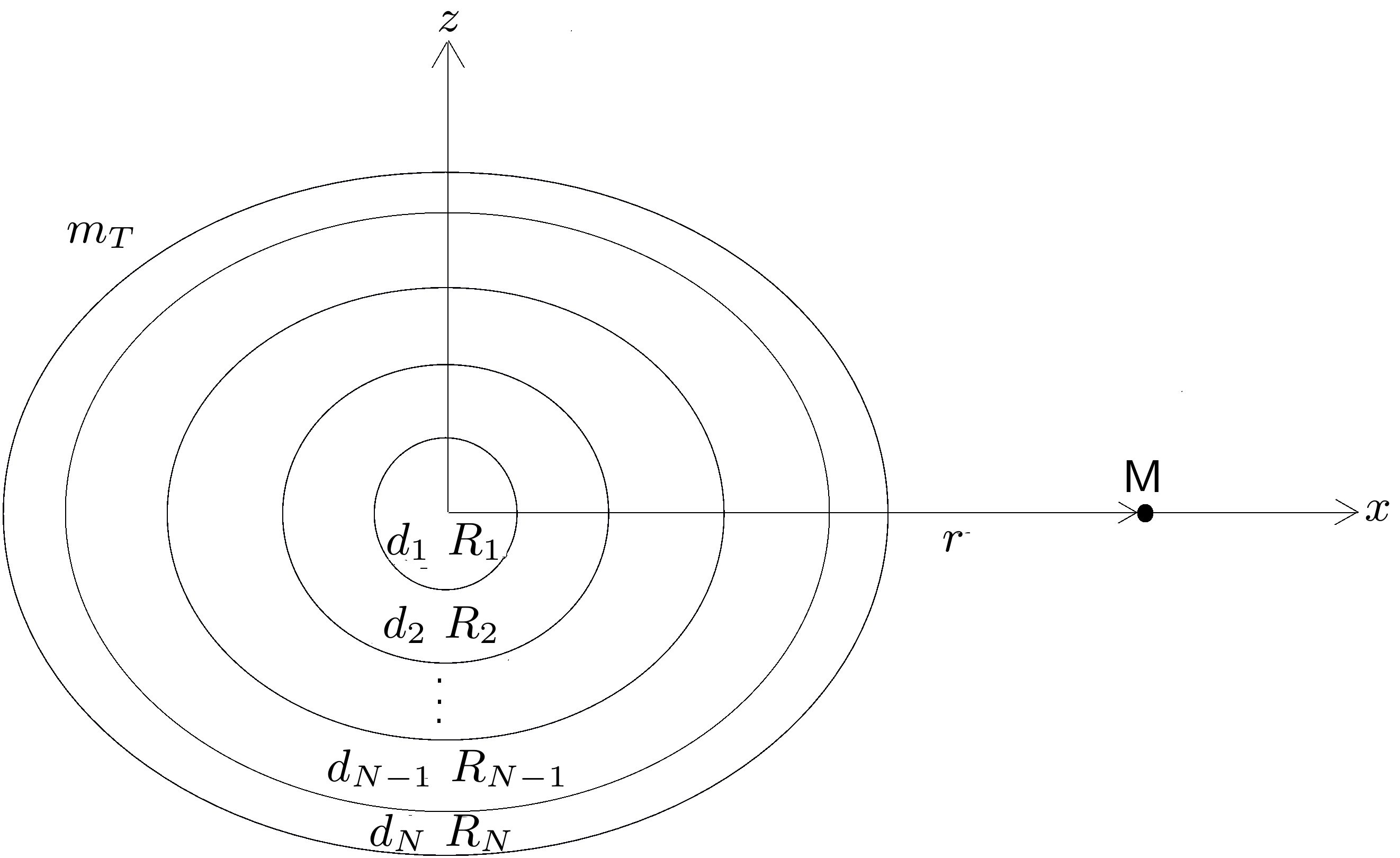}
\caption{Body $\tens{m}$ formed of $N$ homogeneous layers of densities $d_i$ and mean outer radii $R_i$ rotating with angular velocities $\vec{\Omega}_i$ 
and a point mass $\tens{M}$ orbiting at a distance $r$ from the center of $\tens{m}$ in a plane perpendicular to the rotation axis.}
\label{fig16}
\end{center}
\end{figure}

Let us consider one body $\tens{m}$ of mass $m_T$ and one mass point $\tens{M}$ of mass $M$ orbiting at a distance $r$ from the center of $\tens{m}$. We 
assume that the body is composed of $N$ homogeneous layers of density $d_i$ ($i=1,\ldots,N$) and angular velocity $\vec{\Omega}_i=\Omega_i\hat{\vec{k}}$, 
perpendicular to the orbital plane (Fig. \ref{fig16}). We also assume that each layer has an ellipsoidal shape with outer semiaxes $a_i$, $b_i$ and $c_i$; 
the axis $a_i$ is pointing towards $\tens{M}$ while $c_i$ is along the axis of rotation. 

The equatorial prolateness and polar oblateness of the $i$th ellipsoidal surface, respectively, are
\begin{equation}
\epsilon_{\rho}^{(i)} = \frac{a_i-b_i}{R_i}; \ \ \ \ \ \ \ \ \ \epsilon_{z}^{(i)} = \frac{b_i-c_i}{R_i},
\end{equation}
where $R_i=\sqrt{a_ib_i}$ is the outer equatorial mean radius of the $i$th layer.

Following Folonier et al. (2015), and carrying out modifications to account for the different velocities of rotations, the equilibrium equations can be
written as
\begin{eqnarray}
         0  &=& -\frac{3GM}{r^3} + \sum_{j=1}^{i-1} \frac{Gm'_j}{R_i^3} \left(2\epsilon_{\rho}^{(i)} - \frac{6\epsilon_{\rho}^{(j)}}{5}\left(\frac{R_j}{R_i}\right)^2\right) + \frac{Gm'_i}{R_i^3} \frac{4\epsilon_{\rho}^{(i)}}{5} + \sum_{j=i+1}^N \frac{Gm'_j}{R_j^3} \left(2\epsilon_{\rho}^{(i)}-\frac{6\epsilon_{\rho}^{(j)}}{5}\right)\nonumber\\
 \Omega_i^2 &=&                    \sum_{j=1}^{i-1} \frac{Gm'_j}{R_i^3} \left(2\epsilon_{z}^{(i)}    - \frac{6\epsilon_{z}^{(j)}}{5}   \left(\frac{R_j}{R_i}\right)^2\right) + \frac{Gm'_i}{R_i^3} \frac{4\epsilon_{z}^{(i)}}{5}    + \sum_{j=i+1}^N \frac{Gm'_j}{R_j^3} \left(2\epsilon_{z}^{(i)}   -\frac{6\epsilon_{z}^{(j)}}{5}\right),
 \label{eq:eq-equations}
\end{eqnarray}
where $G$ is the gravitation constant and
\begin{equation}
 m'_k = \frac{4\pi}{3}(d_k-d_{k+1}) R^3_k.
\end{equation}

If we assume that
\begin{equation}
 \epsilon_{\rho}^{(i)} = \mathcal{H}_i\epsilon_J; \ \ \ \ \ \ \ \ \  \epsilon_{z}^{(i)} = \mathcal{G}_i\overline{\epsilon}_M,
\label{eq:ep_z-i}
\end{equation}
where $\overline{\epsilon}_M$ is the flattening of the equivalent MacLaurin homogeneous spheroid in synchronous rotation and $\epsilon_J$ is the flattening 
of the equivalent Jeans homogeneous spheroids:
\begin{equation}
\overline{\epsilon}_M = \frac{5R_N^3n^2}{4m_TG} \ \ \ \ \ \ \ \ \ \ \epsilon_J = \frac{15MR_N^3}{4m_Tr^3},
\end{equation}
the Eq. (\ref{eq:eq-equations}) can be written as
\begin{eqnarray}
 \gamma_i \mathcal{H}_i &=& x_i^3 + \displaystyle\sum_{j=1}^{i-1} \alpha_{i j}\mathcal{H}_j + \displaystyle\sum_{j=i+1}^N \beta_{i j}\mathcal{H}_j\nonumber\\
 \gamma_i \mathcal{G}_i &=& \left(\frac{\Omega_i}{n}\right)^2x_i^3 + \displaystyle\sum_{j=1}^{i-1} \alpha_{i j}\mathcal{G}_j + \displaystyle\sum_{j=i+1}^N \beta_{i j}\mathcal{G}_j,
\label{eq:achata1}
\end{eqnarray}
where $x_i=R_i/R_N$ is the normalized mean equatorial radius and the coefficients $\alpha_{i j}$, $\beta_{i j}$ and $\gamma_{i}$ are
\begin{eqnarray}
 \alpha_{i j} &=& \frac{3m'_j}{2m_T}\left(\frac{R_j}{R_i}\right)^2 \nonumber\\
 \beta_{i j}  &=& \frac{3m'_j}{2m_T}\left(\frac{R_i}{R_j}\right)^3\nonumber\\
 \gamma_i     &=& 1 + \frac{3(m_T-m'_i)}{2m_T} - \sum_{k=i+1}^N \frac{5m'_k}{2m_T} \frac{(R_k^3-R_i^3)}{R_k^3}.
\label{eq.para}
\end{eqnarray}

Then, the Clairaut's coefficients $\mathcal{H}_i$ and $\mathcal{G}_i$ are
\begin{equation}
\mathcal{H}_i = \sum_{j=1}^N (\tens{E}^{-1})_{ij}x_j^3\ \ \ \ \ \ \ \ \ \ \ \mathcal{G}_i = \sum_{j=1}^N (\tens{E}^{-1})_{ij}x_j^3\left(\frac{\Omega_i}{n}\right)^2,
\label{eq:Clairaut-numbers}
\end{equation}
where $(\tens{E}^{-1})_{ij}$ are the elements of the inverse of the matrix $\tens{E}$, whose elements are
\begin{equation}
(\tens{E})_{ij} = \left\{\begin{array}{lll}\alpha_{ij}=\displaystyle \frac{3}{2f_N}(\widehat{d}_j-\widehat{d}_{j+1})\frac{x^5_j}{x^2_i},                                              & \ \ \ \ \ i>j\\
                                            \  \gamma_i=\displaystyle \frac{3}{2f_N}(\widehat{d}_i-\widehat{d}_{i+1})x^3_i + \frac{5}{2}- \frac{5}{2f_N}\sum_{k=i+1}^N (\widehat{d}_k-\widehat{d}_{k+1}) (x_k^3-x_i^3), & \ \ \ \ \ i=j \\
                                             \beta_{ij}=\displaystyle \frac{3}{2f_N}(\widehat{d}_j-\widehat{d}_{j+1})x^3_i,                                              & \ \ \ \ \ i<j \end{array}\right.
\end{equation}
where $\widehat{d}_i=d_i/d_1$ is the normalized density of the $i$th layer and $f_N=3\int_0^1\widehat{d}(z)z^2\ dz$.

\section*{Appendix 3: Gravitational coupling}{\label{ap_C}}

\begin{figure}[h]
\begin{center}
\includegraphics[scale=0.4]{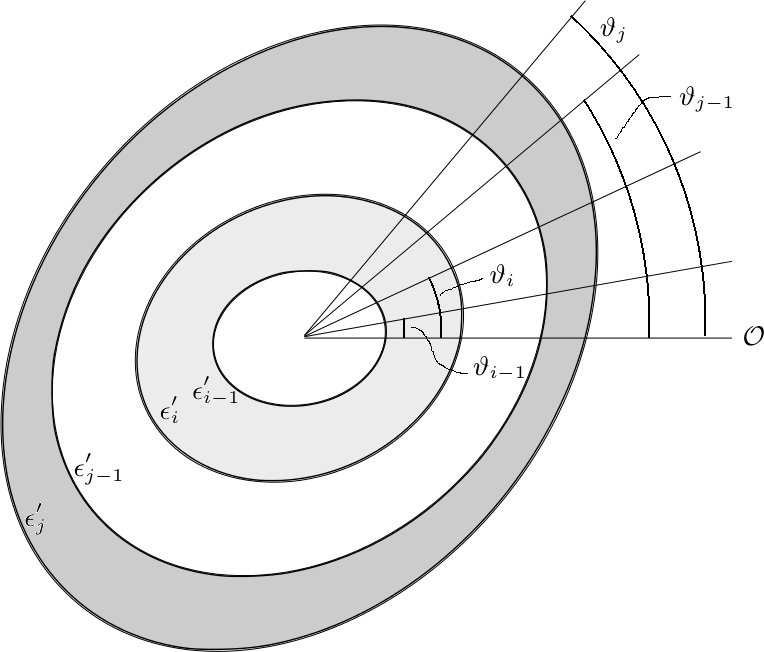}
\caption{Equatorial section of the $i$th and $j$th layers. $\epsilon'_i$ and $\epsilon'_{i-1}$ are the outer and the inner equatorial flattenings of the 
$i$th layer and the angles $\vartheta_i$ and $\vartheta_{i-1}$ are its outer and inner geodetic lags. Similarly, $\epsilon'_j$ and $\epsilon'_{j-1}$ are 
the outer and the inner equatorial flattenings and $\vartheta_j$ and $\vartheta_{j-1}$ are the outer and the inner geodetic lags of the $j$th layer.}
\label{fig17}
\end{center}
\end{figure}

When the principal axes of inertia of two layers (of one body composed by $N$ homogeneous layers) are not aligned, a restoring gravitational torque, which 
tends to align these axes appears. The torque acting on the inner $j$th layer due to the outer $i$th layer (not necessarily contiguous) is
\begin{equation}
\mathbf{\Gamma}_{ji} = -\int_{m_j} (\vec{r}\times \nabla\delta  U_i)\ dm_j =-\int_0^{2\pi}\int_0^\pi\int_{\zeta'_{j-1}}^{\zeta'_j}d_j\ ( \vec{r}\times\nabla\delta  U_i)\ r^2\sin{\theta}\ dr\ d\theta\ d\varphi,
\label{eq:proto-ag}
\end{equation}
where $d_j$, $m_j$ are the density and the mass in the $j$th layer and $\delta U_i$ is the disturbing potential of the $i$th layer at an external point (Fig. 
\ref{fig17}).

The limits of the integral in Eq. (\ref{eq:proto-ag}), $\zeta'_j$ and $\zeta'_{j-1}$, are the \emph{real} outer and inner boundaries of the $j$th layer, 
respectively. In our model we have to consider the actual flattening of the surfaces, which is the composition of the main elastic and anelastic tidal 
components (see Sec. 10 of Ferraz-Mello, 2013). The addition of the two components is virtually equivalent to the use the Maxwell viscoelastic model ab 
initio as done by Correia et al. (2014) (Ferraz-Mello, 2015b).

\begin{figure}[h]
\begin{center}
\includegraphics[scale=0.4]{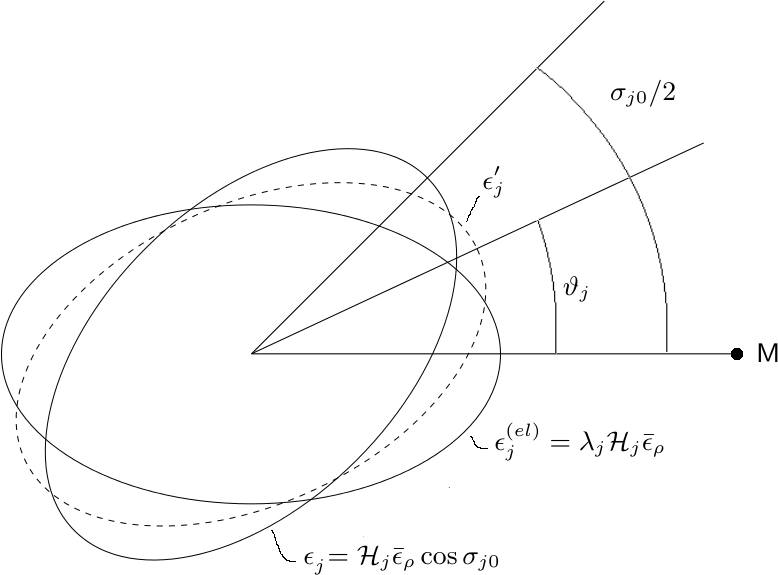}
\caption{Scheme of the composition of the elastic and anelastic tides of the outer boundary of the $j$th layer. $\epsilon_j^{(el)}$ and $\epsilon_j$ are 
the equatorial flattenings of the main elastic and anelastic tides, respectively, and $\epsilon'_j$ is the equatorial flattening of the ellipsoidal surface, 
which result of this composition (\textit{dashed curve}). The semi-major axis of the elastic ellipsoid is oriented towards \tens{M}.}
\label{fig18}
\end{center}
\end{figure}

Assuming that the elastic and the anelastic components have ellipsoidal surfaces (not aligned), the resulting surface can be approximated by a prolate ellipsoid 
with equatorial flattening $\epsilon'$ and rotated by an angle $\vartheta$ with respect to $\tens{M}$. For the sake of simplicity, we also assume that the 
relative motion of the outer body \tens{M} is circular. Then, neglecting the axial term does not contribute to the calculation of the gravitational coupling, 
the height of the outer surface of the $j$th layer with respect to the one sphere of radius $R_j$, in polar coordinates, rotated by an angle $\vartheta_j$ 
with respect to $\tens{M}$ and to first order in the flattenings (see Fig. \ref{fig18}), is
\begin{equation}
 \delta \zeta'_j = \frac{1}{2}R_j\epsilon'_j\sin^2{\theta}\cos{(2\varphi-2\vartheta_j)}=\frac{1}{2}R_j\mathcal{H}_j\overline{\epsilon}_\rho\lambda_j\sin^2{\theta}\cos{2\varphi}+\frac{1}{2}R_j\mathcal{H}_j\overline{\epsilon}_\rho \cos{\sigma_{j0}}\sin^2{\theta}\cos{(2\varphi-\sigma_{j0})},
\end{equation}
where $0<\lambda_j<1$ is a relative measurement of the maximum height of the elastic tides of the outer boundary of the $j$th layer. The angle $\vartheta_j$ 
is often called the \emph{geodetic lag} of the surface.

If we open the trigonometric functions, by identification of the terms with same trigonometric arguments, the resulting equatorial flattening of the outer 
boundary of the $j$th layer is
\begin{equation}
 \epsilon'_j = \mathcal{H}_j\overline{\epsilon}_\rho\sqrt{\lambda_j^2 + \cos^2{\sigma_{j0}}(1 + 2\lambda_j)},
\end{equation}
and the geodetic lag is
\begin{equation}
 \vartheta_j = \frac{1}{2}\tan^{-1}{\left(\frac{\sin{2\sigma_{j0}}}{1+2\lambda_j+\cos{2\sigma_{j0}}} \right)}.
\label{eq:vartheta_j}
\end{equation}

The height of the inner boundary of the $j$th layer, taking into account the composition of the main elastic and anelastic tides has an identical 
expression:
\begin{eqnarray}
 \delta \zeta'_{j-1} &=& \frac{1}{2}R_{j-1}\epsilon'_{j-1}\sin^2{\theta}\cos{(2\varphi-2\vartheta_{j-1})} \nonumber\\
                     &=& \frac{1}{2}R_{j-1}\mathcal{H}_{j-1}\overline{\epsilon}_\rho\lambda_{j-1}\sin^2{\theta}\cos{2\varphi}+\frac{1}{2}R_{j-1}\mathcal{H}_{j-1}\overline{\epsilon}_\rho \cos{\sigma_{j-1,0}}\sin^2{\theta}\cos{(2\varphi-\sigma_{j-1,0})},
\end{eqnarray}
where  $0<\lambda_{j-1}<1$ is the relative measurement of the maximum height of the elastic tides of the inner boundary of the $j$th layer. Then, the 
resulting equatorial flattening is
\begin{equation}
 \epsilon'_{j-1} = \mathcal{H}_{j-1}\overline{\epsilon}_\rho\sqrt{\lambda_{j-1}^2 + \cos^2{\sigma_{j-1,0}}(1 + 2\lambda_{j-1})},
\end{equation}
and the geodetic lag is
\begin{equation}
 \vartheta_{j-1} = \frac{1}{2}\tan^{-1}{\left(\frac{\sin{2\sigma_{j-1\ 0}}}{1+2\lambda_{j-1}+\cos{2\sigma_{j-1,0}}} \right)}.
\end{equation}

In the same way, we assume that the ellipsoidal shape of this layer is also given by the composition of the main elastic and anelastic tidal components. 
Then, the inner and outer equatorial flattenings, respectively, are
\begin{equation}
\epsilon'_{i-1} = \mathcal{H}_{i-1}\overline{\epsilon}_\rho\sqrt{\lambda_{i-1}^2 + \cos^2{\sigma_{i-1,0}}(1 + 2\lambda_{i-1})}; \ \ \ \ \ \ \ \ \epsilon'_i = \mathcal{H}_i\overline{\epsilon}_\rho\sqrt{\lambda_i^2 + \cos^2{\sigma_{i0}}(1 + 2\lambda_i)},
\end{equation}
and the corresponding geodetic lags are
\begin{equation}
 \vartheta_i = \frac{1}{2}\tan^{-1}{\left(\frac{\sin{2\sigma_{i0}}}{1+2\lambda_i+\cos{2\sigma_{i0}}} \right)}; \ \ \ \ \ \ \ \ \vartheta_{i-1} = \frac{1}{2}\tan^{-1}{\left(\frac{\sin{2\sigma_{i-1,0}}}{1+2\lambda_{i-1}+\cos{2\sigma_{i-1,0}}} \right)},
\label{eq:vartheta_i}
\end{equation}
where  $0<\lambda_i,\lambda_{i-1}<1$ are the relative measurements of the maximum heights of the elastic tides of the outer and inner boundaries of the 
$i$th layer.

Using the expression of the disturbing potential, given in Section A in the Online Supplement, and neglecting the axial term, we obtain
\begin{equation}
\delta U_i = -\frac{3GC_i}{4r^3}\sin^2{\theta}\frac{\Delta\big(R_i^5\epsilon'_i\cos{(2\varphi-2\vartheta_i)}\big)}{R_i^5-R_{i-1}^5},
\end{equation}
where $\Delta(f_i) = f_i-f_{i-1}$, denotes the increment of one function $f_i$ between the inner and the outer boundaries of this layer. Then, the 
vectorial product in Eq. (\ref{eq:proto-ag}) is
\begin{equation}
 \vec{r} \times \nabla\delta  U_i = -\frac{2\pi Gd_i}{5r^3}\big(2\sin{\theta}\Delta\big(R_i^5\epsilon'_i\sin{(2\varphi-2\vartheta_i)}\big) \widehat{\vec{\theta}}+\sin{2\theta}\Delta\big(R_i^5\epsilon'_i\cos{(2\varphi-2\vartheta_i)}\big) \widehat{\vec{\varphi}}\big).
\end{equation}

Using the polar unitary vectors in Cartesian coordinates
\begin{eqnarray}
\widehat{\vec{\theta}}  &=& \cos{\theta} \cos{\varphi}\ \widehat{\vec{x}} + \cos{\theta} \sin{\varphi}\ \widehat{\vec{y}} - \sin{\theta}\ \widehat{\vec{z}} \nonumber\\
\widehat{\vec{\varphi}} &=& -\sin{\varphi}\ \widehat{\vec{x}} + \cos{\varphi}\ \widehat{\vec{y}},
\end{eqnarray}
and the approximation of $\ln{\zeta'_j/\zeta'_{j-1}}$ to first order in the flattenings
\begin{equation}
 \ln{\frac{\zeta'_j}{\zeta'_{j-1}}} \approx \ln{\frac{R_j}{R_{j-1}}} + \frac{1}{2}\epsilon'_j\sin^2{\theta}\cos{(2\varphi-2\vartheta_j)} - \frac{1}{2}\epsilon'_{j-1}\sin^2{\theta}\cos{(2\varphi-2\vartheta_{j-1})},
\end{equation}
then, we may perform the integrals of Eq. (\ref{eq:proto-ag}) and obtain the torque acting on the inner $j$th layer due to the outer $i$th layer
\begin{eqnarray}
 \mathbf{\Gamma}_{ji} &=& -\frac{32\pi^2 G}{75} d_id_j \Delta_{ij}\big(R_i^5\epsilon'_i\epsilon'_j\sin{(2\vartheta_j-2\vartheta_i)}\big)\widehat{\vec{z}},
 \label{eq:Gamma_ij}
\end{eqnarray}
where $\Delta_{ij}(f_{ij})\defeq \Delta(f_{ij})-\Delta(f_{i,j-1})=f_{ij}-f_{i-1,j}-f_{i,j-1}+f_{i-1,j-1}$.

As the torque acting on the outer $i$th layer, due to the inner $j$th layer, is the reaction
\begin{eqnarray}
 \mathbf{\Gamma}_{ij} &=& -\mathbf{\Gamma}_{ji},
\end{eqnarray}
then, the total gravitational coupling, acting on the $j$th layer can be written as
\begin{equation}
 \mathbf{\Gamma}_j = \sum_{p=1;\ p\neq j}^N \mathbf{\Gamma}_{jp} = \sum_{p=1}^{j-1}\mathbf{\Gamma}_{jp} - \sum_{p=j+1}^N \mathbf{\Gamma}_{pj}.
\end{equation}

If we consider the two-layer model, the torque acting on the core and the shell, are
\begin{eqnarray}
 \Gamma_c &=& K\sin{(2\vartheta_s-2\vartheta_c)}\nonumber\\
 \Gamma_s &=& -K\sin{(2\vartheta_s-2\vartheta_c)},
\end{eqnarray}
where the gravitational coupling parameter $K$ is
\begin{equation}
 K=\frac{32\pi^2 G}{75} d_cd_s\epsilon'_c\epsilon'_sR_c^5.
 \label{eq:K-ap}
\end{equation}

The equatorial flattenings are
\begin{equation}
\epsilon'_c = \mathcal{H}_c\overline{\epsilon}_\rho\sqrt{\lambda_c^2 + \cos^2{\sigma_{c0}}(1 + 2\lambda_c)}; \ \ \ \ \ \ \ \  \epsilon'_s = \mathcal{H}_s\overline{\epsilon}_\rho\sqrt{\lambda_s^2 + \cos^2{\sigma_{s0}}(1 + 2\lambda_s)},
\label{eq:achata-comp}
\end{equation}
and the geodetic lags are
\begin{equation}
 \vartheta_c = \frac{1}{2}\tan^{-1}{\left(\frac{\sin{2\sigma_{c0}}}{1+2\lambda_c+\cos{2\sigma_{c0}}} \right)}; \ \ \ \ \ \ \ \ \vartheta_s = \frac{1}{2}\tan^{-1}{\left(\frac{\sin{2\sigma_{s0}}}{1+2\lambda_s+\cos{2\sigma_{s0}}} \right)}.
\label{eq:geolag-comp}
\end{equation}
The parameters $0<\lambda_c,\lambda_s<1$ are relative measurements of the heights of the elastic tides of the outer surfaces of the core and the shell, 
respectively. The trigonometric functions in (\ref{eq:achata-comp})-(\ref{eq:geolag-comp}) are frequency functions. Using Eq. (\ref{eq:sigma_i}), an 
elementary calculation shows that
\begin{equation}
\cos^2{\sigma_{i0}} = \frac{\gamma_i^2}{\gamma_i ^2+\nu_i^2}; \ \ \ \ \ \ \ \ \sin{2\sigma_{i0}} = \frac{2\gamma_i\nu_i}{\gamma_i ^2+\nu_i^2}; \ \ \ \ \ \ \ \ \cos{2\sigma_{i0}} = \frac{\gamma_i^2-\nu_i^2}{\gamma_i ^2+\nu_i^2}. 
\end{equation}

It is important to note that some works, as Karatekin et al. (2008), use a different gravitational coupling parameter $K$. When applied to a two-layer model, 
their $K$ differs from Eq. (\ref{eq:K-ap}) by a multiplicative factor $(1-d_s/d_c)$. The reason for this difference is simple: while here we calculate the 
torque due to the mutual gravitational attraction of two layers, through Eq. (\ref{eq:Gamma_ij}), they calculate the gravitational coupling between regions 
that involve various layers simultaneously (see Fig. 2 and Eq. 16 of Van Hoolst et al., 2008).

\section*{Appendix 4: Linear drag}{\label{ap_D}}
 
The model considered here also assumes that a linear friction occurs between two contiguous layers. We assume that between two contiguous layers (for 
instance, the inner boundary of the $i$th layer and the outer boundary of the $(i+1)$th layer) exists a thin liquid boundary with viscosity 
$\widehat{\eta}_i$ and thickness $h_i$.

We assume that the torque, along the axis $z$, acting on the inner $i$th layer due to the outer $(i+1)$th layer is 
\begin{equation}
\Phi_{i,i+1} = \mu_i (\Omega_{i+1}-\Omega_i),
\end{equation}
and vice-versa. The friction coefficient $\mu_i$ of the $i$th boundary is an undetermined constant.
 
Let $d\mathbf{F}_{i,i+1}$ be the force acting tangentially on the area element of an sphere of radius $R_i$. If the fluid in contact with the surface of the 
sphere is a Newtonian fluid, and the thickness of the liquid boundary is thin enough to allow us to consider a plane-parallel geometry (a plane Couette 
flow), the modulus of the force is (Papanastasious et al., 2000, Chap. 6, Eq. 6.15)
\begin{equation}
 dF_{i,i+1}=\displaystyle\frac{\widehat{\eta}_i}{h_i} V_i R_i^2\sin\theta\ d\phi\ d\theta
\end{equation}
where $V_i=R_i\sin\theta(\Omega_{i+1}-\Omega_i)$ is the relative velocity of the $(i+1)$th layer with respect to the $i$th layer at the latitude $\theta$ 
and $R_i,\phi,\theta$ are the spherical coordinates of the center of the area element. The modulus of the torque of the force $d\mathbf{F}_{i,i+1}$, along 
the axis $z$, is 
\begin{equation}
d\Phi_{i,i+1}=R_i\sin\theta\ dF_{i,i+1}. 
\end{equation}
The element of area is $R_id\theta \times R_i\sin\theta d\phi$. The integral of $d\Phi_{i,i+1}$ over the sphere is easy to calculate giving
\begin{equation}
 \Phi_{i,i+1}=\int_0^{2\pi} \int_0^\pi \frac{\widehat{\eta}_i}{h_i} R_i^4\sin^3{\theta} (\Omega_{i+1}-\Omega_i) d\theta\  d\phi = -\frac{8\pi}{3}\ \frac{\widehat{\eta}_i}{h_i}R_i^4 (\Omega_{i+1}-\Omega_i)
\end{equation}
If we compare with the law used to introduce the friction, we obtain
\begin{equation}
 \mu_i = \frac{8\pi}{3}\ \frac{\widehat{\eta}_i}{h_i}R_i^4.
\end{equation}
This is the law corresponding to a liquid-solid boundary for low speeds.
 
The torque, along the axis $z$, acting on the inner $(i+1)$th layer due to the outer $i$th layer is
\begin{eqnarray}
 \Phi_{i+1,i} &=& -\Phi_{i,i+1} = -\mu_i (\Omega_{i+1}-\Omega_i).
\end{eqnarray}

Then, the total torque, due to the friction, acting on the $i$th layer is the sum of the torque due to the outer $(i+1)$th layer plus the the torque due 
to the inner $(i-1)$th layer
\begin{equation}
 \Phi_i = \Phi_{i,i-1} + \Phi_{i,i+1} = \mu_{i-1} (\Omega_{i-1}-\Omega_i) - \mu_i (\Omega_i-\Omega_{i+1}).
\end{equation}

In the two-layer model, the torque acting on the core due to the shell and the torque acting on the shell due to the core are, respectively
\begin{eqnarray}
 \Phi_c &=& \mu (\Omega_s-\Omega_c)\nonumber\\
 \Phi_s &=& -\mu  (\Omega_s-\Omega_c),
\end{eqnarray}
where $\eta_o$ and $h$ are the viscosity and the thickness, respectively, of the core-shell boundary and
\begin{equation}
 \mu = \frac{8\pi}{3}\ \frac{\eta_o}{h}R_c^4.
\end{equation}

\clearpage
\vfill\eject
\appendix
\section*{Online Supplement}
\pagenumbering{arabic}

\renewcommand{\theequation}{A.\arabic{equation}}
\setcounter{equation}{0}  
\section{Shape and gravitational potential of one ellipsoid and one ellipsoidal layer}{\label{online_sup_A}}

\subsection{Homogeneous ellipsoid}

Let us consider a homogeneous triaxial ellipsoid with density $d$, semi axes $a>b>c$, equatorial mean radius $R=\sqrt{ab}$ and equatorial and polar 
flattenings are
\begin{equation}
\epsilon_\rho = \frac{a-b}{R}; \ \ \ \ \ \ \ \epsilon_z=\frac{b-c}{R}.
\end{equation}
Then, the semi axes of this ellipsoid, to first order in the flattenings, can be written as
\begin{equation}
 a = R\left(1+\frac{\epsilon_\rho}{2}\right);  \ \ \ \ \ \ \  b = R\left(1-\frac{\epsilon_\rho}{2}\right);  \ \ \ \ \ \ \  c = R\left(1-\frac{\epsilon_\rho}{2}-\epsilon_z\right).
\label{eq:semiaxes-homo}
\end{equation}

Let us consider the equation of surface of this homogeneous triaxial ellipsoid, in a reference system where the semi axes $a$ and $c$ are aligned to the 
coordinates axes $x$ and $z$, respectively:
\begin{equation}
 \frac{x^2}{a^2}+\frac{y^2}{b^2}+\frac{z^2}{c^2}=1.
\end{equation}
If we use the semi axes (\ref{eq:semiaxes-homo}), the spherical coordinates
\begin{equation}
 x=\rho\sin{\theta}\cos{\varphi};\ \ \ \ \ \ \  y=\rho\sin{\theta}\sin{\varphi};\ \ \ \ \ \ \  z=\rho\cos{\theta},
 \label{eq:coordinate-spherical}
\end{equation}
and expand to first order in the flattenings, we obtain
\begin{equation}
 \rho= R\left(1+\frac{\epsilon_\rho}{2}\sin^2{\theta}\cos{2\varphi}-\left(\frac{\epsilon_\rho}{2}+\epsilon_z\right)\cos^2{\theta}\right).
\end{equation}

The mass of this ellipsoids is
\begin{equation}
 m = \frac{4\pi}{3}d\ abc \approx\frac{4\pi}{3}d\ R^3\left(1-\frac{\epsilon_\rho}{2}-\epsilon_z\right).
\end{equation}

The principal moments of inertia are
\begin{eqnarray}
 A &=& \frac{1}{5}m(b^2+c^2) \approx \frac{2}{5}mR^2 \left(1-\epsilon_\rho-\epsilon_z\right)\nonumber\\
 B &=& \frac{1}{5}m(a^2+c^2) \approx \frac{2}{5}mR^2 \left(1-\epsilon_z\right)\nonumber\\
 C &=& \frac{1}{5}m(a^2+b^2) \approx \frac{2}{5}mR^2,
\end{eqnarray}
and its differences are
\begin{eqnarray}
 C-A &\approx&  \frac{2}{5}mR^2 (\epsilon_\rho+\epsilon_z) \approx C (\epsilon_\rho+\epsilon_z) \nonumber\\
 C-B &\approx&  \frac{2}{5}mR^2 \epsilon_z \approx C\epsilon_z\nonumber\\
 B-A &\approx&  \frac{2}{5}mR^2 \epsilon_\rho \approx C \epsilon_\rho.
\label{eq:C-A}
\end{eqnarray}

The corresponding gravitational potential of this homogeneous triaxial ellipsoid, at an external point $\vec{r^*}$, is
\begin{eqnarray}
U(\vec{r^*}) &=& -\displaystyle\frac{Gm}{r^*} - \frac{G(B-A)}{2r^{*5}}(3x^{*2}-r^{*2}) + \frac{G(C-B)}{2r^5}(3z^{*2}-r^{*2}) \nonumber\\
       &\approx& -\displaystyle\frac{Gm}{r^*} - \frac{GC}{2r^{*3}}\epsilon_\rho(3\cos^2{\varphi}\sin^2{\theta}-1) + \frac{G C}{2r^{*3}}\epsilon_z(3\cos^2{\theta}-1),
\label{eq:U-homo1}
\end{eqnarray}
or
\begin{eqnarray}
U(\vec{r^*}) &=& -\displaystyle\frac{Gm}{r^*} - \frac{3GC}{4r^{*3}}\epsilon_\rho\sin^2{\theta}\cos{2\varphi} + \frac{G C}{2r^{*3}}\left(\frac{\epsilon_\rho}{2}+\epsilon_z\right)(3\cos^2{\theta}-1).
\label{eq:U-homo2}
\end{eqnarray}

\subsection{Ellipsoidal layer}

Let us consider a homogeneous triaxial ellipsoidal shell with density $d_i$, outer semi axes $a_i>b_i>c_i$, outer equatorial mean radius $R_i= \sqrt{a_ib_i}$ 
and outer equatorial and polar flattenings
\begin{equation}
\epsilon_\rho^{(i)} = \frac{a_i-b_i}{R_i}; \ \ \ \ \ \ \ \epsilon_z^{(i)} = \frac{b_i-c_i}{R_i}.
\end{equation}

At the inner ellipsoidal boundary, the semi axes are $a_{i-1}>b_{i-1}>c_{i-1}$ (not necessarily aligned with the axes of the outer surface). The inner 
equatorial mean radius is $R_{i-1}= \sqrt{a_{i-1}b_{i-1}}$ and the inner equatorial and polar flattenings are
\begin{equation}
\epsilon_\rho^{(i-1)} = \frac{a_{i-1}-b_{i-1}}{R_{i-1}}; \ \ \ \ \ \ \ \epsilon_z^{(i-1)} = \frac{b_{i-1}-c_{i-1}}{R_{i-1}}.
\end{equation}

The semi axes of the outer boundary, to first order in flattenings, are
\begin{equation}
 a_i = R_i \left(1+\frac{\epsilon_\rho^{(i)}}{2}\right); \ \ \ \ \ \ \  b_i = R_i \left(1-\frac{\epsilon_\rho^{(i)}}{2}\right); \ \ \ \ \ \ \  c_i = R_i \left(1-\frac{\epsilon_\rho^{(i)}}{2}-\epsilon_z^{(i)}\right),
\label{eq:semis_i}
\end{equation}
and the semi axes of the inner boundary are
\begin{equation}
 a_{i-1} = R_{i-1} \left(1+\frac{\epsilon_\rho^{(i-1)}}{2}\right); \ \ \ \ \ \ \  b_{i-1} = R_{i-1} \left(1-\frac{\epsilon_\rho^{(i-1)}}{2}\right); \ \ \ \ \ \ \  c_{i-1} = R_{i-1} \left(1-\frac{\epsilon_\rho^{(i-1)}}{2}-\epsilon_z^{(i-1)}\right).
\label{eq:semis_i-1}
\end{equation}

Let us consider the equation of the surface of the outer triaxial ellipsoidal layer, in a reference system where the semi axes $a_i$ and $c_i$ are aligned 
to the coordinates axes $x$ and $z$, respectively:
\begin{equation}
 \frac{x^2}{a_i^2}+\frac{y^2}{b_i^2}+\frac{z^2}{c_i^2}=1.
\end{equation}
If we use the semi axes (\ref{eq:semis_i}), the spherical coordinates (\ref{eq:coordinate-spherical}) and expand to first order in the flattenings, we 
obtain
\begin{equation}
 \rho_i = R_i\left(1+\frac{\epsilon_\rho^{(i)}}{2}\sin^2{\theta}\cos{2\varphi}-\left(\frac{\epsilon_\rho^{(i)}}{2}+\epsilon_z^{(i)}\right)\cos^2{\theta}\right).
 \label{eq:rho_i}
\end{equation}

For the inner boundary we have the same expression, when the reference system is again such that the semi axes $a_{i-1}$ and $c_{i-1}$ are aligned to the 
coordinate axes $x$ and $z$, respectively. If we use the semi axes (\ref{eq:semis_i-1}), the spherical coordinates (\ref{eq:coordinate-spherical}) and 
expand to first order in the flattenings, we obtain
\begin{equation}
 \rho_{i-1} = R_{i-1}\left(1+\frac{\epsilon_\rho^{(i-1)}}{2}\sin^2{\theta}\cos{2\varphi}-\left(\frac{\epsilon_\rho^{(i-1)}}{2}+\epsilon_z^{(i-1)}\right)\cos^2{\theta}\right).
 \label{eq:rho_i-1}
\end{equation}

\begin{figure}[h]
\begin{center}
\includegraphics[scale=0.5]{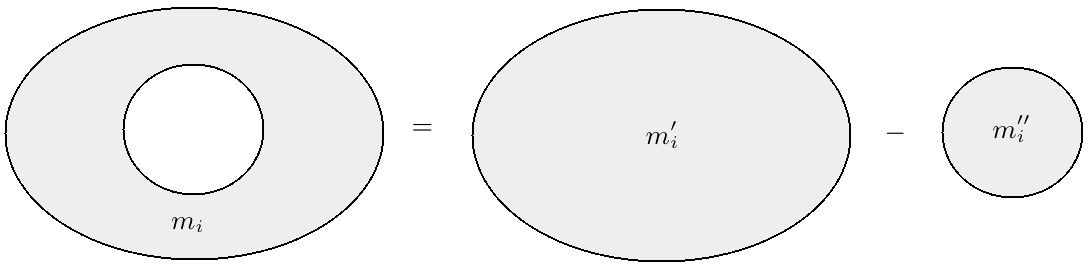}
\caption{Scheme for the calculation of the mass and principal moments of inertia of a homogeneous ellipsoidal layer as the subtraction of two homogeneous 
ellipsoids of same density $d_i$.}
\label{fig19}
\end{center}
\end{figure}

The mass $m_i$ of this layer, can be written as the subtraction of the masses of the two homogeneous ellipsoids of same density $d_i$: the homogeneous 
ellipsoid of mass $m'_i$ and same surface as the outer boundary of the layer, less the homogeneous ellipsoid of mass $m''_i$ and same surface as the inner 
boundary of the layer Fig. \ref{fig19}. The total mass of the layer then is
\begin{equation}
 m_i=m'_i-m''_i\approx\frac{4\pi}{3}d_i\left(R_i^3\left(1-\frac{\epsilon_\rho^{(i)}}{2}-\epsilon_z^{(i)}\right)-R_{i-1}^3\left(1-\frac{\epsilon_\rho^{(i-1)}}{2}-\epsilon_z^{(i-1)}\right)\right).
\end{equation}
Note that this result is independent of the orientation of the ellipsoidal boundaries semi axes. The masses $m'_i$ and $m''_i$ are
\begin{eqnarray}
m'_i  &=& d_i\frac{4\pi}{3}a_ib_ic_i \approx \frac{m_iR_i^3 \left(1-\frac{\epsilon_\rho^{(i)}}{2}-\epsilon_z^{(i)}\right)}{R_i^3\left(1-\frac{\epsilon_\rho^{(i)}}{2}-\epsilon_z^{(i)}\right)-R_{i-1}^3\left(1-\frac{\epsilon_\rho^{(i-1)}}{2}-\epsilon_z^{(i-1)}\right)}\nonumber\\
m''_i &=& d_i\frac{4\pi}{3}a_{i-1}b_{i-1}c_{i-1}\approx \frac{m_iR_{i-1}^3\left(1-\frac{\epsilon_\rho^{(i-1)}}{2}-\epsilon_z^{(i-1)}\right)}{R_i^3\left(1-\frac{\epsilon_\rho^{(i)}}{2}-\epsilon_z^{(i)}\right)-R_{i-1}^3\left(1-\frac{\epsilon_\rho^{(i-1)}}{2}-\epsilon_z^{(i-1)}\right)}.
\label{eq:m'i y m''i}
\end{eqnarray} 

To calculate the principal moments of inertia $A_i,B_i,C_i$ of a homogeneous triaxial ellipsoidal layer when the inner and the outer boundaries are not 
aligned is particularly complicated because the orientation of the principal axes of inertia do not coincide with the axes of symmetry of both boundaries. 
In the sequence we focus in the particular case in which the inner and the outer boundaries are aligned.

In this case, we can use the same scheme used to calculate the mass of the layer. The principal moments of inertia of the layer, can be written as the 
subtraction of the principal moments of inertia of two homogeneous ellipsoids of same density $d_i$: the principal moments of inertia of one homogeneous 
ellipsoid of mass $m'_i$ and the same surface as the outer boundary of the layer, less the principal moments of inertia of the homogeneous ellipsoid of mass 
$m''_i$  and the same surface as the inner boundary of the layer. Using the semi axes (\ref{eq:semis_i}) and the masses (\ref{eq:m'i y m''i}), the principal 
moments of inertia can be approximated to first order in the flattenings as
\begin{eqnarray}
  A_i &=& \frac{1}{5}m'_i\left(b_i^2+c_i^2\right)-\frac{1}{5}m''_i\left(b_{i-1}^2+c_{i-1}^2\right)\nonumber\\
&\approx& \frac{2}{5}m_i\frac{R_i^5-R_{i-1}^5}{R_i^3-R_{i-1}^3}\left(1+\frac{\Delta(R_i^3\epsilon_\rho^{(i)})}{R_i^3-R_{i-1}^3}+\frac{\Delta(R_i^3\epsilon_z^{(i)})}{R_i^3-R_{i-1}^3}-\frac{3}{2}\frac{\Delta(R_i^5\epsilon_\rho^{(i)})}{R_i^5-R_{i-1}^5}-2\frac{\Delta(R_i^5\epsilon_z^{(i)})}{R_i^5-R_{i-1}^5}\right) \nonumber\\
  B_i &=& \frac{1}{5}m'_i\left(a_i^2+c_i^2\right)-\frac{1}{5}m''_i\left(a_{i-1}^2+c_{i-1}^2\right)\nonumber\\ 
&\approx& \frac{2}{5}m_i\frac{R_i^5-R_{i-1}^5}{R_i^3-R_{i-1}^3}\left(1+\frac{\Delta(R_i^3\epsilon_\rho^{(i)})}{R_i^3-R_{i-1}^3}+\frac{\Delta(R_i^3\epsilon_z^{(i)})}{R_i^3-R_{i-1}^3}-\frac{1}{2}\frac{\Delta(R_i^5\epsilon_\rho^{(i)})}{R_i^5-R_{i-1}^5}-2\frac{\Delta(R_i^5\epsilon_z^{(i)})}{R_i^5-R_{i-1}^5}\right) \nonumber\\
 C_i &=& \frac{1}{5}m'_i\left(a_i^2+b_i^2\right)-\frac{1}{5}m''_i\left(a_{i-1}^2+b_{i-1}^2\right)\nonumber\\ 
&\approx& \frac{2}{5}m_i\frac{R_i^5-R_{i-1}^5}{R_i^3-R_{i-1}^3}\left(1+\frac{\Delta(R_i^3\epsilon_\rho^{(i)})}{R_i^3-R_{i-1}^3}+\frac{\Delta(R_i^3\epsilon_z^{(i)})}{R_i^3-R_{i-1}^3}-\frac{1}{2}\frac{\Delta(R_i^5\epsilon_\rho^{(i)})}{R_i^5-R_{i-1}^5}-\frac{\Delta(R_i^5\epsilon_z^{(i)})}{R_i^5-R_{i-1}^5}\right),
 \label{eq:moment-of-inertia}
\end{eqnarray}
and its differences are
\begin{eqnarray}
 C_i-A_i &\approx& C_i \left(\frac{\Delta(R_i^5\epsilon_\rho^{(i)})}{R_i^5-R_{i-1}^5}+\frac{\Delta(R_i^5\epsilon_z^{(i)})}{R_i^5-R_{i-1}^5}\right)\nonumber\\
 C_i-B_i &\approx& C_i \frac{\Delta(R_i^5\epsilon_z^{(i)})}{R_i^5-R_{i-1}^5}\nonumber\\
 B_i-A_i &\approx& C_i \frac{\Delta(R_i^5\epsilon_\rho^{(i)})}{R_i^5-R_{i-1}^5},
\label{eq:moment_i1}
\end{eqnarray}
where $\Delta(f_i) = f_i-f_{i-1}$, denotes the increment of one function $f_i$, between the inner and the outer boundaries of this layer.

Using the same scheme used to calculate the mass and the principal moments of inertia, the corresponding gravitational potential of this homogeneous 
triaxial layer at an external point $\vec{r}^*$ is
\begin{eqnarray}
U_i(\vec{r}^*) &=& -\displaystyle\frac{Gm_i}{r^*} - \frac{G(B_i-A_i)}{2r^{*5}}(3x^{*2}-r^{*2}) + \frac{G(C_i-B_i)}{2r^{*5}}(3z^{*2}-r^{*2}) \nonumber\\
         &\approx& -\displaystyle\frac{Gm_i}{r^*} - \frac{GC_i}{2r^{*3}}\frac{\Delta(R_i^5\epsilon_\rho^{(i)})}{R_i^5-R_{i-1}^5}(3\cos^2{\varphi^*}\sin^2{\theta^*}-1) + \frac{G C_i}{2r^{*3}}\frac{\Delta(R_i^5\epsilon_z^{(i)})}{R_i^5-R_{i-1}^5}(3\cos^2{\theta^*}-1),
\end{eqnarray}
or
\begin{eqnarray}
U_i(\vec{r}^*) &=& -\displaystyle\frac{Gm_i}{r^*} -\frac{3GC_i}{4r^{*3}}\frac{\Delta(R_i^5\epsilon_\rho^{(i)})}{R_i^5-R_{i-1}^5}\sin^2{\theta^*}\cos{2\varphi^*} +\frac{G C_i}{2r^{*3}}\frac{\Delta\big(R_i^5(\frac{\epsilon_\rho^{(i)}}{2}+\epsilon_z^{(i)})\big)}{R_i^5-R_{i-1}^5}(3\cos^2{\theta^*}-1).
\end{eqnarray}

If we consider the static equilibrium figure, the flattenings can be written as
\begin{equation}
 \epsilon_\rho^{(k)} = \mathcal{H}_k\epsilon_\rho; \ \ \ \ \ \ \ \epsilon_z^{(k)} = \mathcal{G}_k\overline{\epsilon}_z,
\end{equation}
where $\mathcal{H}_k$ and $\mathcal{G}_k$ are the Clairaut numbers. Then, the difference of the principal moments of inertia can be approximated to first 
order in the flattenings as
\begin{eqnarray}
 C_i-A_i &\approx& C_i (\mathcal{L}_i\epsilon_\rho+\mathcal{L}'_i\overline{\epsilon}_z)\nonumber\\
 C_i-B_i &\approx& C_i \mathcal{L}'_i\overline{\epsilon}_z\nonumber\\
 B_i-A_i &\approx& C_i \mathcal{L}_i\epsilon_\rho,
\label{eq:moment_i2}
\end{eqnarray}
where the parameters $\mathcal{L}_i$ and $\mathcal{L}'_i$ are
\begin{equation}
\mathcal{L}_i  = \frac{\mathcal{H}_iR_i^5-\mathcal{H}_{i-1}R_{i-1}^5}{R_i^5-R_{i-1}^5}; \ \ \ \ \ \ \mathcal{L}'_i = \frac{\mathcal{G}_iR_i^5-\mathcal{G}_{i-1}R_{i-1}^5}{R_i^5-R_{i-1}^5}.
\label{eq:Li}
\end{equation}
The coefficients $\mathcal{L}_i$ and $\mathcal{L}'_i$ play a role equivalent to the coefficients $\mathcal{H}_i$ and $\mathcal{G}_i$ for the quantities 
$C_i-A_i$, $C_i-B_i$ and $B_i-A_i$. In this case, the moments of inertia $B_i-A_i$ (resp. $C_i-B_i$) of the $i$th layer can be written as the homogeneous 
moments multiplied by the coefficients $\mathcal{L}_i$ (resp. $\mathcal{L}'_i$), characteristics of this layer. The difference between $\mathcal{L}_i$ and 
$\mathcal{L}'_i$ comes from the fact that the body has a differential rotation. If we assume a rigid rotation, then $\mathcal{L}'_i=\mathcal{L}_i(\Omega/n)^2$.

The corresponding gravitational potential of this homogeneous triaxial layer at an external point $\vec{r}^*$ is
\begin{eqnarray}
U_i(\vec{r}^*) &=& -\displaystyle\frac{Gm_i}{r^*} - \frac{G(B_i-A_i)}{2r^{*5}}(3x^{*2}-r^{*2}) + \frac{G(C_i-B_i)}{2r^{*5}}(3z^{*2}-r^{*2}) \nonumber\\
         &\approx& -\displaystyle\frac{Gm_i}{r^*} - \frac{GC_i\mathcal{L}_i}{2r^{*3}}\epsilon_\rho(3\cos^2{\varphi^*}\sin^2{\theta^*}-1) + \frac{G C_i\mathcal{L}'_i}{2r^{*3}}\overline{\epsilon}_z(3\cos^2{\theta^*}-1),
\end{eqnarray}
or
\begin{eqnarray}
U_i(\vec{r}^*) &=& -\displaystyle\frac{Gm_i}{r^*} - \frac{3GC_i\mathcal{L}_i}{4r^{*3}}\epsilon_\rho\sin^2{\theta^*}\cos{2\varphi^*}+\frac{GC_i}{2r^{*3}}\left(\mathcal{L}_i\frac{\epsilon_\rho}{2}+\mathcal{L}'_i\overline{\epsilon}_z\right)(3\cos^2{\theta^*}-1).
\end{eqnarray}

Although we do not calculate the principal moments of inertia when the inner and the outer boundaries are not aligned, it is possible to calculate easily 
the gravitational potential with the same scheme used to calculate the mass of the layer and the principal moments of inertia. The potential of the layer, 
can be written as the subtraction of the potential of two homogeneous ellipsoids of same density $d_i$: the potential of one homogeneous ellipsoid of mass 
$m'_i$ and the same surface as the outer boundary of the layer, given by the Eq. (\ref{eq:rho_i}), less the potential of the homogeneous ellipsoid of mass 
$m''_i$ and the same surface as the inner boundary of the layer, given by the Eq. (\ref{eq:rho_i-1}).

The corresponding gravitational potential is
\begin{eqnarray}
U_i(\vec{r}^*) &=& -\displaystyle\frac{Gm_i}{r^*} -\frac{3GC_i}{4r^{*3}}\frac{\Delta(R_i^5\epsilon_\rho^{(i)}\cos{(2\varphi^*-2\phi_i)})}{R_i^5-R_{i-1}^5}\sin^2{\theta^*}+\frac{G C_i}{2r^{*3}}\frac{\Delta\big(R_i^5(\frac{\epsilon_\rho^{(i)}}{2}+\epsilon_z^{(i)})\big)}{R_i^5-R_{i-1}^5}(3\cos^2{\theta^*}-1).\nonumber\\
\label{eq:U-hete1}
\end{eqnarray}

\renewcommand{\theequation}{B.\arabic{equation}}
\setcounter{equation}{0}  
\section{Near-synchronous solution of the rotational equations}{\label{online_sup_B}}

Using the convention $1=core$ and $2=shell$, the rotational system of the two-layer model, given by Eq. (\ref{eq:system-rotation}), can be written as
\begin{eqnarray}
 \dot{y}_1 &=& - T_{11}^*\mathcal{T}_1+\mathcal{K}_1\sin{2\xi}+\mathcal{F}_1(\gamma_2y_2-\gamma_1y_1)\nonumber\\
 \dot{y}_2 &=&   T_{21}^*\mathcal{T}_1-T_{22}^*\mathcal{T}_2-\mathcal{K}_2\sin{2\xi}-\mathcal{F}_2(\gamma_2y_2-\gamma_1y_1),
 \label{eq:rot-sys1}
\end{eqnarray}
where, the rotational variables are
\begin{equation}
 y_1=\frac{\nu_1}{\gamma_1}=\frac{2\Omega_1}{\gamma_1}-\frac{2n}{\gamma_1}; \ \ \ \ \ \ y_2=\frac{\nu_2}{\gamma_2}=\frac{2\Omega_2}{\gamma_2}-\frac{2n}{\gamma_2},
\end{equation}
the tidal function $\mathcal{T}_i$ is
\begin{equation}
 \mathcal{T}_i = \sum_{k,j\in \mathbb{Z}} E_{2,k}E_{2,k+j}\frac{(y_i+P_{ik})\cos{(jnt)}+\sin{(jnt)}}{1+(y_i+P_{ik})^2}.
\end{equation}
The constants are
\begin{equation}
 T_{ij}^* = \frac{2\mathcal{T}}{\gamma_i}\frac{\mathcal{H}_jR_j^5}{R_i^5-R_{i-1}^5}; \ \ \ \ \ \  \mathcal{K}_i=\frac{2K}{\gamma_iC_i}; \ \ \ \ \ \ \mathcal{F}_i=\frac{\mu}{\gamma_iC_i}; \ \ \ \ \ \ P_{ik} = \frac{kn}{\gamma_i} = kp_i,
 \label{eq:cte1}
\end{equation}
and the tidal parameter $\mathcal{T}$, is defined as
\begin{equation}
\mathcal{T} = \frac{45GM^2R_s^3}{8m_Ta^6} \approx \frac{3n^2\overline{\epsilon}_\rho}{2}.
\end{equation}

We assume that the solution, to second order in eccentricity, can be written as
\begin{eqnarray}
 y_1 &=& b_{10}e^2 + c_{11}e\cos{\ell} + s_{11}e\sin{\ell} + c_{12}e^2\cos{2\ell} + s_{12}e^2\sin{2\ell}\nonumber\\
 y_2 &=& b_{20}e^2 + c_{21}e\cos{\ell} + s_{21}e\sin{\ell} + c_{22}e^2\cos{2\ell} + s_{22}e^2\sin{2\ell},
\label{eq:yprop}
\end{eqnarray}
where $b_{i0}$, $c_{ij}$ and $s_{ij}$ are undetermined coefficients. Introducing the solution (\ref{eq:yprop}) into the rotational system (\ref{eq:rot-sys1}) 
and expanding to second order in eccentricity, by identification of the terms with same trigonometric argument, we can calculate these coefficients.

The derivatives of (\ref{eq:yprop}) are
\begin{eqnarray}
 \dot{y}_1 &=& n s_{11}e\cos{\ell} - n c_{11}e\sin{\ell} + 2n s_{12}e^2\cos{2\ell} - 2n c_{12}e^2\sin{2\ell}\nonumber\\
 \dot{y}_2 &=& n s_{21}e\cos{\ell} - n c_{21}e\sin{\ell} + 2n s_{22}e^2\cos{2\ell} - 2n c_{22}e^2\sin{2\ell},
\label{eq:dot-yprop}
\end{eqnarray}

The tidal function can be approximated by
\begin{eqnarray}
 \mathcal{T}_i &\simeq& \left(b_{i0}-\frac{12p_i}{1+p_i^2}+q_ic_{i1}+r_is_{i1}\right)e^2+\left(c_{i1}-\frac{4p_i}{1+p_i^2}\right)e\cos{\ell}+\left(s_{i1}-\frac{4p_i^2}{1+p_i^2}\right)e\sin{\ell}\nonumber\\
               &      & +\left(c_{i2}-\frac{17p_i}{1+4p_i^2}+q_ic_{i1}-r_is_{i1}\right)e^2\cos{2\ell}+\left(s_{i2}-\frac{34p_i^2}{1+4p_i^2}+r_ic_{i1}+q_is_{i1}\right)e^2\sin{2\ell},
\label{eq:apro-fi}
\end{eqnarray}
where the coefficients $q_i$ and $r_i$ are
\begin{equation}
 q_i = \frac{3(2+p_i^2+p_i^4)}{2(1+p_i^2)^2}; \ \ \ \ \ \ r_i = \frac{3p_i}{(1+p_i^2)^2}.
\end{equation}

In the same way, the trigonometric function of the gravitational coupling can be approximated by
\begin{eqnarray}
 \sin{2\xi} &=& \sin{\left[\tan^{-1}\left(\frac{y_2}{1+\lambda_2(1+y_2^2)} \right)-\tan^{-1}\left(\frac{y_1}{1+\lambda_1(1+y_1^2)} \right)\right]}\nonumber\\
       &\simeq& \left(\frac{b_{20}}{1+\lambda_2} -\frac{b_{10}}{1+\lambda_1}\right)e^2+\left(\frac{c_{21}}{1+\lambda_2}-\frac{c_{11}}{1+\lambda_1}\right)e\cos{\ell}+\left(\frac{s_{21}}{1+\lambda_2}-\frac{s_{11}}{1+\lambda_1}\right)e\sin{\ell}\nonumber\\
       &      & +\left(\frac{c_{22}}{1+\lambda_2}-\frac{c_{12}}{1+\lambda_1}\right)e^2\cos{2\ell}+\left(\frac{s_{22}}{1+\lambda_2}-\frac{s_{12}}{1+\lambda_1}\right)e^2\sin{2\ell},
\label{eq:apro-ac}
\end{eqnarray}
and the amplitude of oscillation is
\begin{eqnarray}
 K&=&\frac{32\pi^2 G}{75}\mathcal{H}_1\mathcal{H}_2\overline{\epsilon}_\rho^2 d_1 d_2  R_1^5E_{2,0}^2\sqrt{\lambda_1^2 +\frac{1 + 2\lambda_1}{1+y_1^2}}\sqrt{\lambda_2^2 +\frac{1 + 2\lambda_2}{1+y_2^2}}\nonumber\\
  &\simeq&\frac{32\pi^2 G}{75}\mathcal{H}_1\mathcal{H}_2\overline{\epsilon}_\rho^2 d_1 d_2  R_1^5 (1+\lambda_1)(1+\lambda_2)+\mathcal{O}(e^2).
\end{eqnarray}

The friction term is
\begin{eqnarray}
\gamma_2y_2-\gamma_1y_1 &\simeq& \left(\gamma_2b_{20}-\gamma_1b_{10}\right)e^2+\left(\gamma_2c_{21}-\gamma_1c_{11}\right)e\cos{\ell}+\left(\gamma_2s_{21}-\gamma_1s_{11}\right)e\sin{\ell}\nonumber\\
                        &      & +\left(\gamma_2c_{22}-\gamma_1c_{12}\right)e^2\cos{2\ell}+\left(\gamma_2s_{22}-\gamma_1s_{12}\right)e^2\sin{2\ell}.
\label{eq:apro-fric}
\end{eqnarray}

Replacing (\ref{eq:dot-yprop})-(\ref{eq:apro-fric}) into (\ref{eq:rot-sys1}) and colecting the terms with same trigonometric argument, we can find three 
linear sub-systems for the undetermined $b_{i0}$, $c_{ij}$ and $s_{ij}$, which can be written in vectorial notation as
\begin{eqnarray}
 \tens{D}_1 \tens{\Lambda}_1 &=& \tens{T}_1\tens{P}_1\nonumber\\
 \tens{D}_2 \tens{\Lambda}_2 &=& \tens{T}_1\tens{P}_2-\tens{T}_1\tens{R}_2\nonumber\\
   \tens{D} \tens{\Lambda}_0 &=& \tens{T}\tens{P}-\tens{T}\tens{R},
\end{eqnarray}
where
\begin{equation}
\tens{\Lambda}_0=\left[ \begin{array}{c} b_{10} \\
                                         b_{20} \end{array}\right]; \ \ \ \ \ \ 
\tens{\Lambda}_1=\left[ \begin{array}{c} c_{11} \\
                                         c_{21} \\
                                         s_{11} \\ 
                                         s_{21} \end{array}\right]; \ \ \ \ \ \ 
\tens{\Lambda}_2=\left[ \begin{array}{c} c_{12} \\
                                         c_{22} \\
                                         s_{12} \\ 
                                         s_{22} \end{array}\right],
\end{equation}
are the undetermined coefficients vectors. The constants matrices are defined as
\begin{equation}
\tens{T}=\left[ \begin{array}{rr} T_{11}^* & 0 \\
                                 -T_{21}^* &  T_{22}^* \end{array}\right]; \ \ \ \ \ \  
\tens{T}_1=\left[ \begin{array}{cc} \tens{T} & \tens{0} \\
                                  \tens{0} & \tens{T} \end{array}\right]; \ \ \ \ \ \  
\tens{D}=\left[ \begin{array}{rr} d_{11} & d_{12} \\
                                  d_{21} & d_{22} \end{array}\right]; \ \ \ \ \ \  
\tens{D}_1=\left[ \begin{array}{cc}  \tens{D} & n\tens{I} \\
                                   -n\tens{I} &  \tens{D} \end{array}\right]; \ \ \ \ \ \  
\tens{D}_2=\left[ \begin{array}{cc}   \tens{D} & 2n\tens{I} \\
                                   -2n\tens{I} &   \tens{D} \end{array}\right],
\end{equation}
where $\tens{I}$ is the identity matrix and the coefficients $d_{ij}$ are
\begin{align}
 d_{11} &= T_{11}^*+\frac{\mathcal{K}_1}{1+\lambda_1}+\mathcal{F}_1\gamma_1;& d_{12} &=-\frac{\mathcal{K}_1}{1+\lambda_2}-\mathcal{F}_1\gamma_2; \nonumber\\
 d_{21} &=-T_{21}^*+\frac{\mathcal{K}_2}{1+\lambda_1}+\mathcal{F}_2\gamma_1;& d_{22} &= T_{22}^*-\frac{\mathcal{K}_2}{1+\lambda_2}-\mathcal{F}_2\gamma_2,
\end{align}
and the vectors $\tens{P}_i$ and $\tens{R}_i$ are
\begin{equation*}
 \tens{P}=12\left[ \begin{array}{c} p_1/(1+p_1^2) \\
                                      p_2/(1+p_2^2)  \end{array}\right]; \ \ \ \ \ \ 
\tens{R}=\left[ \begin{array}{c} q_1c_{11}+r_1s_{11} \\
                                 q_2c_{21}+r_2s_{21} \end{array}\right];
\end{equation*}
\begin{equation}
\tens{P}_1=4\left[ \begin{array}{l} p_1/(1+p_1^2) \\
                                    p_2/(1+p_2^2)\\
                                    p_1^2/(1+p_1^2) \\ 
                                    p_2^2/(1+p_2^2) \end{array}\right]; \ \ \ \ \ \ 
\tens{P}_2=17\left[ \begin{array}{c} p_1/(1+4p_1^2) \\
                                     p_2/(1+4p_2^2)\\
                                    2p_1^2/(1+4p_1^2) \\ 
                                    2p_2^2/(1+4p_2^2) \end{array}\right]; \ \ \ \ \ \ 
 \tens{R}_2=\left[ \begin{array}{l} q_1c_{11}-r_1s_{11} \\
                                    q_2c_{21}-r_2s_{21} \\
                                    r_1c_{11}+q_1s_{11} \\ 
                                    r_2c_{21}+q_2s_{21} \end{array}\right].
\end{equation}

The solution of these sub-systems are
\begin{eqnarray}
  \tens{\Lambda}_1 &=& \tens{D}_1^{-1}\tens{T}_1\tens{P}_1\nonumber\\
  \tens{\Lambda}_2 &=& \tens{D}_2^{-1}\tens{T}_1\tens{P}_2-\tens{D}_2^{-1}\tens{T}_1\tens{R}_2\nonumber\\
  \tens{\Lambda}_0 &=& \tens{D}^{-1}\tens{T}\tens{P}-\tens{D}^{-1}\tens{T}\tens{R}.
\end{eqnarray}

Finally, the rotational solutions can be written as
\begin{eqnarray}
\nu_1 &=& B_{10} + B_{11}\cos{(\ell+\phi_{11})} + B_{12}\cos{(2\ell+\phi_{12})}\nonumber\\
\nu_2 &=& B_{20} + B_{21}\cos{(\ell+\phi_{21})} + B_{22}\cos{(2\ell+\phi_{22})},
\label{eq:aprox-nui}
\end{eqnarray}
where the constants $B_{ij}$ and the phases $\phi_{ij}$ are
\begin{eqnarray}
   B_{i0} = \gamma_ib_{i0}e^2; \ \ \ \ \ \ B_{ij} = \gamma_i\sqrt{c_{ij}^2+s_{ij}^2}\ e^j; \ \ \ \ \ \ \phi_{ij} = -\tan^{-1}{(s_{ij}/c_{ij})}.
\label{eq:B_ij}
\end{eqnarray}

\subsection{Tidal drift and the periodic terms}

The tidal drift is the term $B_{i0}$ of the solution (\ref{eq:aprox-nui}). It is
\begin{eqnarray}
 \nu_1^{(stat)} = B_{10} = \gamma_1b_{10}e^2+\mathcal{O}(e^3) \nonumber\\
 \nu_2^{(stat)} = B_{20} = \gamma_2b_{20}e^2+\mathcal{O}(e^3).
\label{eq:stat-solution1}
\end{eqnarray}

This result can be rewritte as
\begin{eqnarray}
 \nu_1^{(stat)}&=&\frac{12n\kappa_{11}\gamma_1^2e^2}{\gamma_1^2+n^2} +\frac{12n\kappa_{12}\gamma_2^2e^2}{\gamma_2^2+n^2}-\kappa_{11}\gamma_1(q_1c_{11}+r_1s_{11})e^2 -\kappa_{12}\gamma_2(q_2c_{21}+r_2s_{21})e^2+\mathcal{O}(e^3)\nonumber\\
 \nu_2^{(stat)}&=&\frac{12n\kappa_{21}\gamma_1^2e^2}{\gamma_1^2+n^2} +\frac{12n\kappa_{22}\gamma_2^2e^2}{\gamma_2^2+n^2}-\kappa_{21}\gamma_1(q_1c_{11}+r_1s_{11})e^2 -\kappa_{22}\gamma_2(q_2c_{21}+r_2s_{21})e^2+\mathcal{O}(e^3).
 \label{eq:excess}
\end{eqnarray}
The coefficient $\kappa_{ij}$ can be written as $\kappa_{ij}=f_{ij}/g$, where $f_{ij}$ is
\begin{equation}
 f_{ij} = \delta_{i,j}\frac{\mathcal{T}C_1C_2}{C}\frac{\mathcal{H}_1\mathcal{H}_2R_2^5}{R_2^5-R_1^5}+\frac{\mathcal{D}_j\gamma_i}{\gamma_j}\frac{(1+\lambda_i)K}{(1+\lambda_1)(1+\lambda_2)}+\mathcal{D}_j\gamma_i\frac{\mu}{2},
\end{equation}
$\delta_{i,j}$ is the Kronecker delta ($\delta_{1,1}=\delta_{2,2}=1$ and $\delta_{1,2}=\delta_{2,1}=0$), the parameter $\mathcal{D}_1$ and $\mathcal{D}_2$ 
are defined as
\begin{equation}
\mathcal{D}_1 = \left(C_1-\frac{C_2R_1^5}{R_2^5-R_2^5}\right)\frac{\mathcal{H}_1}{C}; \ \ \ \ \ \ \mathcal{D}_2 = \frac{C_2R_2^5}{R_2^5-R_1^5}\frac{\mathcal{H}_2}{C}
\label{eq:Di}
\end{equation}
and the constant $g$ is
\begin{equation}
g = f_{11}+f_{22}-\frac{\mathcal{T}C_1C_2}{C}\frac{\mathcal{H}_1\mathcal{H}_2R_2^5}{R_2^5-R_2^5}.
\end{equation}

The two first terms of each Eq. (\ref{eq:excess})
\begin{eqnarray}
 N_i &=&\frac{12n\kappa_{i1}\gamma_1^2e^2}{\gamma_1^2+n^2} +\frac{12n\kappa_{i2}\gamma_2^2e^2}{\gamma_2^2+n^2}
\end{eqnarray}
come from the non-periodic terms with $|j|=0$, while the terms that involve $c_{i1}$ and $s_{i1}$.
\begin{eqnarray}
 P_i &=&-\kappa_{i1}\gamma_1(q_1c_{11}+r_1s_{11})e^2-\kappa_{i2}\gamma_2(q_2c_{21}+r_2s_{21})e^2,
\end{eqnarray}
come from the periodic terms with $|j|=1$. The harmonic terms with $|j|=2$, do not contribute to the stationary rotation at order $e^2$.

It is worth emphasizing that in the absence of friction and gravitational coupling, that is, $K=\mu=0$, the coefficient $\kappa_{ij}=\delta_{i,j}$. Then, 
the non-periodic tidal drift of the $i$th layer has the same expression that the excess rotation in the case of a homogeneous body, with $\gamma_i$ instead 
of $\gamma$
\begin{equation}
 \nu_i^{(stat)}=\frac{12n\gamma_i^2e^2}{\gamma_i^2+n^2}+\mathcal{O}(e^3).
 \label{eq:excess2}
\end{equation}

In the case $n/\gamma_1\gg1$, $n/\gamma_2\gg1$, an elementary calculation shows that each coefficient $\kappa_{ij}$ becomes independent of the friction 
parameter $\mu$, depending only on the internal structure and on the relaxation factors $\gamma_1$ and $\gamma_2$, with $f_{ij}$ tending to
\begin{equation}
 f_{ij} = \delta_{i,j}\frac{\mathcal{T}C_1C_2}{C}\frac{\mathcal{H}_1\mathcal{H}_2R_2^5}{R_2^5-R_1^5}+\frac{\mathcal{D}_j\gamma_i}{\gamma_j}\frac{(1+\lambda_i)K}{(1+\lambda_1)(1+\lambda_2)}.
\end{equation}

In the case $n/\gamma_1\ll1$, $n/\gamma_2\ll1$, each coefficient $\kappa_{ij}$ becomes independent of $\mathcal{T}$, $K$ and $\mu$, depending only on the 
internal structure and on the relaxation factors $\gamma_1$ and $\gamma_2$, tending to
\begin{equation}
\kappa_{ij}= \frac{\mathcal{D}_j\gamma_i}{\mathcal{D}_1\gamma_1+\mathcal{D}_2\gamma_2},
\label{eq:kappa-rigid}
\end{equation}
and the stationary solution tends to synchronous rotation.

\begin{figure}[h]
\begin{center}
\includegraphics[scale=0.5]{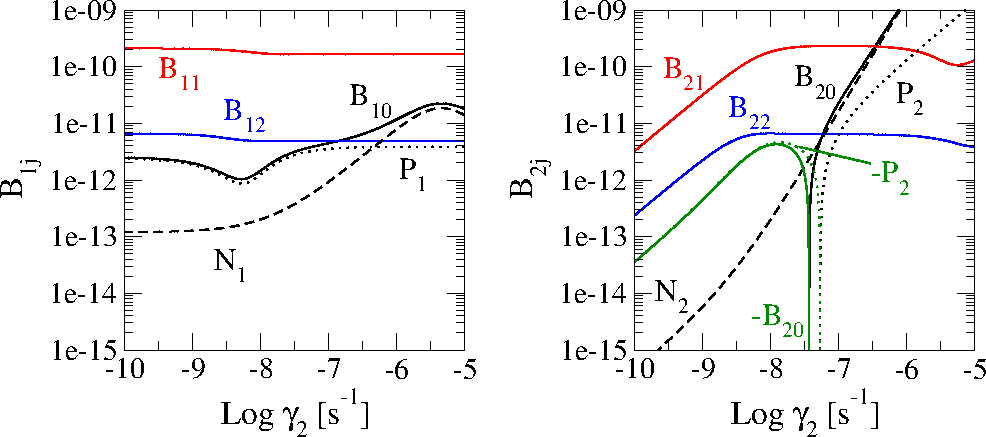}
\caption{The tidal drift $B_{i0}$ (\textit{black solid lines}), the contribution of the non-periodic tidal drift $N_i$ (\textit{black dashed lines}) and the 
periodic tidal drift $P_i$ (\textit{black dotted lines}) to the tidal drift, and the amplitudes of oscillation of the periodic terms $B_{i1}$ (\textit{red 
solid lines}) and $B_{i2}$ (\textit{blue solid lines}), of the Titan's core and shell in function of the shell relaxation factor $\gamma_2$. The core 
relaxation factor is $\gamma_1=10^{-8}$ {\rm s}$^{-1}$ and the ocean's viscosity and thickness are $\eta_o=10^{-3}$ {\rm Pa s} and $h=178$ {\rm km}, 
respectively (see Tables \ref{Data-tab}-\ref{Table-parameters-2l-I}). \textit{Left}: The parameter of the core. \textit{Right}: The parameters of the shell. 
We also plot the negative values of $B_{20}$ (\textit{green solid line}) and $P_{2}$ (\textit{green dotted line}).}
\label{fig20}
\end{center}
\end{figure}

The periodic terms have amplitudes $B_{i1}$ and $B_{i2}$, given by the Eq. (\ref{eq:B_ij}). The coefficients $c_{ij}$ and $s_{ij}$ gives rise to intricate 
analytical expressions, but are easy to calculate numerically. Fig. \ref{fig20} shows one example for the Titan's core and the shell constants $B_{1j}$ and 
$B_{2j}$, respectively, in function of the shell relaxation factor $\gamma_2$ (see Table \ref{Data-tab}-\ref{Table-parameters-2l-I}). We use that the core 
relaxation factor is $\gamma_1=10^{-8}$ {\rm s}$^{-1}$, and fix the ocean's viscosity and thickness values to $\eta_o=10^{-3}$ {\rm Pa s} and $h=178$ {\rm 
km}, respectively. We also plot the non-periodic $N_i$ and periodic $P_i$ terms, separately, and the total tidal drift $B_{i0}=N_i+P_i$. We can observe 
that if $\gamma_2\gtrsim10^{-7.5}$ {\rm s}$^{-1}$, the shell oscillates around the super-synchronous rotation. When $\gamma_2\lesssim10^{-7.5}$ {\rm s}$^{-1}$, 
the tidal drift $B_{20}$ becomes negative and tends to zero, that is, the shell oscillates around the synchronous rotation, with a period of oscillation 
equal to the orbital period. The negative sign of the tidal drift $B_{20}$, is due to the contribution of the periodic terms $P_2$, which becomes negative 
and $|P_2|\gg N_2$. Finally, if $\gamma_2\lesssim10^{-8}$ {\rm s}$^{-1}$, the amplitude of the shell rotation decreases, tending to zero when $\gamma_2$ 
decreases. On the other hand, the core oscillates around the synchronous rotation, with a period of oscillation equal to the orbital period, independently 
of the shell relaxation factor. This behavior is confirmed by the numerical simulations of non-approximate system (see Sec. \ref{sec10}).

\begin{figure}[h]
\begin{center}
\includegraphics[scale=0.6]{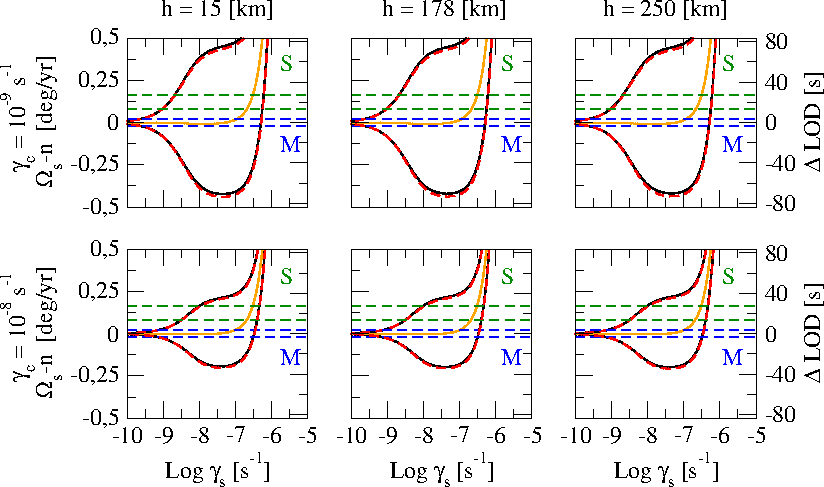}
\caption{Comparison of the amplitudes of the shell rotation and corresponding length-of-day variation of Titan, between the numerical integration of the 
system Eq. (\ref{eq:system-rotation}) (\textit{solid black lines}) and the analytical solution $\nu_i\simeq B_{i0}+B_{i1}\cos{(\ell+\phi_{i1})}$ 
(\textit{dashed red lines}), when $\eta_o=10^{-3}$ {\rm Pa s}. We also plot the stationary solution given by $B_{i0}$ (\textit{solid orange line}). The 
horizontal dashed lines show the confidence interval of the observed values, as determined by Meriggiola (2016) (\textit{blue}) and by Stiles et al. (2010) 
(\textit{green}).}
\label{fig21}
\end{center}
\end{figure}

In Fig. \ref{fig21}, we show the comparison of the Titan's shell rotation in the complete non-linear system given by Eq. (\ref{eq:rot-sys1}) and 
in the approximate analytical solution given by Eq. (\ref{eq:aprox-nui}), for some values of the core's relaxation factor $\gamma_1$ and ocean thickness $h$. 
The dashed red lines show the maximum and minimum values of $\Omega_2-n$ given by the approximate solution, taking into account only the first harmonic 
($|j|\leq1$), while the solid black lines show the maximum and minimum values of $\Omega_2-n$ when the complete non-linear system is integrated (using $|j|\leq7$). 
The approximate solution is in excellent agreement with numerical integration of the equations.

\subsection{Atmospheric influence}

When we consider the effect of the atmosphere, the rotational system becomes 
\begin{eqnarray}
 \dot{y}_1 &=& - T_{11}^*\mathcal{T}_1+\mathcal{K}_1\sin{2\xi}+\mathcal{F}_1(\gamma_2y_2-\gamma_1y_1)\nonumber\\
 \dot{y}_2 &=&   T_{21}^*\mathcal{T}_1-T_{22}^*\mathcal{T}_2-\mathcal{K}_2\sin{2\xi}-\mathcal{F}_2(\gamma_2y_2-\gamma_1y_1)+\mathcal{A}_\odot\sin{2\alpha_\odot}.
 \label{eq:rot-sys1-atm}
\end{eqnarray}
where
\begin{equation}
 \mathcal{A}_\odot=\frac{2A_\odot}{\gamma_2}.
\end{equation}

We assume that the particular solution
\begin{eqnarray}
y_{1\odot} = C_{1\odot}\cos{2\alpha_\odot} + S_{1\odot}\sin{2\alpha_\odot}\nonumber\\
y_{2\odot} = C_{2\odot}\cos{2\alpha_\odot} + S_{2\odot}\sin{2\alpha_\odot},
\label{eq:yatm}
\end{eqnarray}
can be added to (\ref{eq:yprop}) to obtain the general solutions of the complete equation. $C_{j\odot}$ and $S_{j\odot}$ are undetermined coefficients to 
be obtained by substitution of the parts of the solution into Eq. (\ref{eq:rot-sys1-atm}) and identification.

The derivative of (\ref{eq:yatm}) is
\begin{eqnarray}
 \dot{y}_{1\odot} &=& -2n_{\odot} C_{1\odot}\sin{2\alpha_\odot} + 2n_{\odot} S_{1\odot}\cos{2\alpha_\odot}\nonumber\\
 \dot{y}_{2\odot} &=& -2n_{\odot} C_{2\odot}\sin{2\alpha_\odot} + 2n_{\odot} S_{2\odot}\cos{2\alpha_\odot}.
\label{eq:dot-yatm}
\end{eqnarray}

The tidal function can be approximated by
\begin{eqnarray}
 \mathcal{T}_i &\simeq& C_{i\odot}\cos{2\alpha_\odot}+S_{i\odot}\sin{2\alpha_\odot},
\label{eq:apro-fi-atm}
\end{eqnarray}
the trigonometric function of the gravitational coupling can be approximated by
\begin{eqnarray}
 \sin{2\xi} &\simeq& \left(\frac{C_{2\odot}}{1+\lambda_2}-\frac{C_{1\odot}}{1+\lambda_1}\right)\cos{2\alpha_\odot}+\left(\frac{S_{2\odot}}{1+\lambda_2}-\frac{S_{1\odot}}{1+\lambda_1}\right)e^2\sin{2\alpha_\odot},
\label{eq:apro-ac-atm}
\end{eqnarray}
and the friction term is
\begin{eqnarray}
\gamma_2y_2-\gamma_1y_1 &\simeq& \left(\gamma_2C_{2\odot}-\gamma_1C_{1\odot}\right)\cos{2\alpha_\odot}+\left(\gamma_2S_{2\odot}-\gamma_1S_{1\odot}\right)\sin{2\alpha_\odot}.
\label{eq:apro-fric-atm}
\end{eqnarray}

Defining the constant matrix $\tens{D}_\odot$ and the constant vectors $\tens{\Lambda}_\odot$, $\tens{P}_\odot$, as
\begin{equation}
\tens{D}_\odot=\left[ \begin{array}{cc}   \tens{D} & 2n_\odot\tens{I} \\
                                   -2n_\odot\tens{I} &   \tens{D} \end{array}\right]; \ \ \ \ \ \ 
\tens{\Lambda}_\odot=\left[ \begin{array}{c} C_{1\odot} \\
                                             C_{2\odot} \\
                                             S_{1\odot} \\ 
                                             S_{2\odot} \end{array}\right]; \ \ \ \ \ \ 
\tens{P}_\odot=\mathcal{A}_\odot \left[ \begin{array}{c} 0 \\
                                                         0 \\ 
                                                         0 \\
                                                         1 \end{array}\right],
\end{equation}
the undetermined coefficient vector is
\begin{eqnarray}
 \tens{\Lambda}_\odot = \tens{D}^{-1}_\odot\tens{P}_\odot.
\end{eqnarray}

\begin{figure}[h]
\begin{center}
\includegraphics[scale=0.6]{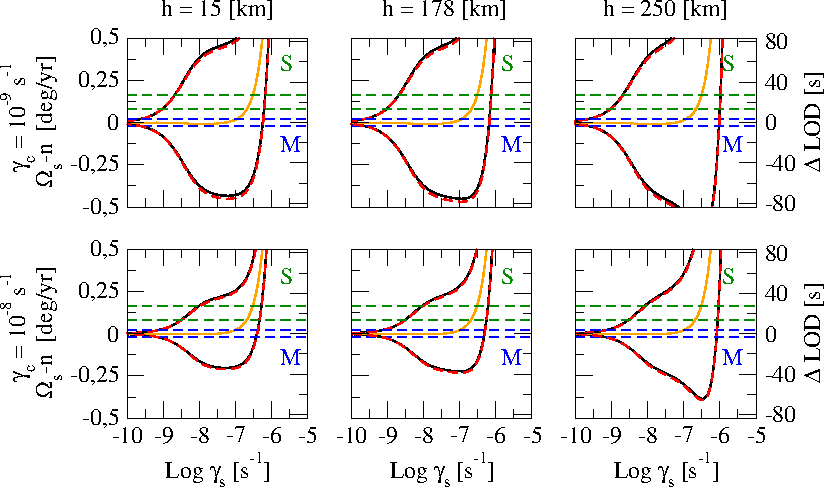}
\caption{Same as Fig. \ref{fig21}, including the atmospheric influence.}
\label{fig22}
\end{center}
\end{figure}

In Fig. \ref{fig22}, we show the same comparison of the Titan's shell rotation in the complete non-linear system and in the approximate 
analytical solution of the above section. The approximate solution, also is in excellent agreement with numerical integration. It is important to note that 
the fact that the approximate solution of the non-linear system (\ref{eq:rot-sys1-atm}) can be expressed as the sum of solutions (\ref{eq:yprop}) and 
(\ref{eq:yatm}), it means that this system has a behavior quasi-linear, at least for the Titan's problem.

\renewcommand{\theequation}{C.\arabic{equation}}
\setcounter{equation}{0}  
\section{The integral of section \ref{sec6}}{\label{online_sup_C}}

\subsubsection*{Proposition:}
\begin{equation}
\frac{1}{2\pi}\int_0^{2\pi}\Omega_j^2\left(\frac{a}{r}\right)^4\sin{v}\ d\ell = 0.
\label{eq:integral} 
\end{equation}
To prove (\ref{eq:integral}), we consider only the tidal force. Introducing the adimensinals variables and time
\begin{equation}
 y_i=\frac{\nu_i}{\gamma_i}; \ \ \ \ \ \ \ \ \ x=nt=\ell,
\end{equation}
the rotational system can be written as
\begin{eqnarray}
 \dot{y}_1 &=& - T_{11}^*\sum_{k,j\in \mathbb{Z}} E_{2,k}E_{2,k+j}\frac{(y_1+P_{1k})\cos{(jx)}+\sin{(jx)}}{1+(y_1+P_{1k})^2}\nonumber\\
 \dot{y}_2 &=& - T_{22}^*\sum_{k,j\in \mathbb{Z}} E_{2,k}E_{2,k+j}\frac{(y_2+P_{2k})\cos{(jx)}+\sin{(jx)}}{1+(y_2+P_{2k})^2}\nonumber\\
             & & + T_{21}^*\sum_{k,j\in \mathbb{Z}} E_{2,k}E_{2,k+j}\frac{(y_1+P_{1k})\cos{(jx)}+\sin{(jx)}}{1+(y_1+P_{1k})^2} \nonumber\\
  \vdots     & & \nonumber\\
 \dot{y}_N &=& - T_{NN}^*\sum_{k,j\in \mathbb{Z}} E_{2,k}E_{2,k+j}\frac{(y_N+P_{Nk})\cos{(jx)}+\sin{(jx)}}{1+(y_N+P_{Nk})^2}\nonumber\\
             & & + T_{NN-1}^*\sum_{k,j\in \mathbb{Z}} E_{2,k}E_{2,k+j}\frac{(y_{N-1}+P_{N-1k})\cos{(jx)}+\sin{(jx)}}{1+(y_{N-1}+P_{N-1k})^2}.
\end{eqnarray}
where the constants
\begin{equation}
 T_{ij}^* =  \frac{2\mathcal{T}}{\gamma_in}\frac{\mathcal{H}_jR_j^5}{R_i^5-R_{i-1}^5}; \ \ \ \ \ \ \  P_{ik} = \frac{kn}{\gamma_i}.
\end{equation}

In low-$\gamma$ approximation ($\gamma_i\ll n$), we can neglet the terms $k\neq 0$. If we consider only the terms $j=0$, the system becomes
\begin{eqnarray}
 \dot{y}_1 &=& - \frac{T_{11}^* E_{2,0}^2y_1}{1+y_1^2}\nonumber\\
 \dot{y}_2 &=& - \frac{T_{22}^* E_{2,0}^2y_2}{1+y_2^2}+\frac{T_{21}^* E_{2,0}^2y_1}{1+y_1^2}\nonumber\\
  \vdots     & & \nonumber\\
 \dot{y}_N &=& - \frac{T_{NN}^* E_{2,0}^2y_N}{1+y_N^2}+\frac{T_{NN-1}^* E_{2,0}^2y_{N-1}}{1+y_{N-1}^2}.
 \label{eq:sys-red}
\end{eqnarray}

In the same way in Ferraz-Mello (2015), each solution of this system tends to zero. The role of the terms $j\neq 0$ are periodic fluctuations which are the 
harmonics of the orbital period are that added to the solution. If we consider the terms $j\neq 0$, we have that $y_i\ll 1$, and the rotational system is
\begin{eqnarray}
 \dot{y}_i &=& - \sum_{j\in \mathbb{Z}\ j\neq 0} K_{ij}\sin{(jx)},
\end{eqnarray}
where $K_{ij}=(T_{ii}^*-T_{ii-1}^*)E_{2,0}E_{2,j}$. The solution of this differential equation is
\begin{equation}
 y_i(x) = y_{i0} -K_{ij} + \sum_{j\in \mathbb{Z}\ j\neq 0} \frac{K_{ij}}{j} \cos{(jx)},
\end{equation}
or, in term of the angular velocity, we obtain
\begin{equation}
 \Omega_i = \Omega_{i0} -\frac{\gamma_iK_{ij}}{2}  + \sum_{j\in \mathbb{Z}\ j\neq 0} \frac{\gamma_iK_{ij}}{2j} \cos{(jnt)}.
\end{equation}

Therefore, the square of the angular velocity of the $j$th layer can be written as
\begin{equation}
 \Omega_j^2 = \sum_{k=0}^\infty A_{jk} \cos{k\ell}.
\end{equation}

Finally, the integral (\ref{eq:integral}) is
\begin{eqnarray}
 && \frac{1}{2\pi}\int_0^{2\pi}\Omega_j^2\left(\frac{a}{r}\right)^4\sin{v}\ d\ell =\nonumber\\
 &&\ \ \ \ \ \ \ \ \ \    \sum_{k=0}^\infty\frac{A_{jk}}{2}\left(\frac{1}{2\pi}\int_0^{2\pi}\left(\frac{a}{r}\right)^4\sin{(v+k\ell)}\ d\ell-\frac{1}{2\pi}\int_0^{2\pi}\left(\frac{a}{r}\right)^4\sin{(v-k\ell)}\ d\ell\right)=0.
\end{eqnarray}

In high-$\gamma$ approximation ($\gamma_i\gg n$),  we can neglet $P_{ik}$, then, the system can be written as
\begin{eqnarray}
 \dot{y}_1 &=& - T_{11}^*\sum_{k,j\in \mathbb{Z}} E_{2,k}E_{2,k+j}\frac{y_1\cos{(jx)}+\sin{(jx)}}{1+y_1^2}\nonumber\\
 \dot{y}_2 &=& - T_{22}^*\sum_{k,j\in \mathbb{Z}} E_{2,k}E_{2,k+j}\frac{y_2\cos{(jx)}+\sin{(jx)}}{1+y_2^2} + T_{21}^*\sum_{k,j\in \mathbb{Z}} E_{2,k}E_{2,k+j}\frac{y_1\cos{(jx)}+\sin{(jx)}}{1+y_1^2} \nonumber\\
  \vdots     & & \nonumber\\
 \dot{y}_N &=& - T_{NN}^*\sum_{k,j\in \mathbb{Z}} E_{2,k}E_{2,k+j}\frac{y_N\cos{(jx)}+\sin{(jx)}}{1+y_N^2} + T_{NN-1}^*\sum_{k,j\in \mathbb{Z}} E_{2,k}E_{2,k+j}\frac{y_{N-1}\cos{(jx)}+\sin{(jx)}}{1+y_{N-1}^2}.\nonumber\\
\end{eqnarray}

If we consider only the terms $j=0$, the system becomes
\begin{eqnarray}
 \dot{y}_1 &=& - T_{11}^*\sum_{k\in \mathbb{Z}} E_{2,k}^2\frac{y_1}{1+y_1^2}\nonumber\\
 \dot{y}_2 &=& - T_{22}^*\sum_{k\in \mathbb{Z}} E_{2,k}^2\frac{y_2}{1+y_2^2}+ T_{21}^*\sum_{k\in \mathbb{Z}} E_{2,k}^2\frac{y_1}{1+y_1^2} \nonumber\\
  \vdots     & & \nonumber\\
 \dot{y}_N &=& - T_{NN}^*\sum_{k\in \mathbb{Z}} E_{2,k}^2\frac{y_N}{1+y_N^2} + T_{NN-1}^*\sum_{k\in \mathbb{Z}} E_{2,k}^2\frac{y_{N-1}}{1+y_{N-1}^2},
\end{eqnarray}
which is identical to the system \ref{eq:sys-red}, with $\sum_{k\in \mathbb{Z}} E_{2,k}^2$ instead of $E_{2,0}^2$. Therefore, each solution of this system 
tends to zero. As in low-$\gamma$ approximation, the role of the terms $j\neq 0$ are periodic fluctuations which are the harmonics of the orbital period are 
added to the solution. If we consider the terms $j\neq 0$, we have that $y_i\ll 1$, and the rotational system is
\begin{eqnarray}
 \dot{y}_i &=& - \sum_{j\in \mathbb{Z}\ j\neq 0} K'_{ij}\sin{(jx)},
\end{eqnarray}
where $K'_{ij}=(T_{ii}^*-T_{ii-1}^*)\sum_{k\in \mathbb{Z}} E_{2,k}E_{2,k+j}$. Using the solution of the low-$\gamma$ approximation, then, the angular velocity is
\begin{equation}
 \Omega_i = \Omega_{i0} -\frac{\gamma_iK'_{ij}}{2}  + \sum_{j\in \mathbb{Z}\ j\neq 0} \frac{\gamma_iK'_{ij}}{2j} \cos{(jnt)},
\end{equation}
and the integral (\ref{eq:integral}) is zero.

\end{document}